\documentclass[12pt]{article}

\usepackage{caption}
\usepackage{amsmath,amsfonts,amsthm,amssymb,dsfont}
\usepackage{subfigure}
\usepackage{setspace}
\usepackage{Tabbing}
\usepackage{lastpage}
\usepackage{extramarks}
\usepackage{chngpage}
\usepackage[usenames,dvipsnames]{color}
\usepackage{graphicx,float,wrapfig}

\newcommand{\lp}{\left(}\newcommand{\rp}{\right)}

\def\IZ{\rlx\hbox{\sf Z\kern-.4em Z}}
\def\IR{\rlx\hbox{\rm I\kern-.18em R}}
\def\IC{\rlx\hbox{\,$\inbar\kern-.3em{\rm C}$}}
\def\one{\hbox{{1}\kern-.25em\hbox{l}}}

\begin{document}
\numberwithin{equation}{section}

\begin{titlepage}

\begin{center}
{\bf Collective coordinate approximation to the scattering of solitons\\ in modified NLS and sine-Gordon models.}
\end{center}

\vspace{.5cm}

\begin{center}
{H. E. Baron~$^{\star}$\\ and\\ W. J. Zakrzewski~$^{\star}$}

\vspace{.2 in}
\small

\par \vskip .2in \noindent
$^{(\star)}$ Department of Mathematical Sciences,\\
  Durham University, Durham DH1 3LE, U.K.\\
h.e.baron@durham.ac.uk,\\
w.j.zakrzewski@durham.ac.uk

\end{center}

\par \vskip 1 in

\begin{abstract}
We investigate the validity of collective coordinate approaximations to the scattering of solitons
in several classes of models in (1+1) dimensional field theory models. We look at models
which are deformations of the sine-Gordon (SG) or the nonlinear Schr\"odinger (NLS) model as they
posses solitons which are topological (SG) or non-topological (NLS). Our deformations
preserve their topology (SG),  but changes their integrability properties, either 
completely or partially (models become `quasi-integrable').

As our collective coordinate approximation does not allow for the radiation of energy out the system we look also, in some detail, at how good this approximation is for models which are `quasi-integrable'. 
 Our results are based on the studies 
of the interactions and scatterings in two soliton systems.

Our results show that a well chosen approximation, based on geodesic motion {\it etc}, works amazingly well in all cases where it is expected to work. This is true for the trajectories of the solitons and even for their quasi-conserved (or not) charges. The only time the approximation is not very reliable (and even then the qualitative features are reasonable, but some details are not reproduced 
well) involves the processes when the solitons, during their scattering, can come very close
together (within one width of each other).
\end{abstract}
\end{titlepage}

\section{Introduction}
Solitons are special solutions of non-linear evolution equations that maintain their shape and energy as they propagate; and when they interact with each other no energy is radiated and a long time after the interaction the only effect is a shift in their positions compared to the positions they would have had if the scattering had not taken place. In (1+1) dimensional models this behaviour is a result of an infinite number of conserved charges constraining the soliton dynamics, and these conserved charges are a consequence of the integrability of the field theory. However, many non-integrable field theories describe processes which are similar to those seen in integrable field theories, for example the scattering of soliton-like structures which do not significantly alter their shape and radiate very little energy during the scattering. These processes prompted the formulation of a concept of quasi-integrability for field theories in (1+1) dimensions in \cite{Zakr2011}, \cite{Zakr2012} and \cite{Zakr2014}. In these papers modifications of the integrable sine-Gordon and non linear Schr\"odinger models were analysed and found to possess,  sometimes,  characteristics similar to their non-perturbed counterparts such as an infinite number of asymptotically conserved charges (\textit{i.e.} charges $Q(t)$ for which $Q(t\rightarrow-\infty)=Q(t\rightarrow\infty)$). The possession of these characteristics is independent of whether the model is topological (like the sine-Gordon model) or non-topological (like the NLS model). However, unlike in the integrable theories, the possession of these characteristics was shown to be dependent on the field configurations and only those configurations with additional symmetries demonstrated integrability-like properties. It was found that these charges are asymptotically conserved in the scattering of two-soliton configurations when the fields are eigenstates of the space-time reflection around a point for some choice of $(x_{\Delta},t_{\Delta})$, given by
\begin{equation}
\label{parity}
P: \,\,\,\,\,\,\,\,(\tilde{x},\tilde{t})\rightarrow (-\tilde{x},-\tilde{t}),\,\,\,\,\,\,\,\,\,\,\,\,\,\,\,\,\,\,\,\,\,\text{with}\,\,\,\,\,\,\, \tilde{x}=x-x_\Delta \,\,\,\, \text{and} \,\,\,\, \tilde{t}=t-t_\Delta.
\end{equation}

In this paper we compare two approaches to investigating soliton behaviour in different systems; namely, the collective coordinate approximation and the full numerical simulation. The collective coordinate approximation has previously been identified as a useful tool when considering perturbed soliton-bearing equations \cite{Sanc1998}. Our work is an extension of our previous study in which we compared these two approaches for bright soliton collisions in the integrable NLS model \cite{Baro2014}. Here we generalise the method used in \cite{Baro2014} to the modified models considered in \cite{Zakr2012} and \cite{Zakr2014}; the quasi-integrability of these models allows us to also compare the scattering anomalies (see equation (1.1) of \cite{Zakr2012}) in addition to the trajectories of the two solitons and this enables us to investigate more thoroughly the extent to which the collective coordinate approximation is useful. For the modified NLS model we are able to return the system to the integrable NLS which we have considered previously by taking our deformation parameter $\epsilon$ equal to zero; this acts as a check of our numerics, particularly in the calculation of the effective Lagrangian which can be computed analytically only in the case $\epsilon=0$.

This paper is organised as follows: in section 2 we  give details of the two approaches we use to investigate the soliton behaviour. In section 3 we describe our modified NLS model and construct an approximation ansatz for a two-soliton solution of this model. We then compare the results of the two approaches in this system starting with the $\epsilon=0$ case, when the model returns to the integrable NLS, before discussing the more general $\epsilon\neq 0$ case. In section 4 we describe our modified sine-Gordon model and construct a two-soliton approximation ansatz in this model; we then compare the two approaches in this system. Finally we present our conclusions in section 5.

\section{The two approaches}

\subsection{Collective coordinate approximation}
\label{cca_sec}
The idea of using collective coordinates to describe the main features of the scattering of solitons and other extended structures is quite old. An early work in this area was performed by Thiele \cite{Thie1973} who suggested an equation which describes the dynamics of solitons. This was further generalised by Tretiakov and others \cite{Tret2008} to a larger systems of variables (see also a recent paper \cite{Mert2010} which uses such an approach to discuss perturbed NLS equations). In our work we use the approach of Manton \cite{book}, \cite{collective}, which can be used to model the dynamics of solitons in a wide variety of systems and generally reproduces the results of the full simulations in such systems with good accuracy. Any collective coordinate approach reduces an infinite-dimensional problem to a finite dimensional system described by a set of ODEs and so is much quicker to implement than a full numerical simulation.  However, the important issue here 
involves choosing the variables that describe, as accurately as possible, the full problem (see for example \cite{Dawe2013}). The main observation that helps here is the realisation that for a system that possesses free parameters a slow change of these parameters has only a minimal effect on the total energy of the full system and so may be a good approximation to its slow dynamics.
Thus one starts with a static solution $\psi(x,q_1,...,q_n)$. The energy of this solution is independent of the parameters but changing the field configuration can only increase the energy, so in the field space there are low energy valleys in the direction of the parameters of the solution. Consider now moving solitons. For small velocities of the solitons the motion is easiest along the valleys described by the parameters of the static solution (as then the increase of the energy is only due to the kinetc energy associated with this change which for very slow changes is very small). Thus, it makes sense to approximate the dynamics of slow moving solitons by allowing these parameters to vary in time, \textit{i.e.} $q_i=q_i(t)$, and assume that these parameters describe most of the solitons' dynamics.

However, such an approximation neglects other modifications of the fields and, in particular, all radiative corrections to the solitons and so is valid only  for very slow motions and when such corrections are small.
In practice, we want to use this approximation not only for infinitesimally small velocities and so
we include some extra parameters and then check whether their inclusion improves the approximation.
Here we are often helped by the physics of the problem, and in integrable models we are sometimes guided by considerations
of their conserved quantities. 

To describe the dynamics of the collective coordinates we proceed as follows \cite{book}.
We start with an approximation ansatz whose form is based on the stationary solution with a suitable choice of parameters which become collective coordinates; these coordinates generally describe physical properties of the soliton such as position, height, {\it etc}. This ansatz is then substituted into the Lagrangian density of the system to obtain a coupled system of ODEs for the coordinates. Solving these ODEs describes the time evolution of the coordinates, which in turn tells us how the field evolves in time. In some cases, and sometimes with further simplifying assumptions, the equations of motion for the collective coordinates can be solved analytically; such is the case in \cite{Zou1994}. In our work the equations of motion need to be solved numerically and for this we use a 4th order Runge-Kutta method.

\subsection{General comments on the numerical approach}

The work described in this paper has involved some numerical methods. They included performing `full numerical
simulations' and the calculations involving collective coordinates. 
The two approaches were then compared to each other to assess the validity of the collective coordinate approximations. Both approaches  used the 4th order Runge-Kutta method of simulating the time evalution, and we used this method both for the modified NLS equation and for the modified Sine-Gordon models. 

For the full simulations the implementations were different in these two classes of models as the NLS equation involved only first derivatives with respect to time and the phase of the complex field $\psi$ performed fast rotation with the increase of time, while the modified Sine-Gordon model involved only a nonlinear wave motion. 

Thus, in the NLS case we chose to perform the simulation in a rotating frame ({\it i.e.} we had to go to the frame in which the phase rotation involved the additional dynamical variation relative to this global rotation). The global rotation was calculated at each value of time and the equation was transformed to that frame. Therefore, in this frame, the further phase variation was small and it was only due to the dynamics of the system of solitons.
Consequently, for a given time and position steps of the program, the changes of the derivatives of $\psi$ were kept small. 

Our approach is a standard procedure for such fields and more discussion of it use in the NLS case can be found in \cite{Zakr2012} where it is shown that it had worked very well in this case. The numerical errors were negligible and the results of our simulations were the essentially the same when we tested the method by varying a little the parameters of our numerical approach. To obtain reliable simulations we experimented by using various lattice sizes, various numbers of points {\it etc}. until we were satisfied that we could `trust' our results; {\it i.e.} when the numerical errors were very small and so insignificant. 

Then we performed many simulations  as described in this paper. In fact, most of the results we are presenting here were obtained on a lattice involving $N=5000$ grid  points   with lattice spacing $dx=0.01$. As in \cite{Zakr2012} the initial configurations involved two one soliton fields, with solitons placed  at $x=\pm x_0$ (as discussed in the text below) and with the fields tied together at $x=0$. Luckily, at small values of $x$ the values of the fields are very close to each other and so the numerical errors due to this joining procedure were negligible (in fact we even smoothed the fields there over 3 lattice points). 

For our calculations, as the equations were first order in time derivative, we had to take a small time step. We varied this too and have found that we could trust our results when $dt=0.00002$ or smaller. Most of our results reported in this
paper were obtained with this value of $dt$.

In the modified sine-Gordon model the equations are of the wave type and so the numerical calculations are simpler than in the NLS case discussed above. We used the fixed boundary conditions with $N=10001$ points, with the lattice spacing $dx=0.01$ and $dt=0.0001$. We absorbed the energy at the boundaries but, in fact, very little energy ever reached the boundaries as the scatterings were very elastic.

In the collective coordinate approximation the ODE's are solved numerically with lattice spacing $dx=0.01$ and $dt=0.005$ (the simulations were run with various values of $dx, dt$ and we were satisfied with the accuracy of the results using the given values). In addition to this we were required to integrate numerically over $x$ in order to obtain the effective Lagrangian from the Lagrangian density. We compared our numerical integrations with the integrations performed analytically (which could only be achieved when the models were integrable, \textit{i.e.} when $\epsilon=0$) and found the results to be sufficiently accurate for our chosen value of $dx=0.01$.  

\section{The modified NLS model}
Here we consider the Lagrangian for a non-relativistic complex scalar field in (1+1) dimensions
\begin{equation}
\label{lag}
\mathcal{L}= \int dx\; \frac{i}{2}\left( \psi^\ast\partial_t \psi - \psi \partial_t \psi^\ast\right) - \partial_x\psi^\ast \partial_x\psi - V \left( | \psi|^2 \right) .
\end{equation}
This Lagrangian has an internal symmetry $\psi \rightarrow e^{i \alpha} \psi$ for $\alpha\equiv \text{constant}$.
The equations of motion are 

\begin{equation}
\label{eom}
i\partial_t \psi = -\partial_x^2\psi +\frac{\delta V}{\delta |\psi|^2} \psi,
\end{equation}
together with its complex conjugate.

The equation (\ref{eom}) admits an anomalous zero curvature representation with the connection given by
\begin{eqnarray}
A_x&=&-i T^{1}_{3}+\sqrt{|\eta |}\, \psi^*\, T^0_++\sqrt{|\eta |} \, \psi \, T^0_- ,\\
A_t&=& iT^2_3+i\frac{\delta  V}{\delta |\psi|^2} T_3^0-\sqrt{|\eta |}\,\left( \psi^* \, T^1_++\psi \, T^1_-\right)-i\sqrt{|\eta |}\,\left( \partial_x \psi^* \,T^0_+-\partial_x \psi \,T^0_- \right), \nonumber
\end{eqnarray}
where $T^n_i$, $i=3,+,-$ and $n$ integer, are generators which satisfy the $SL(2)$ loop algebra commutation relations (for more results see \cite{Zakr2012}) and can be realised in terms of the finite $SL(2)$ algebra generators as $T^n_i\equiv \lambda^n T_i$, where $\lambda$ is an arbitrary complex parameter. The curvature of this connection is
\begin{eqnarray}
\label{curvature}
\partial_t A_x-\partial_x A_t +\left[ A_x, A_t\right] &=&X T_3^0+i \sqrt{|\eta |}\,\left( -i\partial_t \psi^*+\partial^2_x \psi^*-\psi^* \frac{\delta  V}{\delta |\psi|^2}\right) T^0_+\\ \nonumber
&-& i\sqrt{|\eta |}\,\left( i\partial_t \psi+\partial^2_x \psi -\psi \frac{\delta  V}{\delta |\psi|^2}\right) T^0_-
\end{eqnarray}
where $X$ is the anomaly given by
\begin{equation}
\label{ano}
X\equiv -i \partial_x\left(\frac{\delta  V}{\delta |\psi|^2}-2\eta \,|\psi|^2 \right).
\end{equation}
This curvature simplifies to $\partial_t A_x-\partial_x A_t +\left[ A_x, A_t\right] =X T_3^0$ when the equations of motion (\ref{eom}) are imposed. For the NLS potential, $V_{{\text NLS}}=\eta \, |\psi|^4$, the anomaly $X$ vanishes and it is this vanishing of the curvature which makes the theory integrable. For a general potential we can carry out the abelianization technique of the integrable field theories, for full details see \cite{Zakr2012}, gauge transforming the connection such that the curvature (\ref{curvature}) becomes
\begin{equation}
\label{new_curvature}
\partial_t a_x^{(3,-n)} -\partial_x a_t^{(3,-n)}=X \alpha^{(3,-n)}; \,\,\,\,\,\,\,\,\,\,\,\,\,\,\,\,\, n=0,1,2,...
\end{equation}
Explicit expressions for the first few $a_x^{(3,-n)}$ and $\alpha^{(3,-n)}$ are given in appendix A in \cite{Zakr2012}. In the example that we consider $a_t^{(3,-n)}$ satisfies the boundary condition $a_t^{(3,-n)}(x=\infty)=a_t^{(3,-n)}(x=-\infty)$ and so from (\ref{new_curvature}) we have an infinite number of anomalous conservation laws:
\begin{equation}
\label{con_laws}
\frac{d Q^{(n)}}{dt}=\beta_n; \,\,\,\,\,\,\,\,\,\,\,\, \text{with}\,\,\,\,\,\,\, Q^{(n)}=\int^{\infty}_{-\infty}dx\, a_x^{(3,-n)}; \,\,\,\,\,\,\,\,\,\,\,\, \text{where}\,\,\,\,\,\,\,\beta_n=\int^{\infty}_{-\infty}dx\, X\, \alpha^{(3,-n)}
\end{equation}
for $n=0,1,2,...$.
It is clear that when the potential corresponds to the NLS potential, \textit{i.e.} $V_{\text{NLS}}=\eta\, |\psi|^4$, the anomaly $X$ given in (\ref{ano}) vanishes and so does $\beta_n$. Therefore the theory with the potential $V_{\text{NLS}}$ is integrable as it has an infinite number of conserved charges $Q^{(n)}$.

In our modified model we use a perturbation of the NLS potential as in \cite{Zakr2012}
\begin{equation}
\label{pot}
V=\frac{2}{2+\epsilon}\eta \left( |\psi|^2\right)^{2+\epsilon}
\end{equation}
so that it returns to the unperturbed NLS potential in the case $\epsilon=0$.

As shown in \cite{Zakr2012}, for $\eta<0$, this model has a one-soliton solution given by
\begin{equation}
\label{sol_bright}
\psi = \left(\sqrt{\frac{2+\epsilon}{2 \, |\eta|}}\frac{b}{ \cosh{\left[(1+\epsilon) \, b \, \left( x-vt-x_0 \right)\right]}}\right)^{\frac{1}{1+\epsilon}}e^{i\left[ \left(b^2 - \frac{v^2}{4}  \right)t+\frac{v}{2}x \right]},
\end{equation}
where $b$, $v$ and $x_0$ are real parameters of the solution.
 
For convenience we rewrite $\psi$ in terms of new fields $R$ and $\varphi$ as $\psi\equiv\sqrt{R} e^{i \frac{\varphi}{2}}$. When the fields transform under the parity defined in \eqref{parity} as
\begin{equation}
\label{psi_trans}
P: \,\,\,\,\,\,\,\,\,\,\,\,\,\,\,\,\, R\rightarrow R; \,\,\,\,\,\,\,\,\,\,\,\,\,\,\,\,\,\varphi \rightarrow -\varphi + \text{constant}.
\end{equation}
then $\alpha^{(3,-n)}$ is odd under P (see appendix A in \cite{Zakr2012}), and $X$ is even under P. Therefore for field configurations which transform as in (\ref{psi_trans}) we have $\beta_n=0$ and the system has an infinite number of asymptotically conserved charges, \textit{i.e.}
\begin{equation}
Q^{(n)}(t=+\infty)=Q^{(n)}(t=-\infty).
\end{equation}

\subsection{The two-soliton configuration for modified NLS}

Here we construct a set of collective coordinates for the study of the scattering of two solitons with $\eta=-1$ in the NLS system with our modified potential. We use a natural extension of our approximation ansatz in \cite{Baro2014} and so we take our approximation ansatz for two solitons in the modified NLS system to be given by 
\begin{equation}\label{ansatz}
\psi=\psi_1+\psi_2=\varphi_1 e^{i\theta_1}+\varphi_2e^{i\theta_2} 
\end{equation}
where
$$
\varphi_1=\left(\sqrt{\frac{2+\epsilon}{2}} \frac{a_1 (t)}{\cosh{\left[(1+\epsilon)\, a_1 (t)\, (x+\xi_1 (t))\right]}}\right)^{\frac{1}{1+\epsilon}},  \,\,\,\,\,\, \theta_1=-\frac{\mu_1 (t)}{2} \left( x+ \frac{\xi_1 (t)}{2}\right) + a_1  ^2 (t) \, t +\lambda_1 (t) ,
$$
$$
\varphi_2=\left(\sqrt{\frac{2+\epsilon}{2}} \frac{a_2 (t)}{\cosh{\left[(1+\epsilon)\, a_2(t) \, (x+\xi_2 (t))\right]}}\right)^{\frac{1}{1+\epsilon}}, \,\,\,\,\,\, \theta_2=-\frac{\mu_2 (t)}{2} \left( x+ \frac{\xi_2 (t)}{2}\right) + a_2  ^2 (t) \, t +\lambda_2 (t) ,
$$
and $a_{1,2}(t), \ \xi_{1,2}(t), \ \mu_{1,2}(t)$ and $\lambda_{1,2}(t)$ are our collective coordinates.  This approximation ansatz models two lumps with heights $a_{1,2}(t)$, positions $\xi_{1,2}(t)$, velocities $\mu_{1,2}(t)$ and phases $\lambda_{1,2}(t)$. When the two lumps are far apart they resemble two one-soliton solutions akin to \eqref{sol_bright}.

In the case $\epsilon=0$ the system is integrable and this ansatz is similar to the one we used in \cite{Baro2014} with the additional features of a time dependence in the width of the solitons; also the height, position, velocity and phase of each soliton are allowed to vary independently (whereas previously we insisted that $a_1(t)=a_2(t)$, $\xi_1(t)=-\xi_2(t)$, $\mu_1(t)=-\mu_2(t)$ and $\lambda_1(t)=\lambda_2(t)$). In particular this allows a previously static parameter, the phase difference between the solitons $\delta\equiv \lambda_1-\lambda_2$, to vary in time. These changes have been made based on our observations in \cite{Baro2014} and we later have found that this improved approximation ansatz gives more accurate results for the NLS solitons when compared with our results in \cite{Baro2014}.

For $\epsilon\neq 0$ and $\delta= n \pi,$ where $\,n\in \mathds{Z},$  the approximation ansatz (\ref{ansatz}) transforms under the parity defined in (\ref{parity}) as in (\ref{psi_trans});  thus the field configuration possesses the additional symmetries necessary for the system to be quasi-integrable and has asymptotically conserved charges. 

For  $\epsilon\neq 0$ and $\delta\neq n \pi$  the approximation ansatz does not transform under the parity defined in (\ref{parity}) as required  for quasi-integrability and so the system is non-integrable and there are no constraints on the charges.

\subsection{Implementing the approximation in modified NLS}

In order to proceed with the collective coordinate approximation we insert our approximation ansatz (\ref{ansatz}) into the lagrangian (\ref{lag}) to obtain an effective lagrangian:
\begin{equation}
\mathcal{L}=I_{a_1} \dot{a_1}+I_{a_2} \dot{a_2}+I_{\xi_1} \dot{\xi_1}+I_{\xi_2} \dot{\xi_2}+I_{\mu_1} \dot{\mu_1}+I_{\mu_2} \dot{\mu_2}+I_{\lambda_1} \dot{\lambda_1}+I_{\lambda_2} \dot{\lambda_2}-V,
\end{equation}
where the dot denotes differentiation with respect to time; and the $I$'s and $V$ are functions of $a_{1,2}(t), \ \xi_{1,2}(t), \ \mu_{1,2}(t), \ \lambda_{1,2}(t)$ and $t$. These functions are fully described in appendix \ref{appendix}.

From this effective lagrangian we derive equations of motion as a set of coupled ODEs of the form, (for different choices of $q$):
\begin{equation}
I_{\dot{q}}-\dot{a_1}\frac{\partial I_{a_1}}{\partial q}-\dot{a_2}\frac{\partial I_{a_2}}{\partial q} -\dot{\xi_1}\frac{\partial I_{\xi_1}}{\partial q}-\dot{\xi_2}\frac{\partial I_{\xi_2}}{\partial q}-\dot{\mu_1}\frac{\partial I_{\mu_1}}{\partial q}-\dot{\mu_2}\frac{\partial I_{\mu_2}}{\partial q}-\dot{\lambda_1}\frac{\partial I_{\lambda_1}}{\partial q}-\dot{\lambda_2}\frac{\partial I_{\lambda_2}}{\partial q}+\frac{\partial V}{\partial q}=0,
\end{equation}
where $q$ denotes the collective coordinates $q=a_1, a_2, \xi_1, \xi_2, \mu_1, \mu_1, \lambda_1,\lambda_1$. We decouple these equations and solve them using the 4th order Runge-Kutta method.

\subsection{Results for NLS}
Here we describe the results of our analysis of the scattering of two solitons in our collective coordinate approximation for a range of initial values of the collective coordinates, and compare the results against those given by a full numerical simulation. This allows us to determine the effective range of parameters for our choice of the approximation ansatz. First, we consider the cases when $\epsilon=0$ which correspond to the non-perturbed, integrable NLS; in all our studies we use $\eta=-1$.

\begin{figure}
  \centering
  \subfigure[]{\label{fig:k=0}\includegraphics[trim = 2cm 2cm 12cm 9.8cm, width=0.45\textwidth]{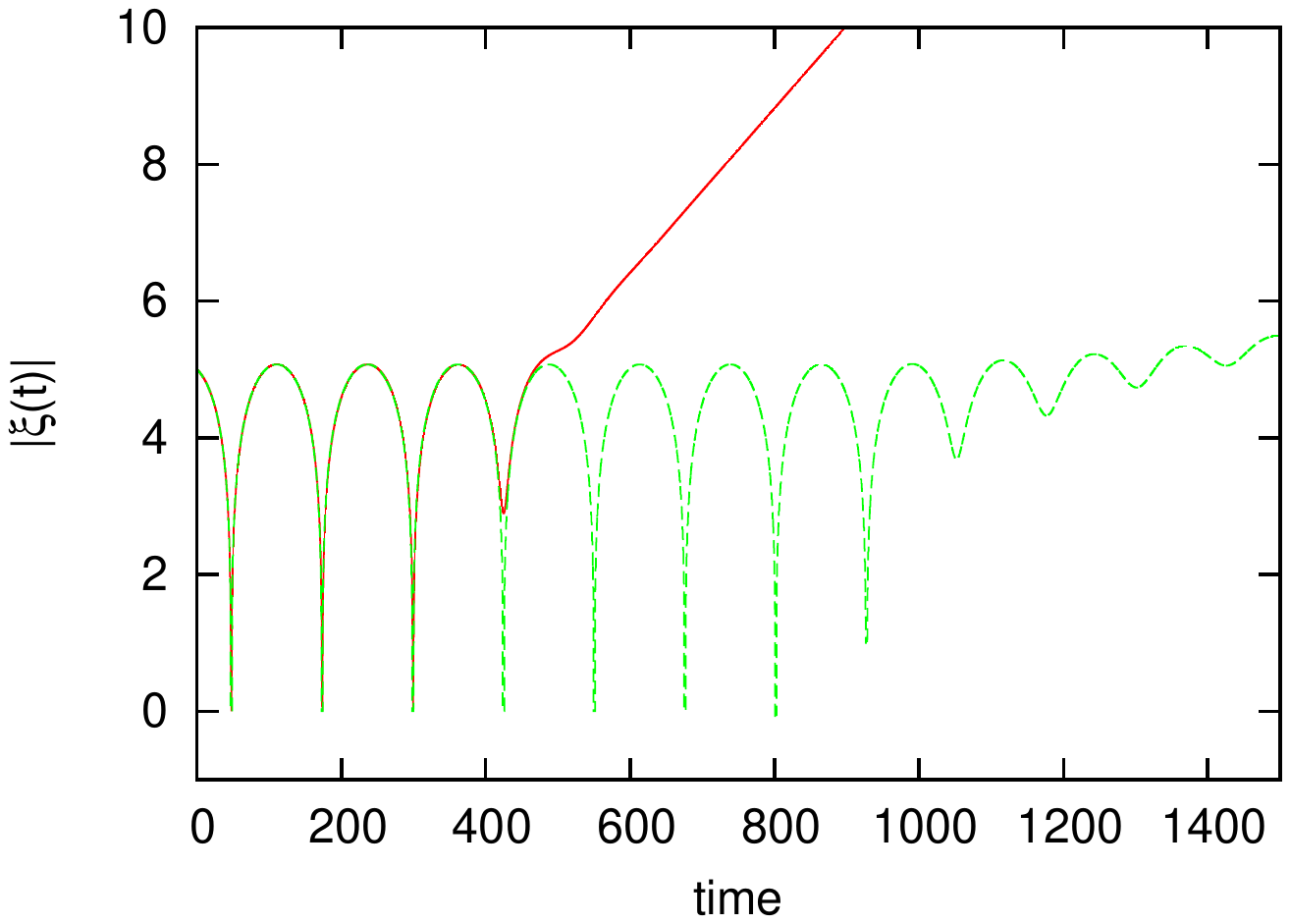}}                
  \subfigure[]{\label{fig:k=05}\includegraphics[trim = 2cm 2cm 12cm 9.8cm, clip, width=0.45\textwidth]{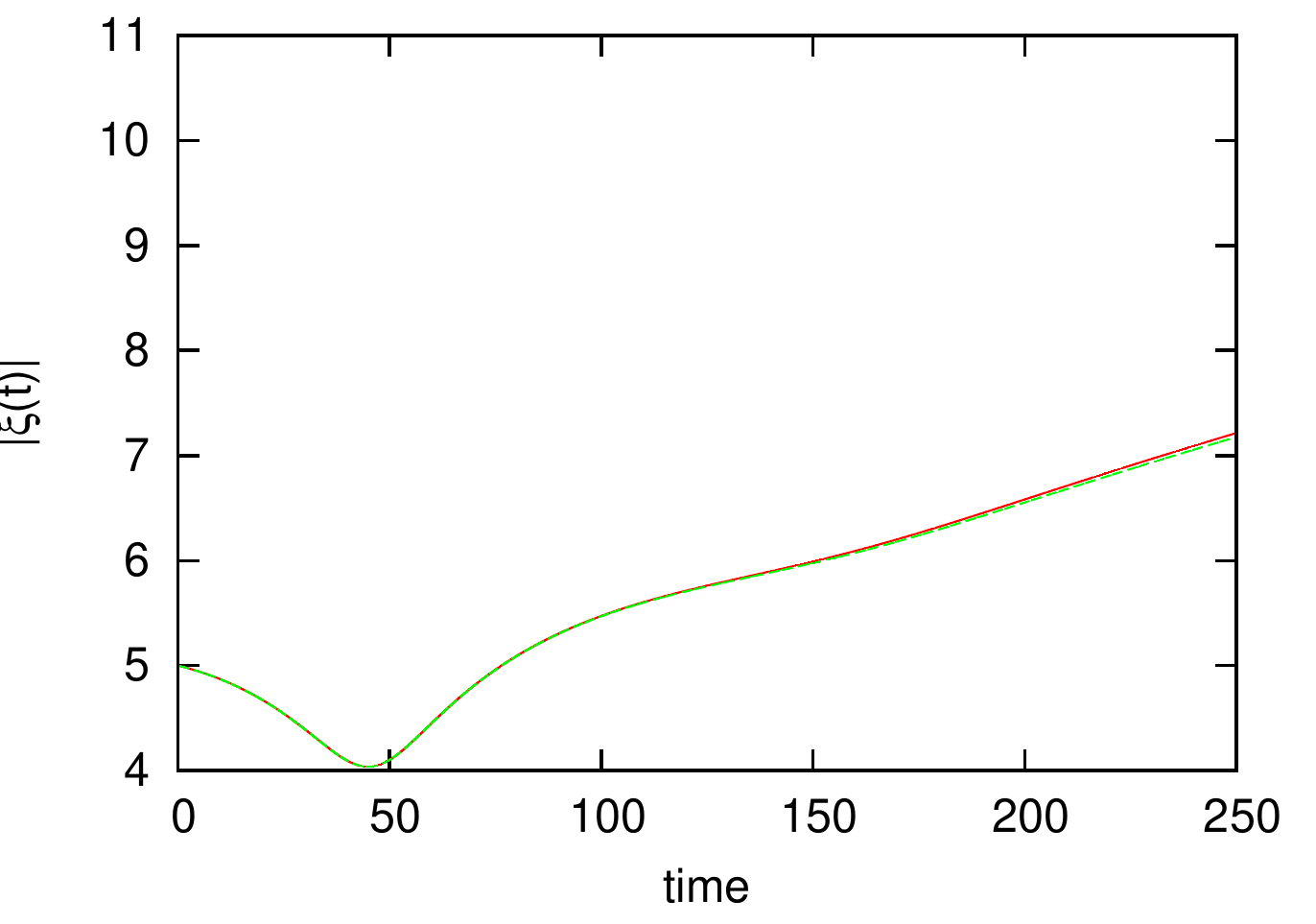}}
  \subfigure[]{\label{fig:k=1}\includegraphics[trim = 2cm 2cm 12cm 9.8cm, clip, width=0.45\textwidth]{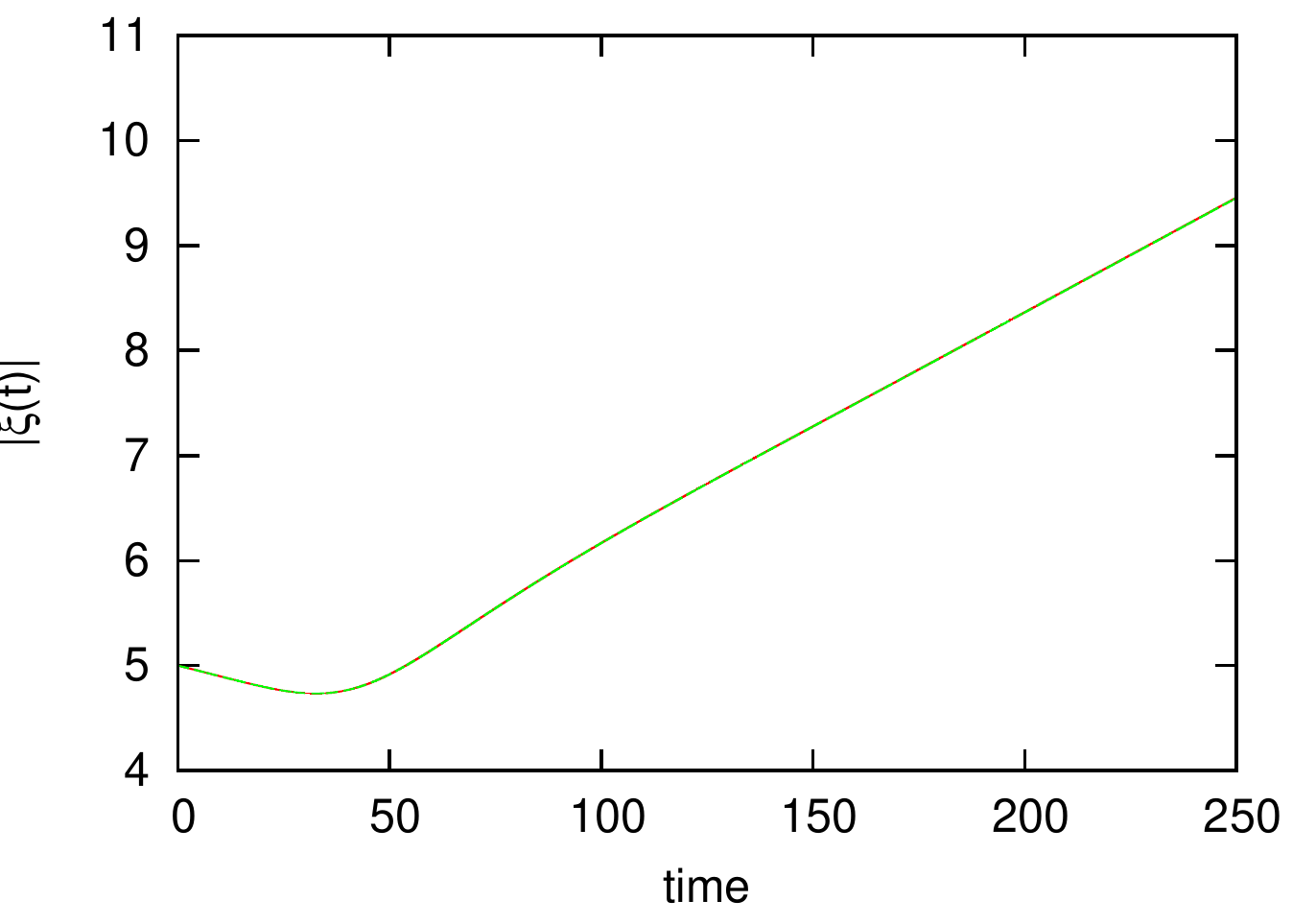}}
    \subfigure[]{\label{fig:k=15}\includegraphics[trim = 2cm 2cm 12cm 9.8cm, width=0.45\textwidth]{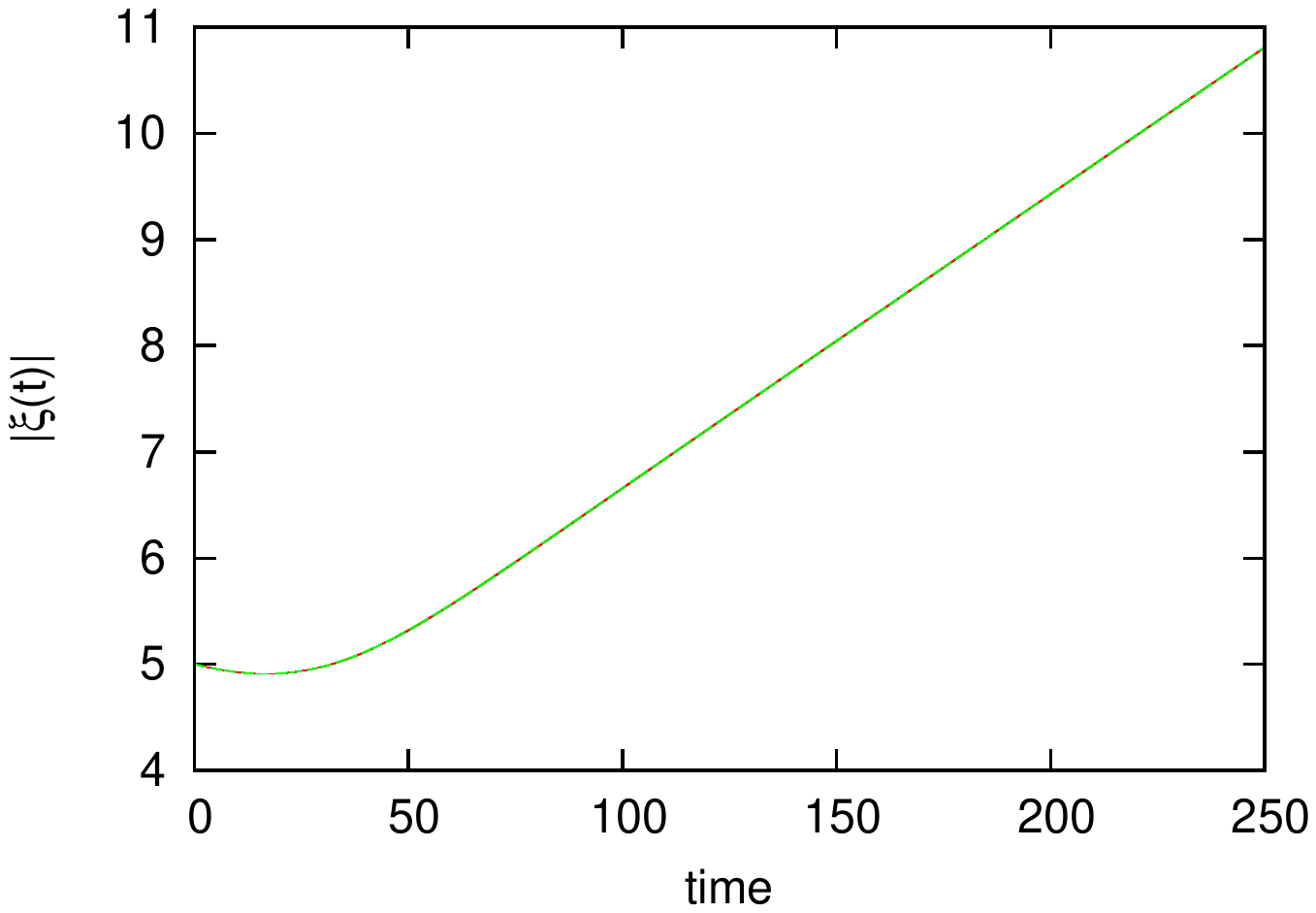}}                
    \subfigure[]{\label{fig:k=2}\includegraphics[trim = 2cm 2cm 12cm 9.8cm, clip, width=0.45\textwidth]{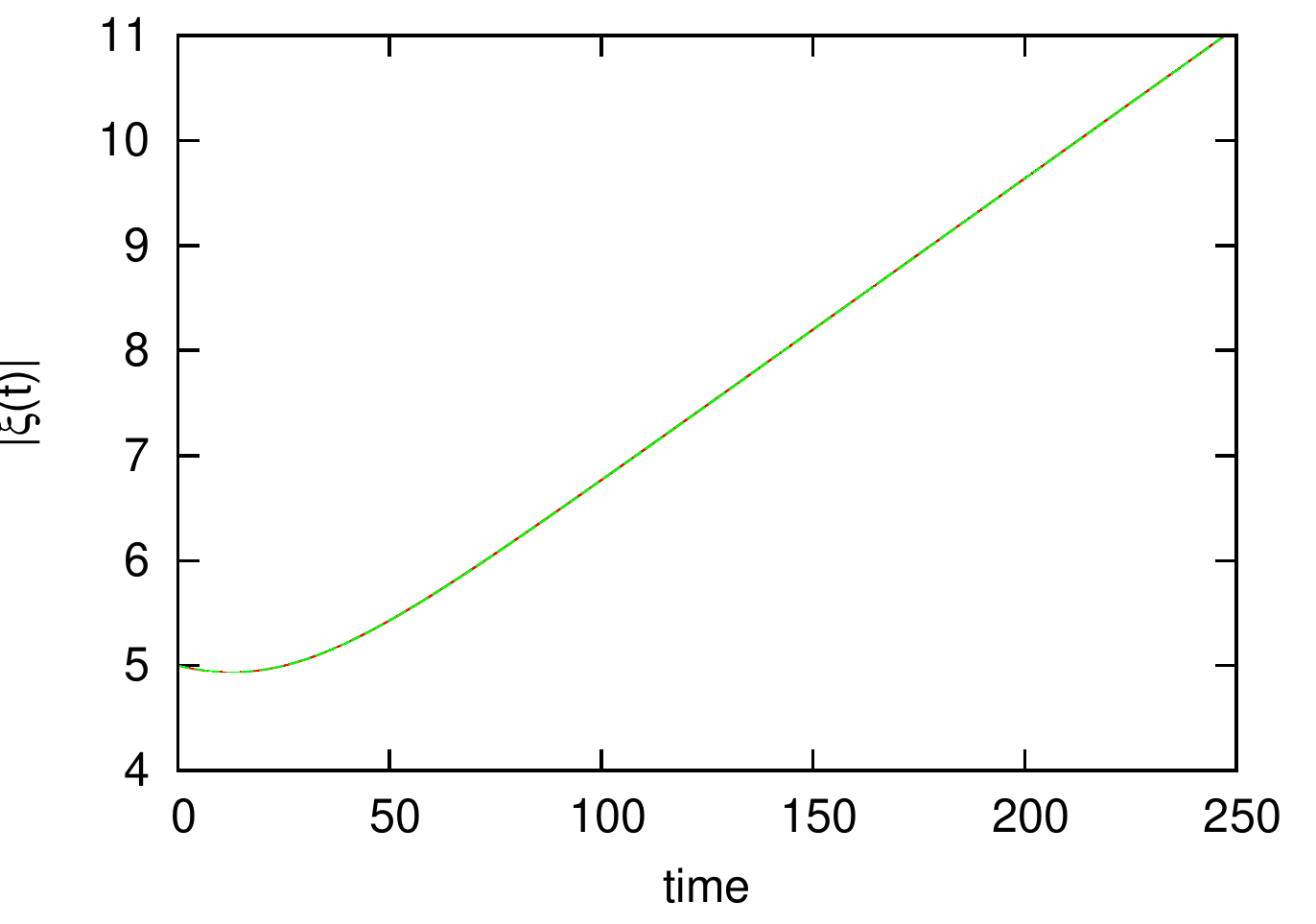}}
  \caption{The distance of a soliton from the centre of mass of a system with time. The system consists of two solitons initially placed at $\pm 5$ and sent towards their centre of mass with an initial velocity $v=0.01$. Initial height/width parameter of each soliton is $1$ and the initial phase difference between them is: (a) $\delta=0$, (b) $\delta=\frac{\pi}{4}$, (c) $\delta=\frac{\pi}{2}$, (d) $\delta=\frac{3\pi}{4}$,(e) $\delta=\pi$. For each plot the solid line has been obtained using the collective coordinate approximation and the dashed line is the result of the full simulation (these are often coincident).}
  \label{fig:e=0}
\end{figure}

As shown in our previous work \cite{Baro2014} the solitons' scattering is highly dependent on the relative phase between them, \textit{i.e.} $\delta\equiv \lambda_1-\lambda_2$; so initially we compare the solitons' dynamics between the collective coordinate approximation and full numerical simulation for a range of $\delta$. Figure \ref{fig:e=0} compares the trajectories given by  the collective coordinate approximation and those given by the full numerical simulation for solitons with initial position $\xi_1=-5, \, \xi_2=5$; initial velocity $\mu_1=0.01, \, \mu_2=-0.01$; initial height/width parameter $a_1=a_2=1$ and initial phase difference $\delta=0, \frac{\pi}{4}, \frac{\pi}{2}, \frac{3\pi}{4}, \pi$ (the results are symmetric around $\pi$ and periodic in $2\pi$). 
 This figure shows that the collective coordinate approximation is remarkably accurate for most values of $\delta$, with both approaches producing almost identical trajectories whenever $\delta\neq 0$ (making it difficult to distinguish all the lines). However, in the case $\delta=0$ the solitons in the collective coordinate approximation break away from oscillating around each other much earlier than in the full simulation. To test this result we reran the collective coordinate approximation simulations with a variety of $dt$'s and found that there are small differences in the trajectories of the solitons due to numerical effects (\emph{i.e.} with larger $dt$ the solitons come together fractionally earlier and ultimately repel slightly earlier). However, the qualitative results remain the same, and we believe that the main cause for any disagreement between the collective coordinate approximation and full simulation is because in the full simulation the solitons deform one another away from the form given by \eqref{ansatz} when they are in close proximity and the collective coordinate approximation does not allow such a deformation.
 
Next we consider the effect of the initial velocity on the accuracy of the collective coordinate approximation. Figure \ref{fig:e=0,vv} compares the trajectories given by  the collective coordinate approximation and those given by the full numerical simulation for solitons with initial position $\xi_1=-5, \, \xi_2=5$; initial height/width parameter $a_1=a_2=1$; initial phase difference $\delta=\frac{\pi}{4}$ and initial velocity $\mu_1=0.1, \mu_2=-0.1$ and $\mu_1=0.2, \mu_2=-0.2$. The effect of the initial velocity on the accuracy of the collective coordinate approximation is difficult to gauge in full generality as varying the initial velocity changes the amount of time the solitons spend close together during their interaction which, as we have already surmised, affects the accuracy of the approximation. Figure \ref{fig:e=0,vv} shows that, as expected, increasing initial velocity decreases the accuracy of the approximation slightly, though for solitons which do not spend much time close together the approximation is still very good up to an initial velocity of at least $v=0.2$. As the collective coordinate approximation assumes slow moving solitons (see section \ref{cca_sec}) our results show that the approximation is extremely reliable for low velocity and more reliable for high velocities than could have been expected.
 
 \begin{figure}
  \centering
  \subfigure[]{\label{fig:v=0.1}\includegraphics[trim = 2cm 2cm 12cm 9.8cm, width=0.45\textwidth]{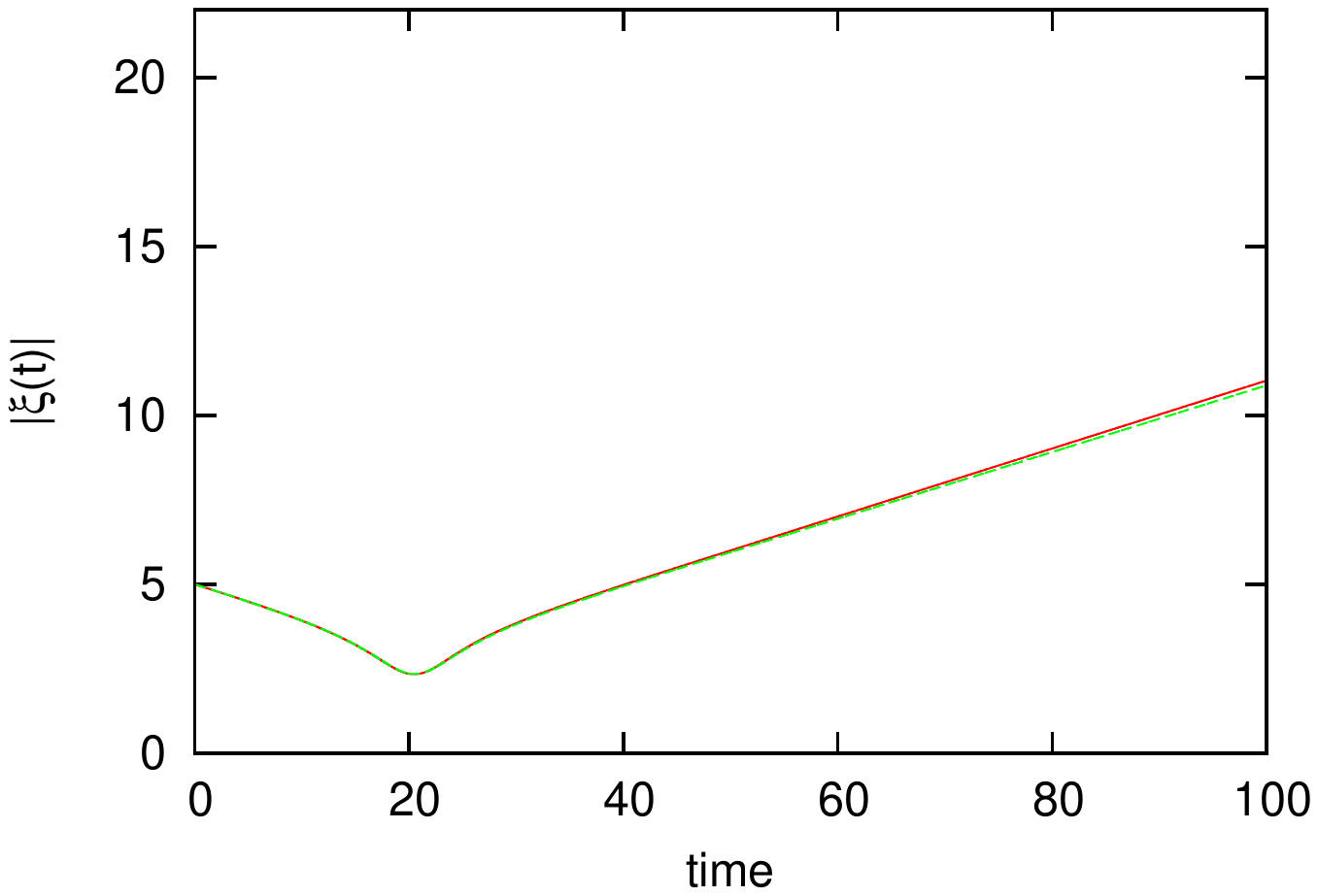}}                
  \subfigure[]{\label{fig:v=0.2}\includegraphics[trim = 2cm 2cm 12cm 9.8cm, clip, width=0.45\textwidth]{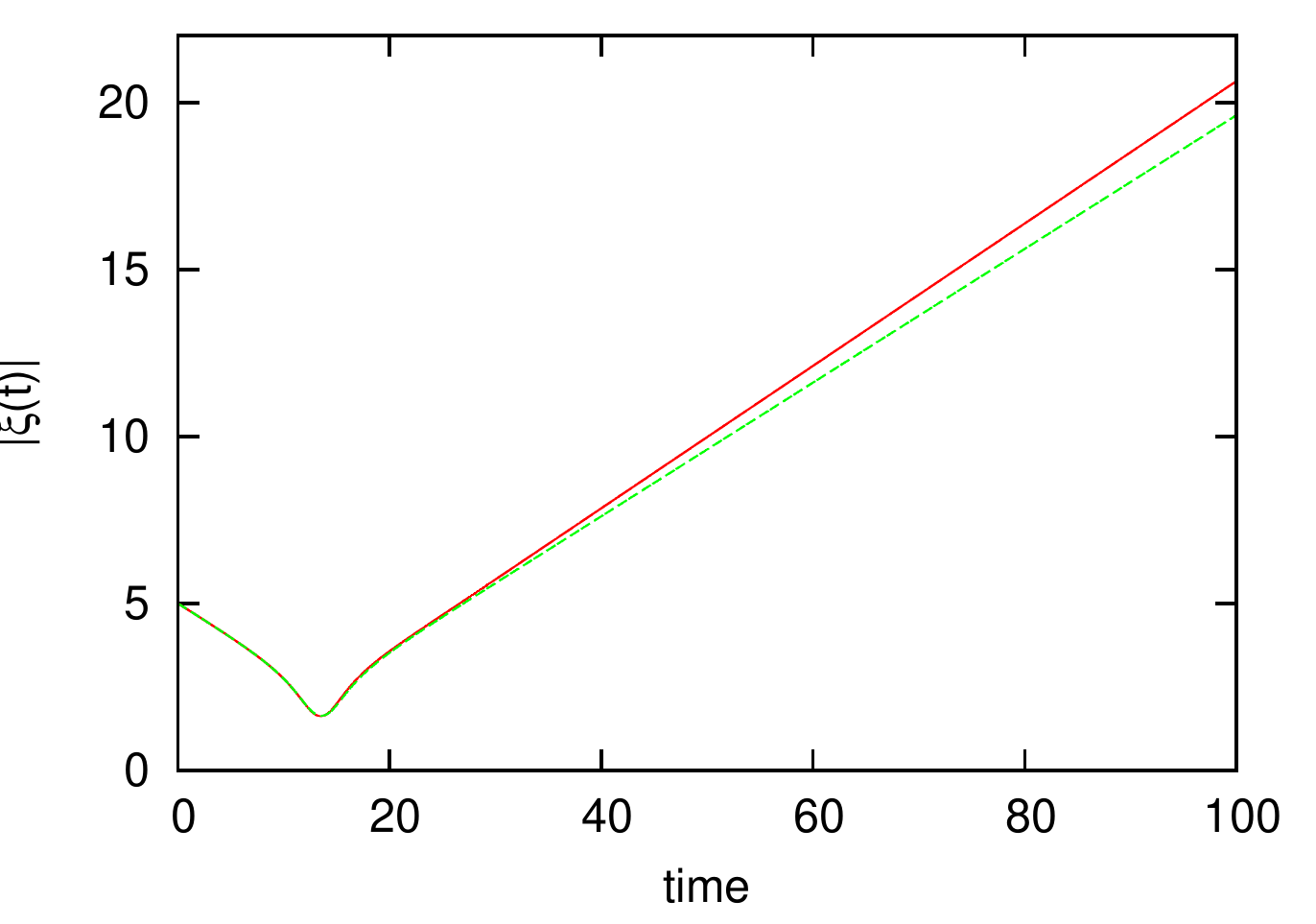}}
  \caption{The distance of a soliton from the centre of mass of a system with time. The system consists of two solitons initially placed at $\pm 5$, with initial height/width parameter of $1$ and the initial phase difference between them of $\delta=\frac{\pi}{4}$. The solitons move towards the centre of mass with initial velocity (a) $v=0.1$, and (b) $v=0.2$. For each plot the solid line describes the outcome obtained in the collective coordinate approximation and the dashed line shows the result of the full simulation (these are often coincident).}
  \label{fig:e=0,vv}
\end{figure}

\subsection{Results for modified NLS}
\begin{figure}
  \centering
    \subfigure[]{\label{fig:e=006,k=0}\includegraphics[trim = 2cm 2cm 12cm 9.8cm, width=0.45\textwidth]{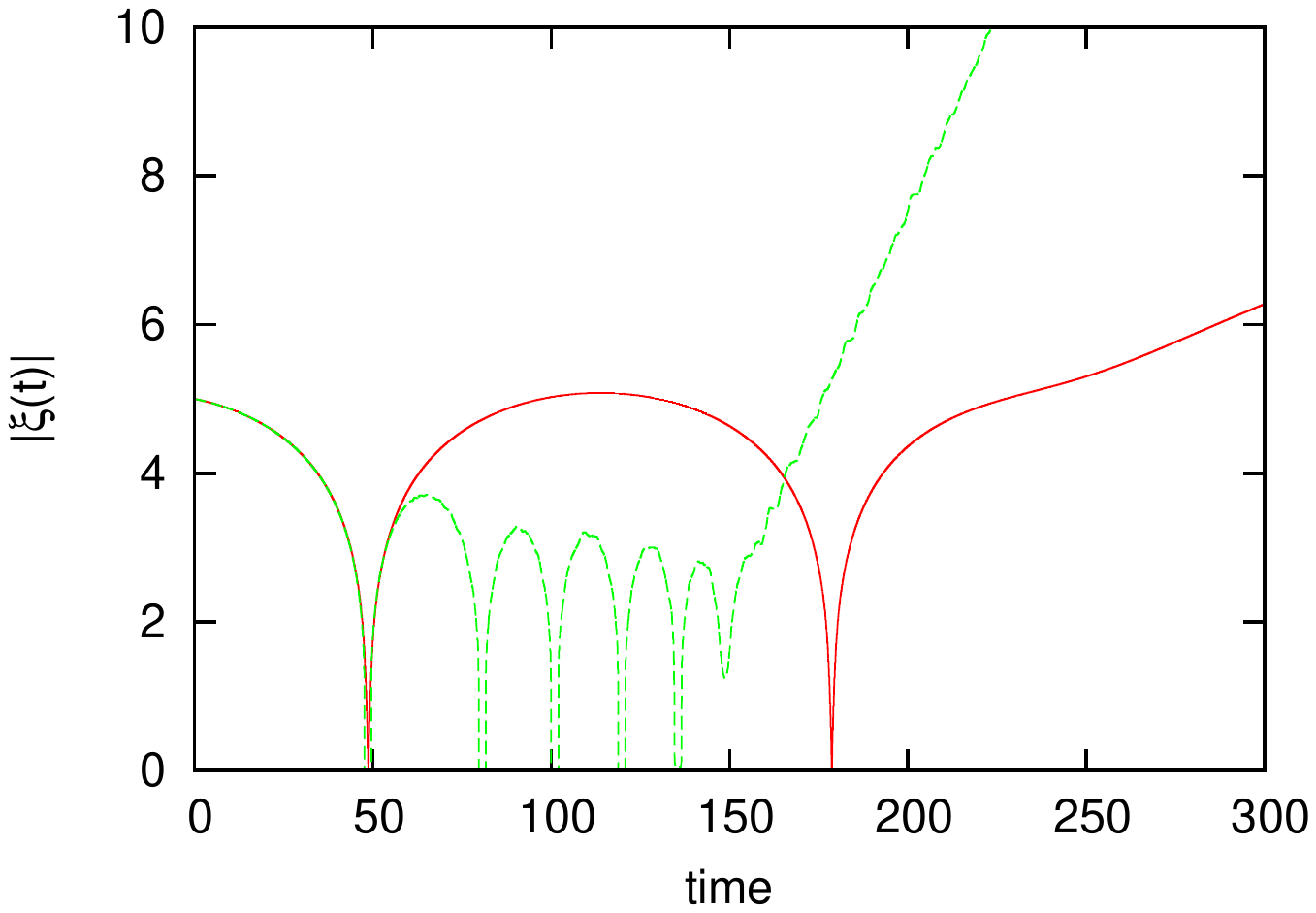}}                
  \subfigure[]{\label{fig:e=-006,k=0}\includegraphics[trim = 2cm 2cm 12cm 9.8cm, clip, width=0.45\textwidth]{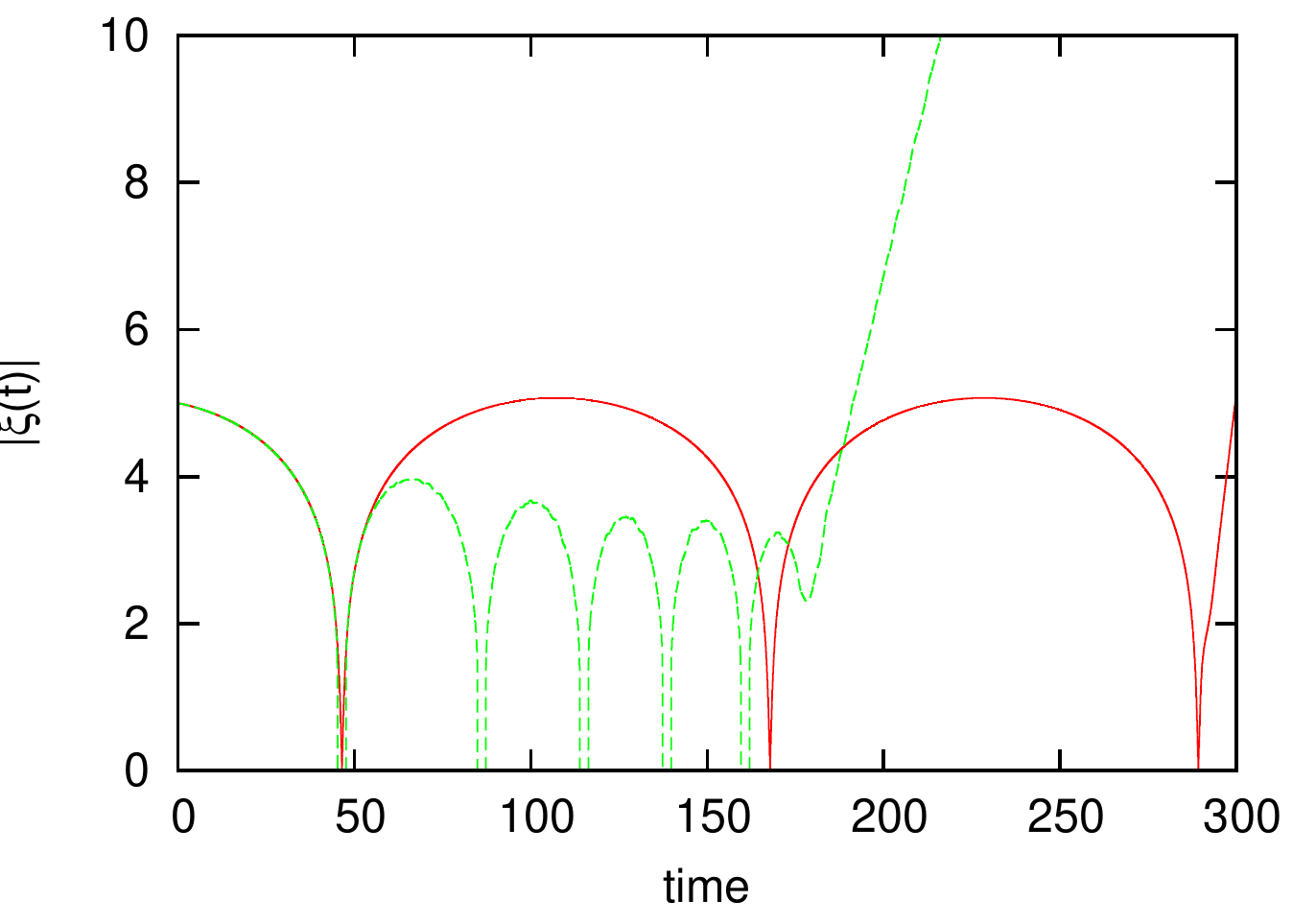}}
    \subfigure[]{\label{fig:e=006,k=0.5}\includegraphics[trim = 2cm 2cm 12cm 9.8cm, width=0.45\textwidth]{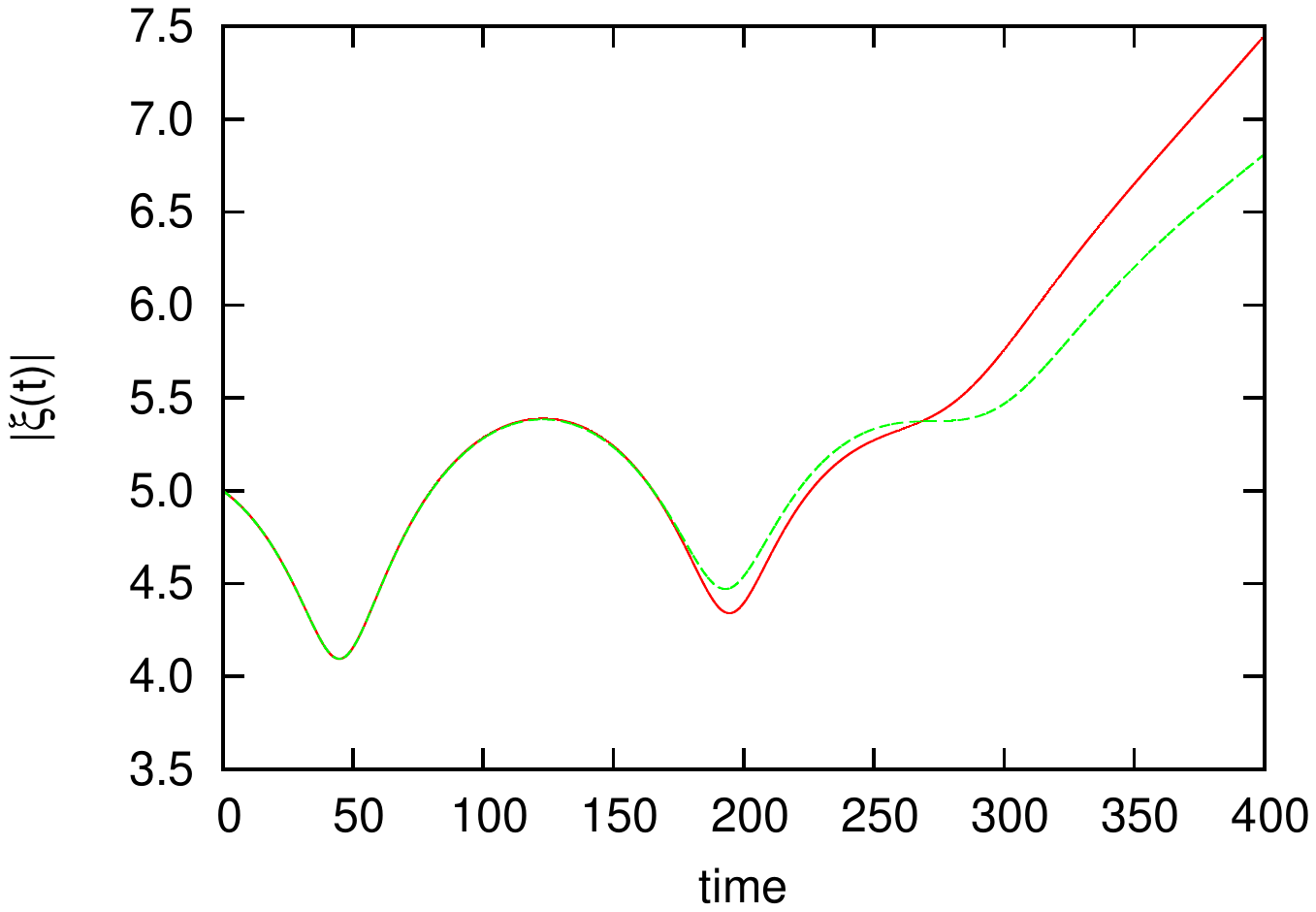}}                
  \subfigure[]{\label{fig:e=-006,k=0.5}\includegraphics[trim = 2cm 2cm 12cm 9.8cm, clip, width=0.45\textwidth]{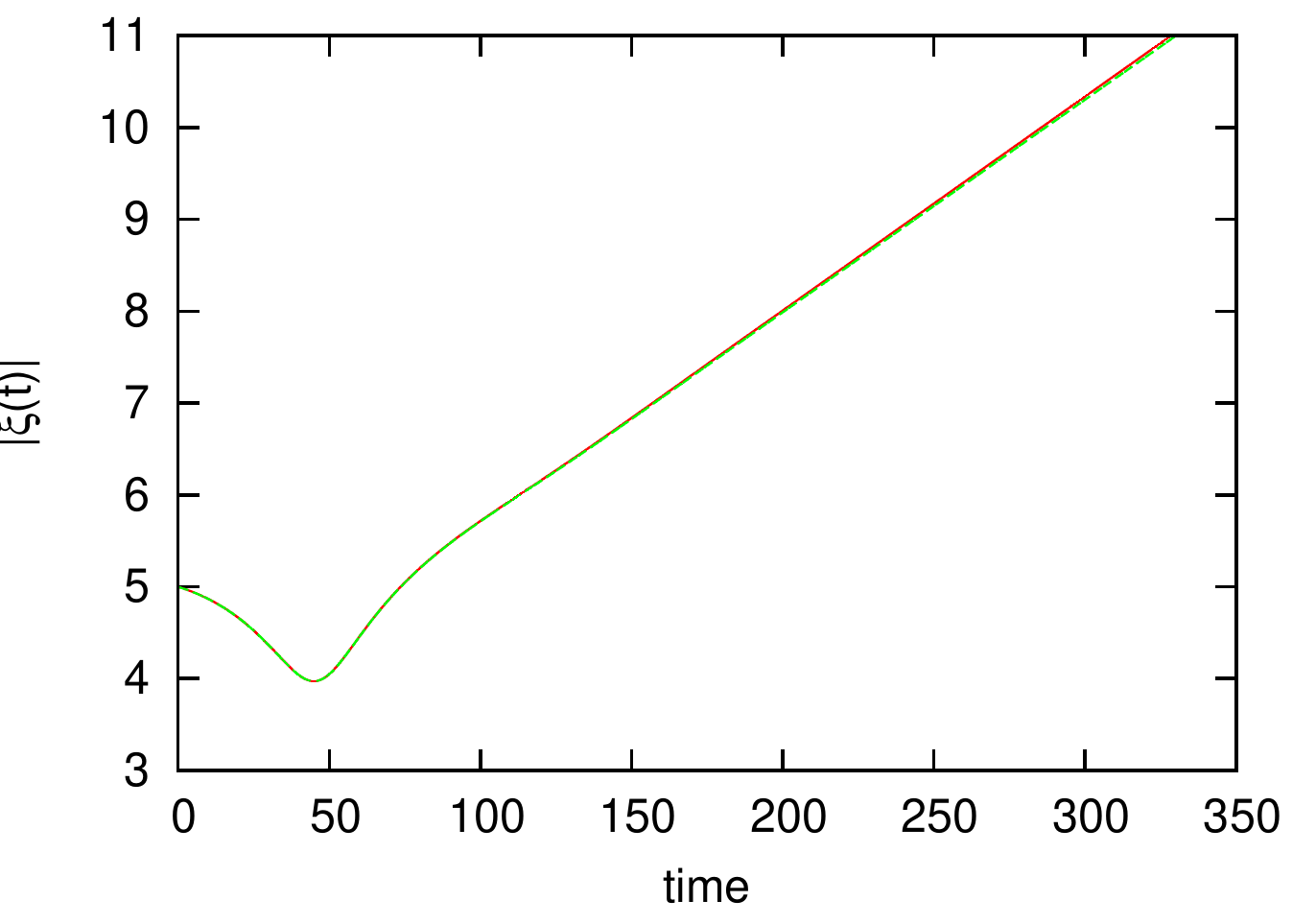}}
  \subfigure[]{\label{fig:e=006,k=1}\includegraphics[trim = 2cm 2cm 12cm 9.8cm, width=0.45\textwidth]{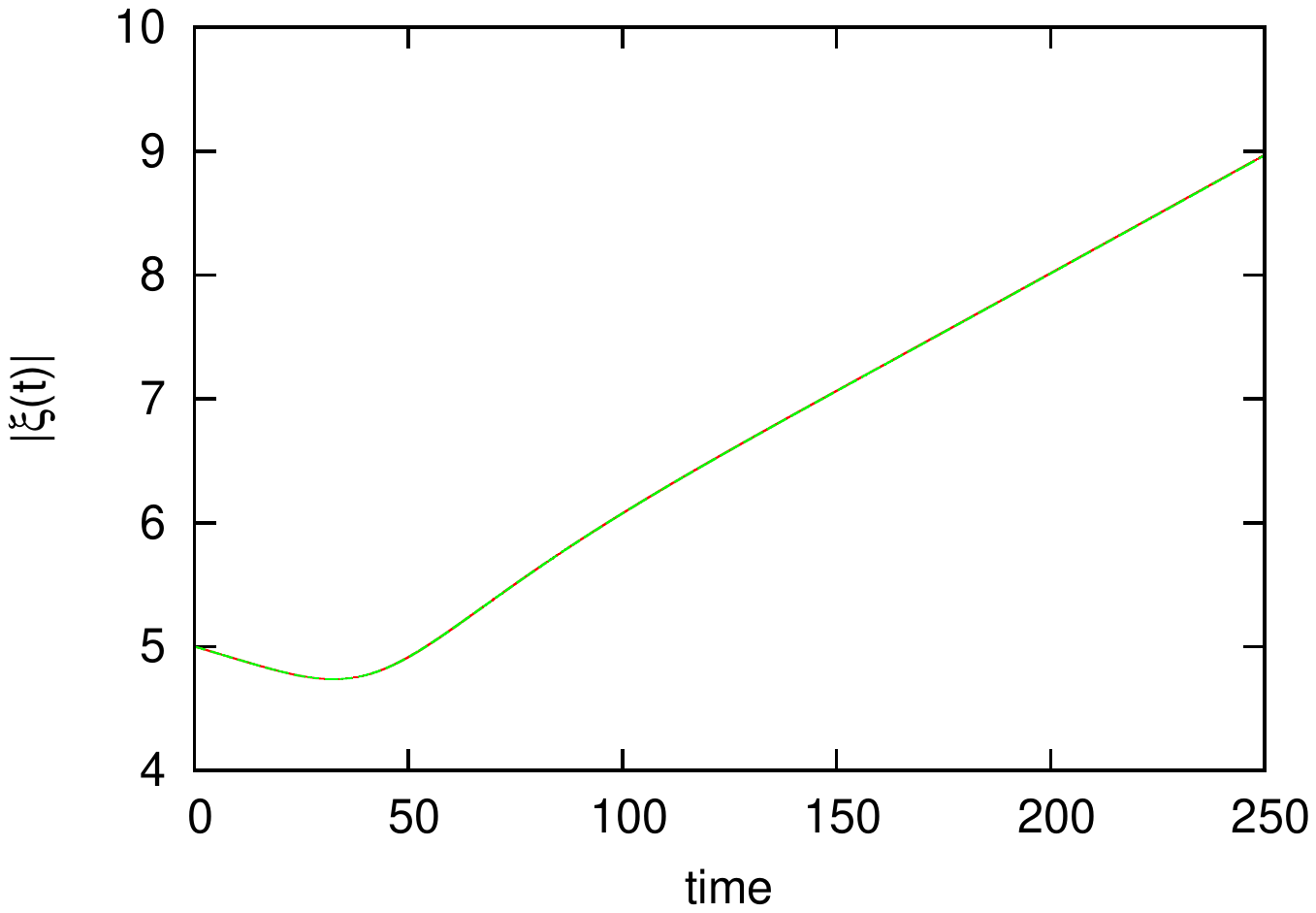}}                
  \subfigure[]{\label{fig:e=-006,k=1}\includegraphics[trim = 2cm 2cm 12cm 9.8cm, clip, width=0.45\textwidth]{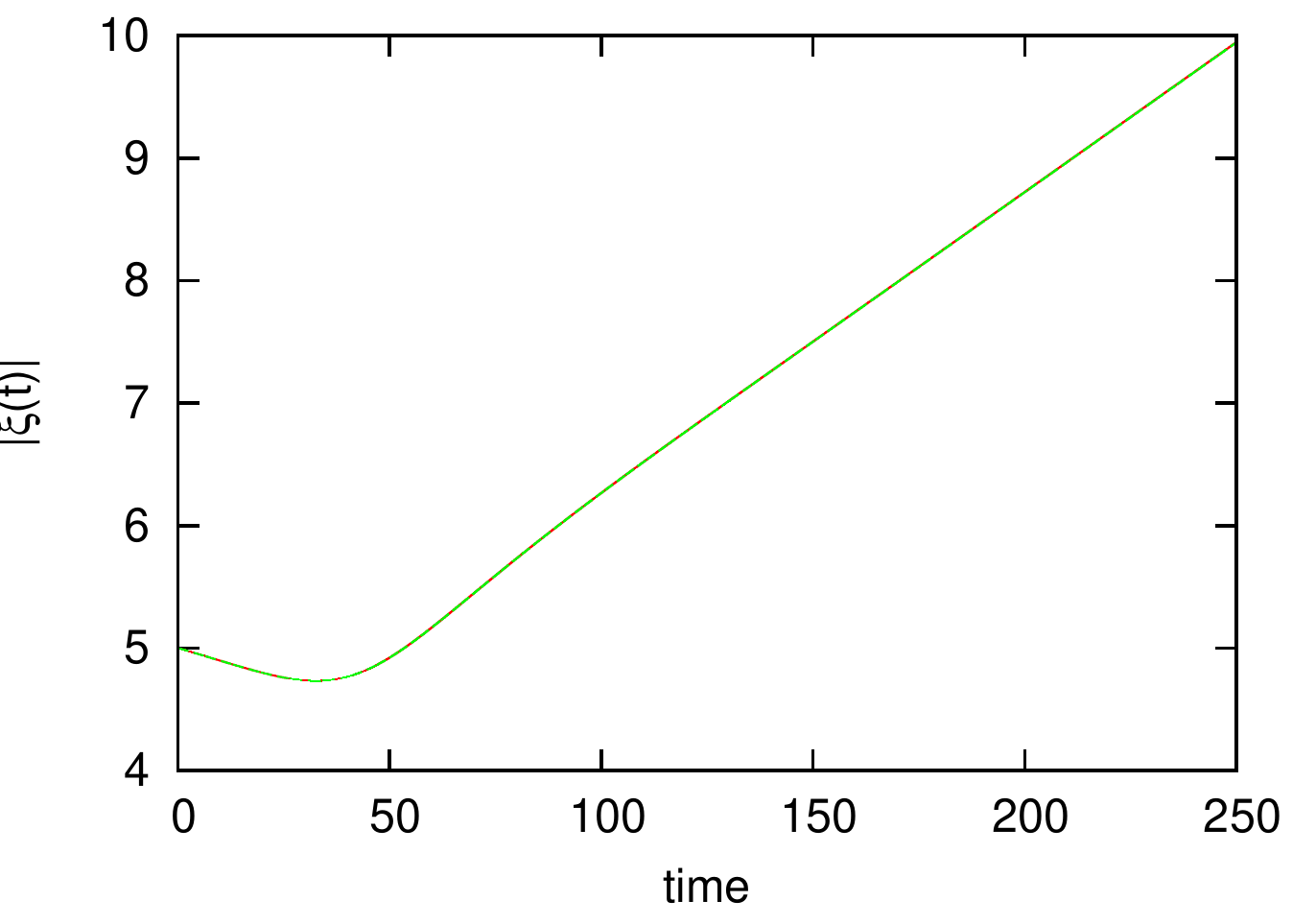}}
  \caption{The distance of a soliton from the centre of mass of a system with time. The system consists of two solitons initially placed at $\pm 5$ each with an initial velocity of $0.01$ towards the centre of mass. Initial height/width parameter of each soliton is $1$ with $\delta=0$ and (a) $\epsilon=0.06$, (b) $\epsilon=-0.06$;  $\delta=\frac{\pi}{4}$ and (c) $\epsilon=0.06$, (d) $\epsilon=-0.06$; $\delta=\frac{\pi}{2}$ and (e) $\epsilon=0.06$, (f) $\epsilon=-0.06$. For each plot the solid line is result of the collective coordinate approximation and the dashed line is the result of the full simulation (these are often coincident).}
  \label{fig:e=pm006}
\end{figure}

In the case $\epsilon\neq 0$ the modified NLS system of two solitons is no longer integrable; this means that the system no longer has an infinite number of conserved quantities and so some energy can be lost as radiation during soliton interactions. As in the $\epsilon=0$ case we see that the accuracy of the approximation depends on the amount of time the solitons spend in close proximity of each other during their interaction. This can be seen in figure \ref{fig:e=pm006} which compares the trajectories of solitons with initial positions $\xi_1=-5, \, \xi_2=5$; initial velocities $\mu_1=0.01, \, \mu_2=-0.01$ and initial height/width parameter $a_1=a_2=1$ as before, and $\epsilon=\pm 0.06$ and $\delta=0, \frac{\pi}{4}, \frac{\pi}{2}$ (plots for $\delta=\frac{3 \pi}{4}, \pi$ show excellent agreement so are not included). For $\epsilon=\pm 0.06$ and $\delta\neq 0$ the results of the collective coordinate approximation still show excellent agreement with the results of the full numerical simulation. However, for $\epsilon=\pm 0.06$ and $\delta=0$ the differences between the approximation and full simulation are more pronounced than in the $\epsilon=0$ case: the collective coordinate approximation accurately describes the initial coming together of the solitons, but it does not capture the  decreasing amplitude and increasing frequency of the oscillations demonstrated by the full simulation before the solitons eventually repel. This increased difference is probably because, for the $\epsilon\neq 0$ case, the solitons deform each other to a greater extent as they approach and some energy is radiated out which is not accounted for in the approximation. 

\begin{figure}
  \centering
{\includegraphics[trim = 2cm 2cm 12cm 9.8cm, width=0.45\textwidth]{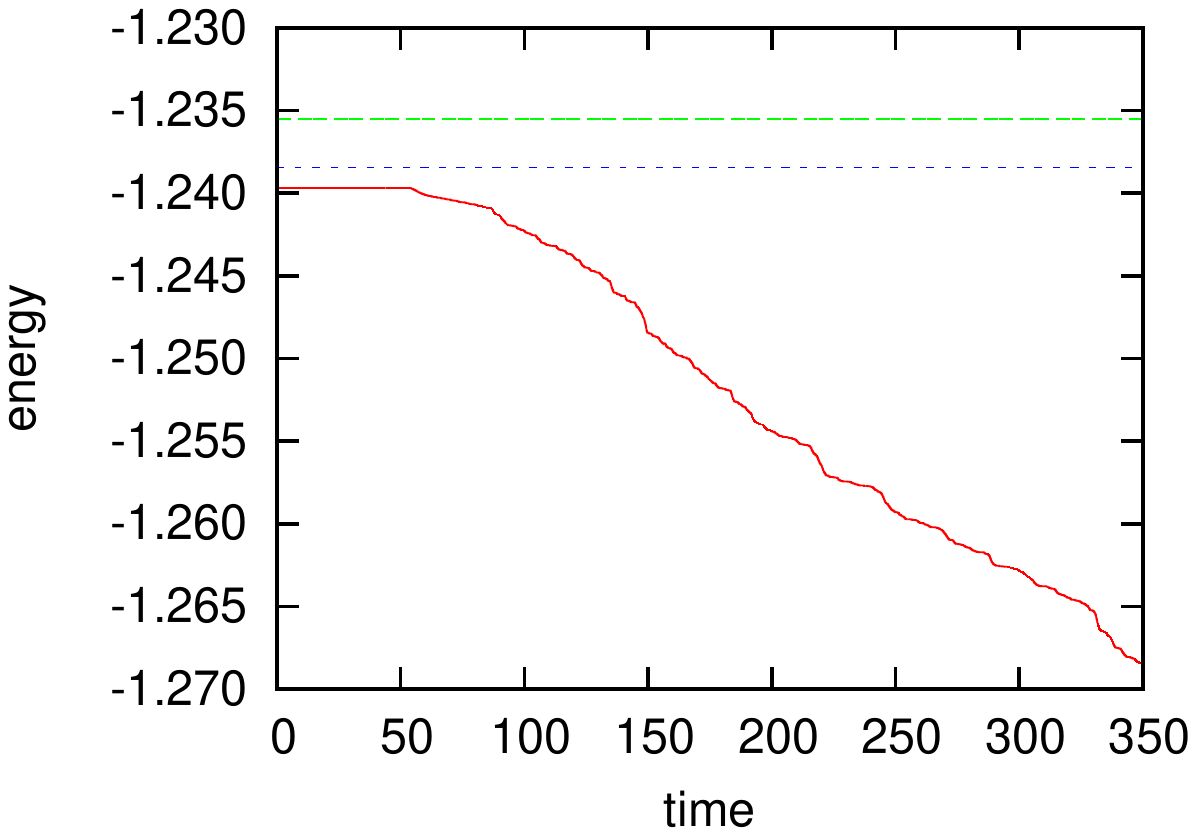}}                

  \caption{The time dependency of the energy of the solitons for $\epsilon=0.06$ placed initially at $\pm5$. Each soliton is of initial height/width parameter of $1$  and is sent towards the centre of mass with initial velocity $0.01$.  $\delta=0$ corresponds to the solid line, $\delta=\frac{\pi}{4}$ the dotted line and $\delta=\frac{\pi}{2}$ the dashed line.}
  \label{fig:engy_006}
\end{figure}

The amount of energy lost by the solitons in the full simulation is shown in figure \ref{fig:engy_006} where we plot the energy of the system during a scattering for $\epsilon=0.06$, and $\delta=0, \frac{\pi}{4}, \frac{\pi}{2} $ and for the same initial conditions as those used in the trajectory plots (plots for $\epsilon=-0.06$ are very similar). In this figure it is clear that in the case $\delta=0$ energy is constant until the solitons come together at which point some energy is radiated out, the system then evolves as two solitons and some small energy waves which we absorb as they reach the boundary so we see that the total energy of the soliton decreases. Over time the $\delta=0$ case sees much more energy radiated out than $\delta=\frac{\pi}{4}, \frac{\pi}{2}$ cases both of which demonstrate an incredibly small energy change, and though the energy loss in the $\delta=0$ case is also small it corresponds to a large deformation of the solitons so we would expect the collective coordinate approximation to be the least reliable for $\delta=0$.

Next we consider the effect of the initial velocity on the accuracy of the collective coordinate approximation when $\epsilon=0.06$. Figure \ref{fig:e=0.06,vv} compares the trajectories obtained in the collective coordinate approximation and those found by the full numerical simulation for solitons with initial values as in figure \ref{fig:e=0,vv} but with $\epsilon=0.06$. This shows that the accuracy of the approximation is still quite good up to $v=0.2$ though slightly less so than in the equivalent simulations with $\epsilon=0$ (this can be seen by comparing figures \ref{fig:e=0.06,vv} and \ref{fig:e=0,vv}).

\begin{figure}
  \centering
  \subfigure[]{\label{fig:e=0.06,v=0.1}\includegraphics[trim = 2cm 2cm 12cm 9.8cm, width=0.45\textwidth]{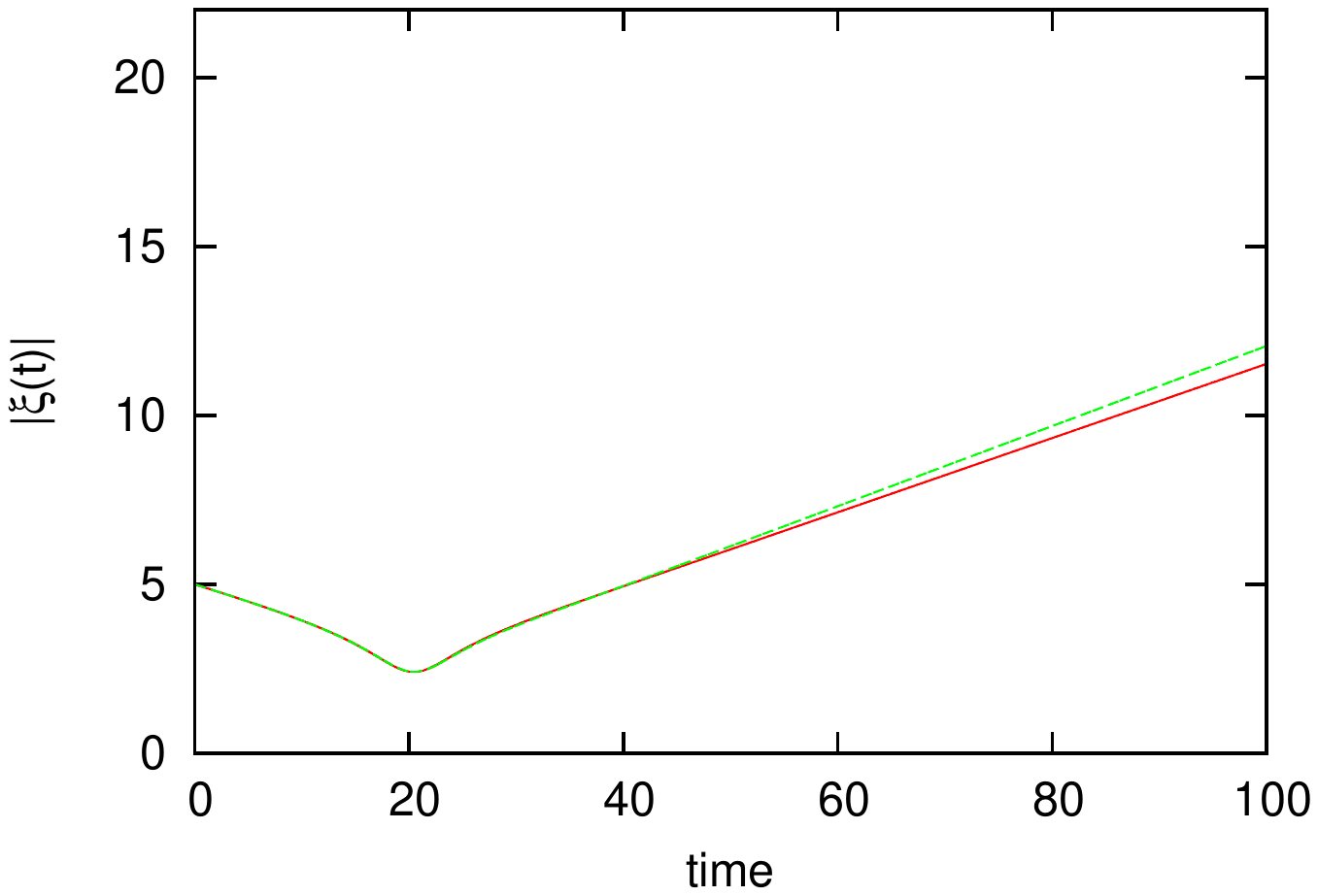}}                
  \subfigure[]{\label{fig:e=0.06,v=0.2}\includegraphics[trim = 2cm 2cm 12cm 9.8cm, clip, width=0.45\textwidth]{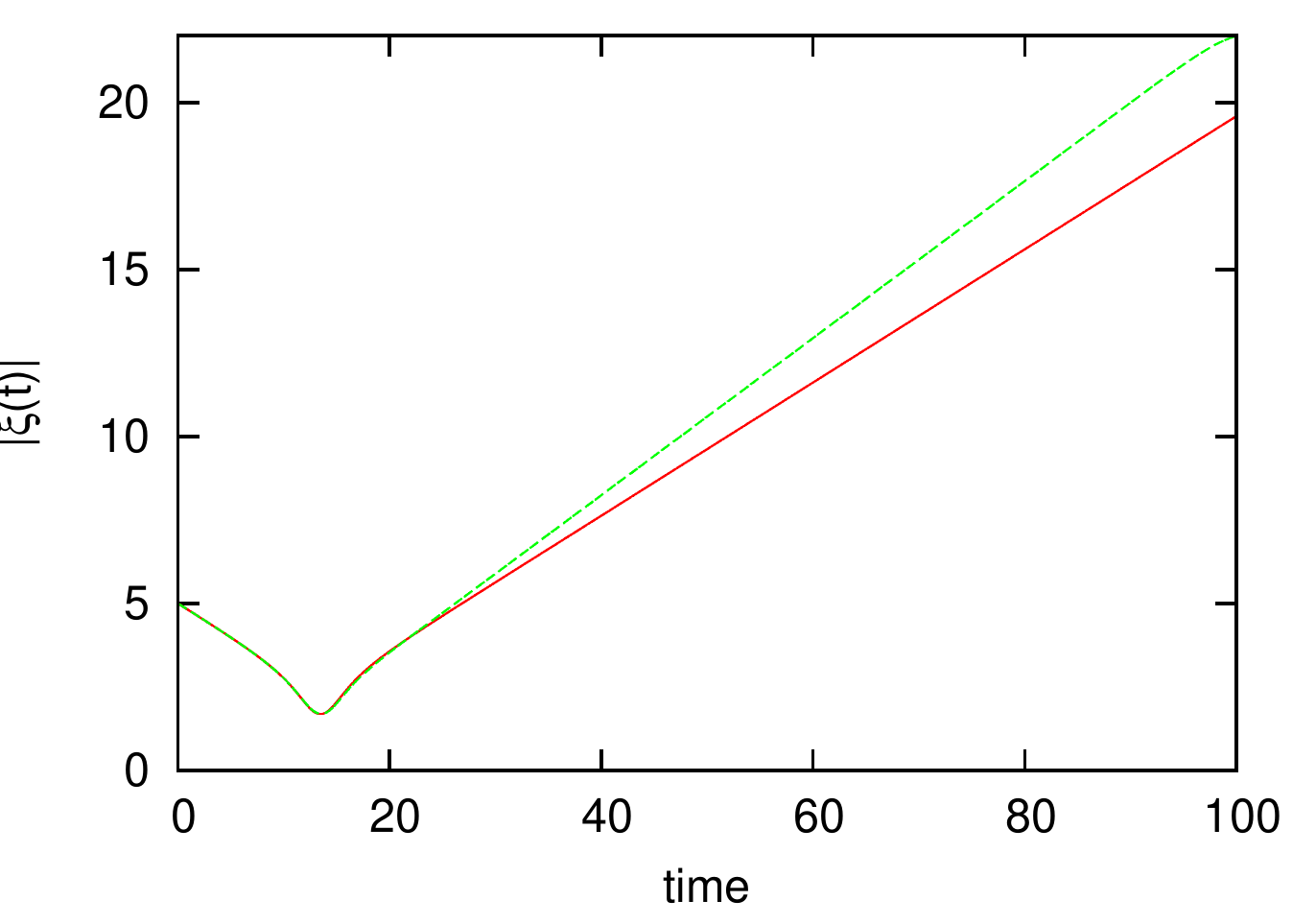}}
  \caption{The distance of a soliton from the centre of mass of a system with time with $\epsilon=0.06$. The system consists of two solitons initially placed at $\pm 5$, with initial height/width parameter of $1$ and the initial phase difference between them of $\delta=\frac{\pi}{4}$. The solitons move towards the centre of mass with initial velocity (a) $v=0.1$, and (b) $v=0.2$. For each plot the solid line is result of the collective coordinate approximation and the dashed line is the result of the full simulation.}
  \label{fig:e=0.06,vv}
\end{figure}

Next we increase the parameter $\epsilon$ to investigate its effect on the accuracy of the approximation. Figure \ref{fig:ev,vv} presents the plots of the trajectories derived in the collective coordinate approximation and the full numerical simulation for solitons  with initial position $\xi_1=-5, \, \xi_2=5$; initial height/width parameter $a_1=a_2=1$; initial phase difference $\delta=\frac{\pi}{4}$, and various values of $\epsilon$ and initial velocity. This figure also shows that, for solitons which do not spend much time in close proximity of each other, increasing the value of $\epsilon$ reduces the accuracy of the approximation very slightly with excellent agreement of  up to at least $\epsilon=0.3$.

\begin{figure}
  \centering
  \subfigure[]{\label{fig:e=0.1,v=0.01}\includegraphics[trim = 2cm 2cm 12cm 9.8cm, width=0.45\textwidth]{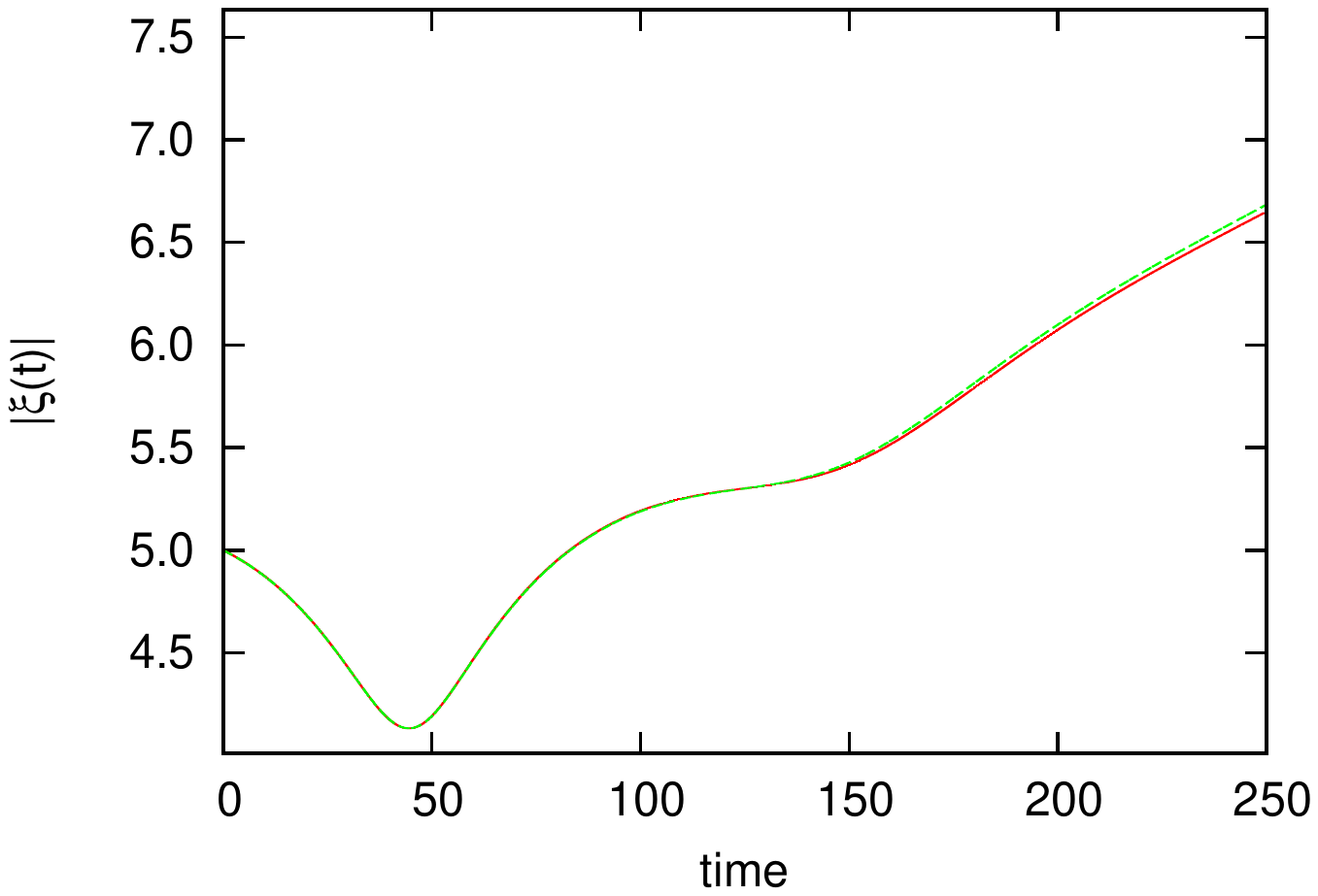}}                
  \subfigure[]{\label{fig:e=0.3,v=0.1}\includegraphics[trim = 2cm 2cm 12cm 9.8cm, clip, width=0.45\textwidth]{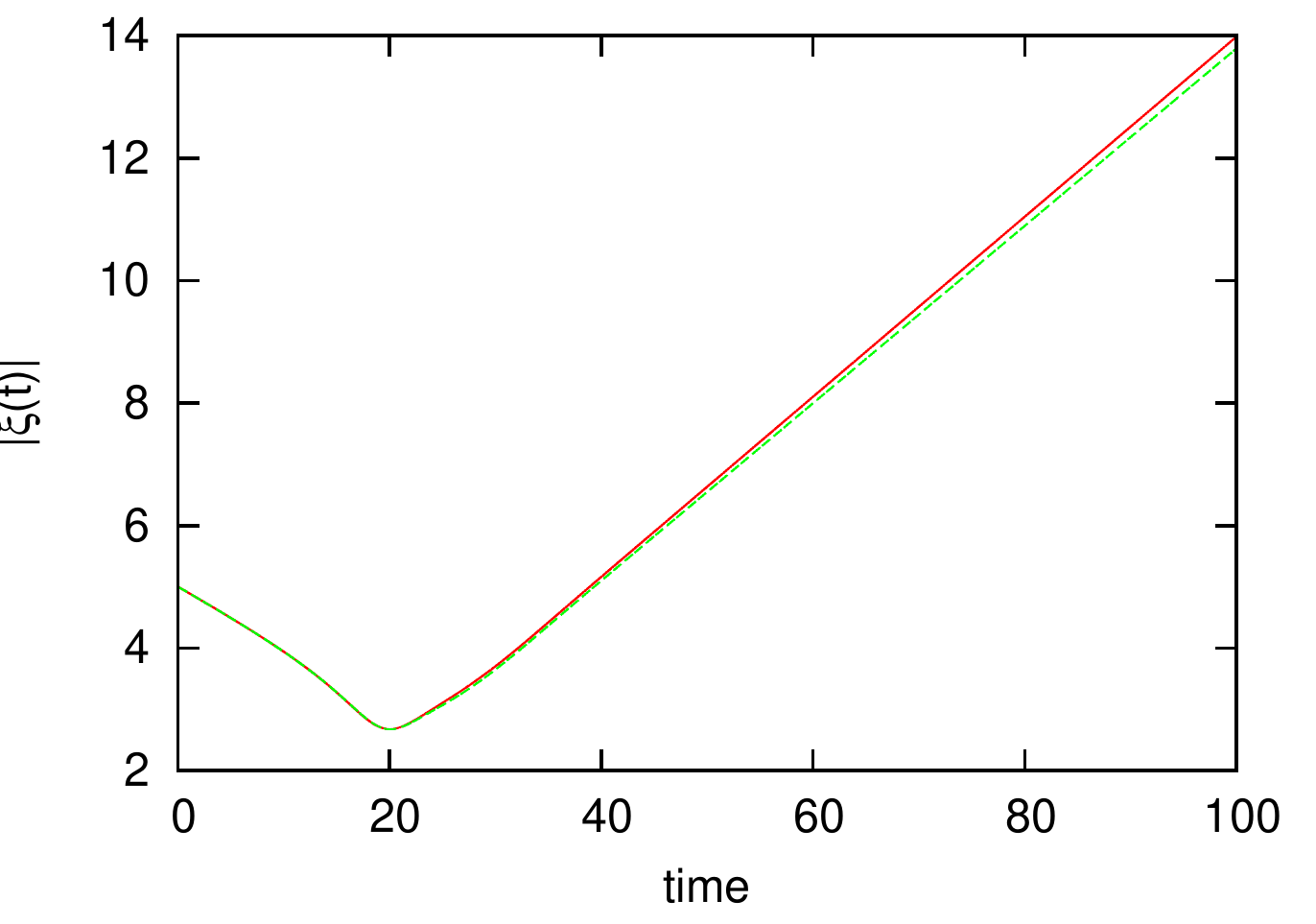}}
  \caption{The distance of a soliton from the centre of mass of a system with time. The system consists of two solitons initially placed at $\pm 5$, with initial height/width parameter of $1$ and the initial phase difference between them of $\delta=\frac{\pi}{4}$. For (a) the solitons have initial velocity $v=0.01$ and $\epsilon=0.1$, and for (b) the solitons have initial velocity $v=0.1$ and $\epsilon=0.3$. For each plot the solid line is result of the collective coordinate approximation and the dashed line is the result of the full simulation.}
  \label{fig:ev,vv}
\end{figure}
In our numerical simulations we calculate and compare the quasi-conservation of the first non-trivial charge beyond the energy and momentum, \textit{i.e.} the charge $Q^{(4)}$ defined in \eqref{con_laws}. We do this by computing the corresponding anomaly $\beta^{(4)}$, also defined in \eqref{con_laws}, and by integrating it over time to get the integrated anomaly:
\begin{eqnarray}
\chi^{(4)}(t)&\equiv & \int^{t}_{-\infty} dt' \, \beta_4 =\int^t_{-\infty}dt'\int^{\infty}_{-\infty} dx \, X \alpha^{(3,-4)}\\
&=&-2i \int^t_{-\infty}dt'\int^{\infty}_{-\infty} dx \left( \left( \epsilon +1 \right) R^{\epsilon} -1\right) \partial_x R \left[ -6 R^2+\frac{3}{2}\left(\partial_x \varphi\right)^2 R-2\,\partial_x^2 R +\frac{3}{2}\frac{\left(\partial_x R\right)^2}{R}\right]. \nonumber
\end{eqnarray}
This is then been computed in terms of the fields $R$ and $\varphi$ which are defined by writing each soliton field $\psi$ in the form $\psi\equiv\sqrt{R} e^{i \frac{\varphi}{2}}$. 

\begin{figure}
  \centering
    \subfigure[]{\label{fig:ano,p,k=0}\includegraphics[trim = 2cm 2cm 12cm 9.9cm, width=0.45\textwidth]{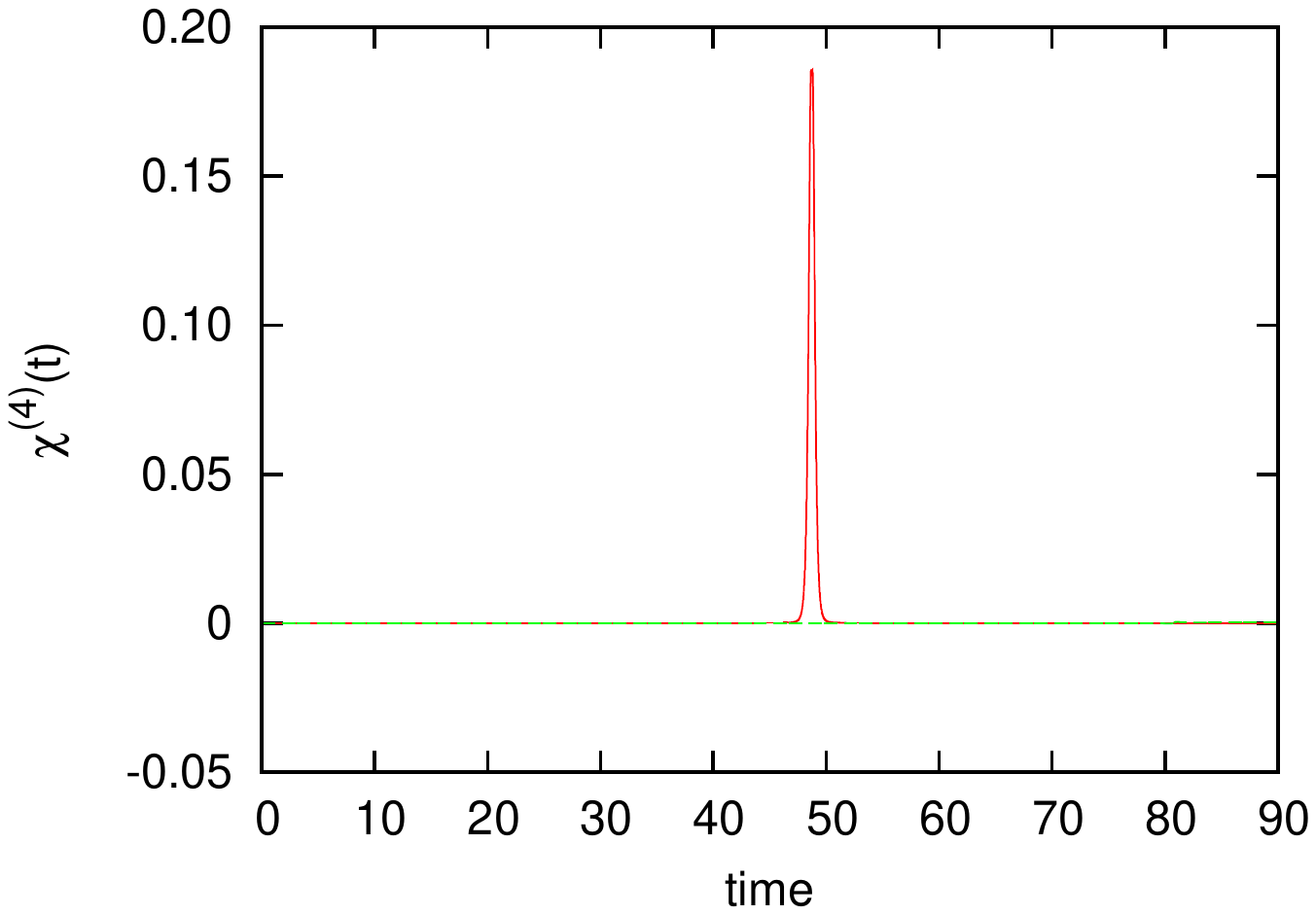}}                
  \subfigure[]{\label{fig:ano,m,k=0}\includegraphics[trim = 1.9cm 2cm 12cm 9.8cm, clip, width=0.45\textwidth]{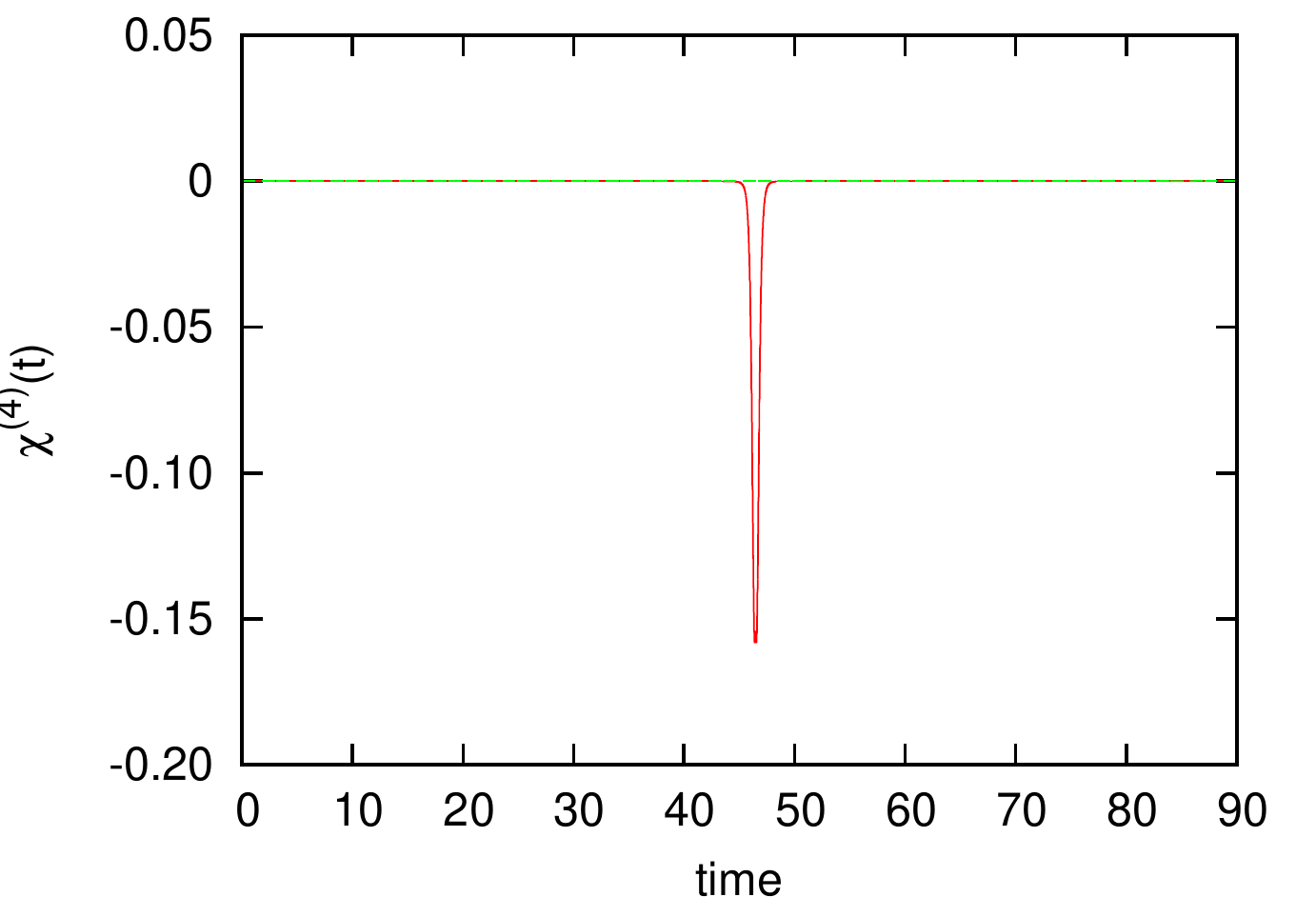}}
    \subfigure[]{\label{fig:ano,p,k=0.5}\includegraphics[trim = 2cm 2cm 12cm 9.9cm, width=0.45\textwidth]{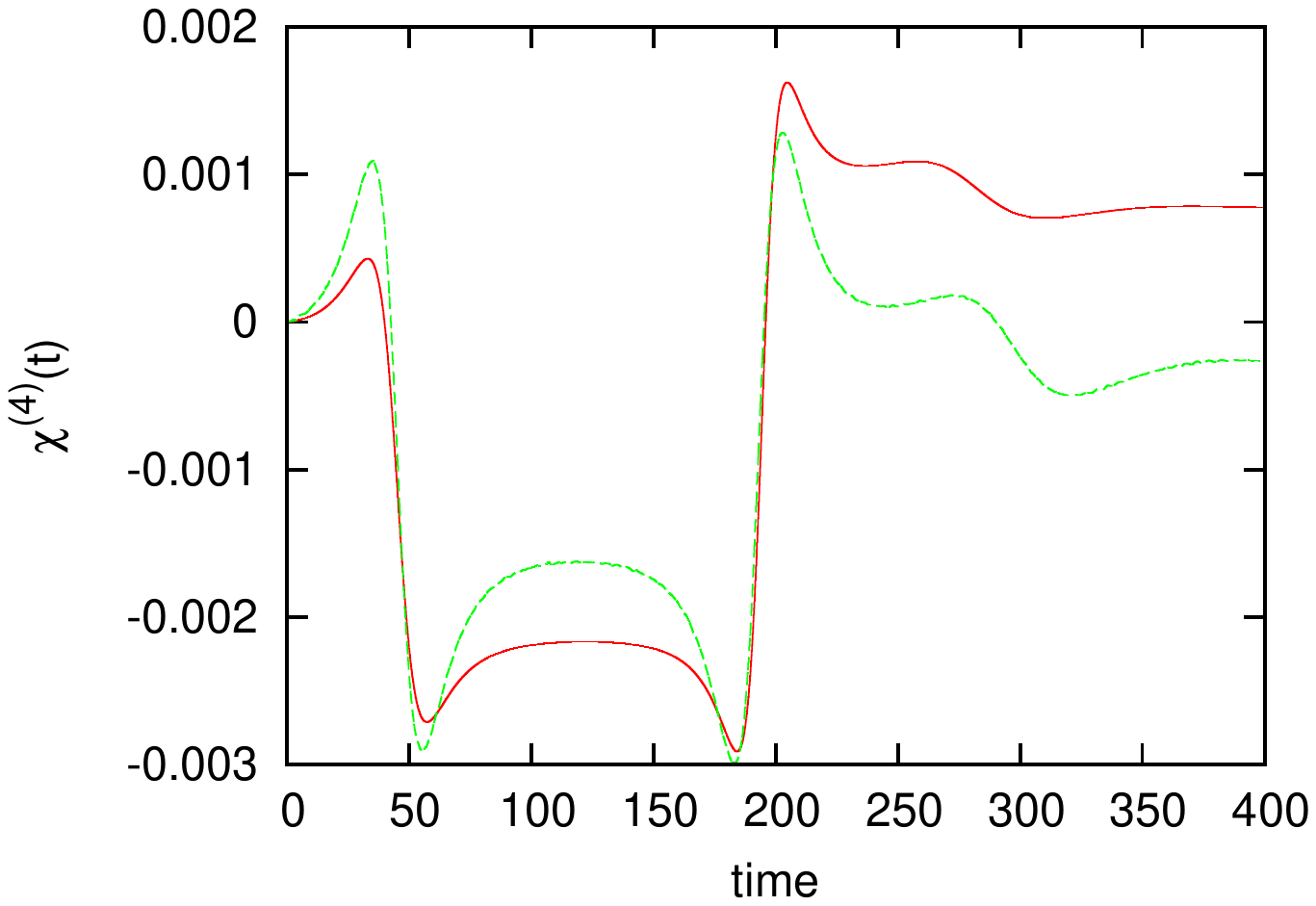}}                
  \subfigure[]{\label{fig:ano,m,k=0.5}\includegraphics[trim = 1.9cm 2cm 12cm 9.8cm, clip, width=0.45\textwidth]{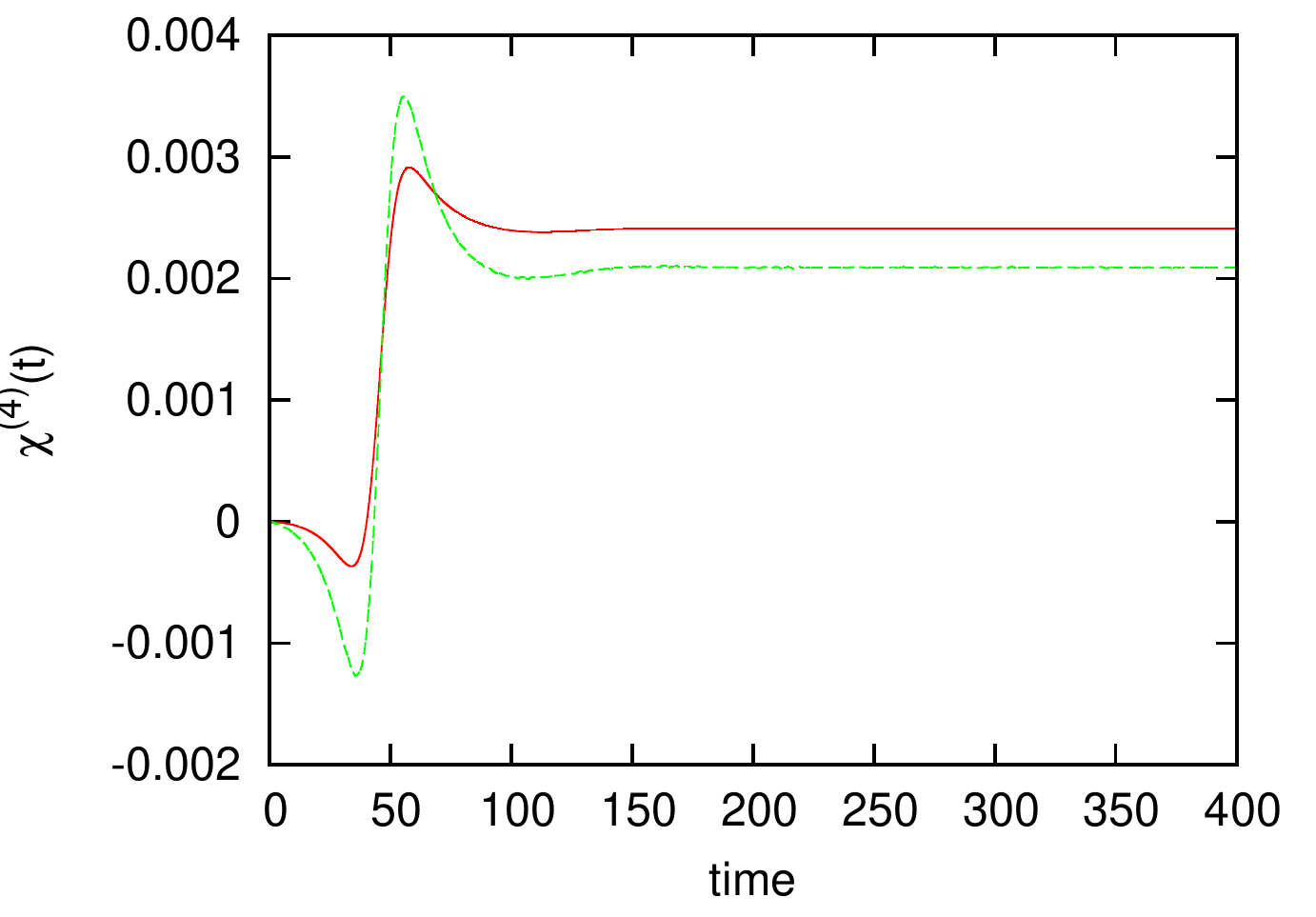}}
 \subfigure[]{\label{fig:ano,p,k=1}\includegraphics[trim = 2cm 2cm 12cm 9.9cm, width=0.45\textwidth]{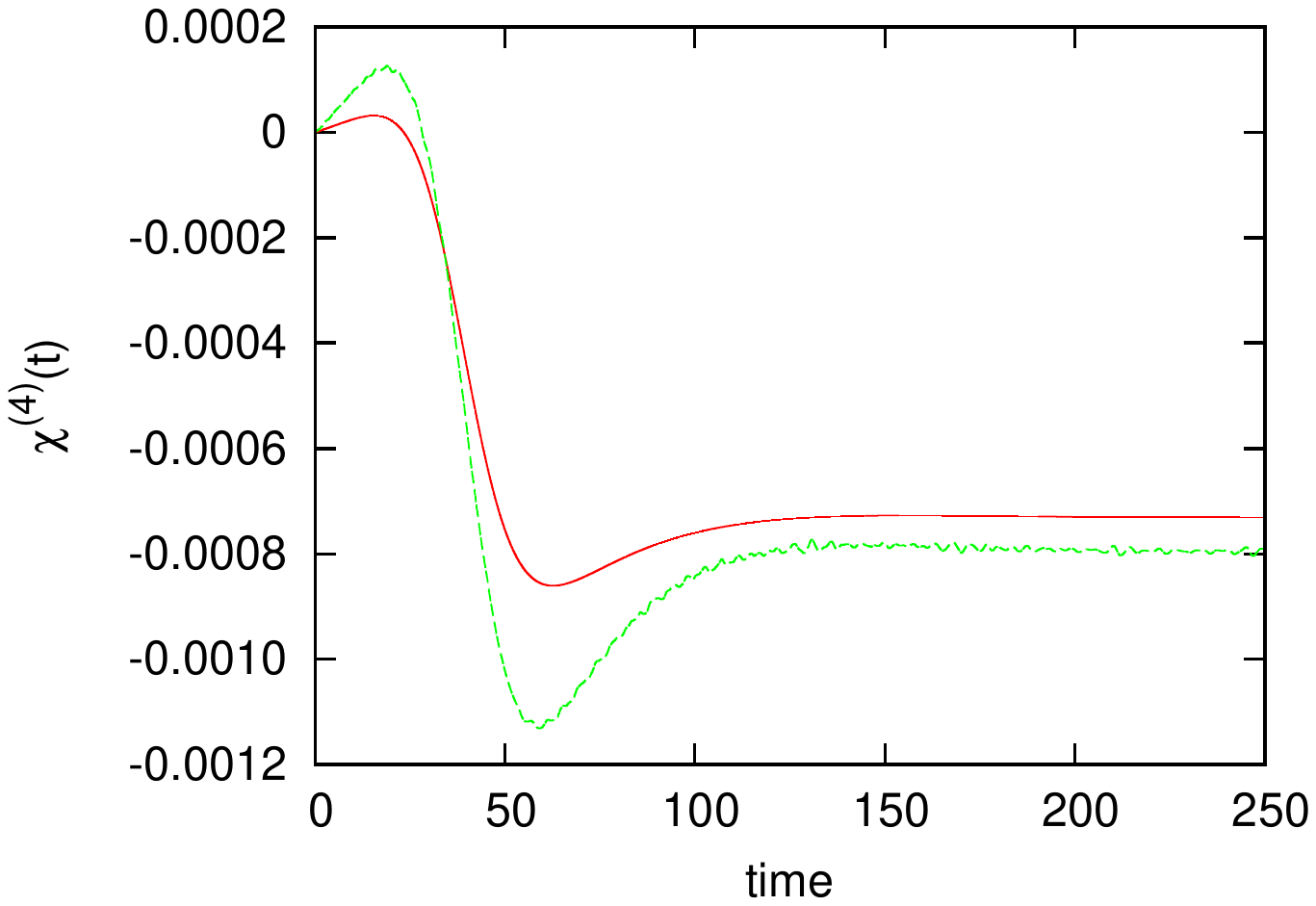}}                
  \subfigure[]{\label{fig:ano,m,k=1}\includegraphics[trim = 1.9cm 2cm 12cm 9.8cm, clip, width=0.45\textwidth]{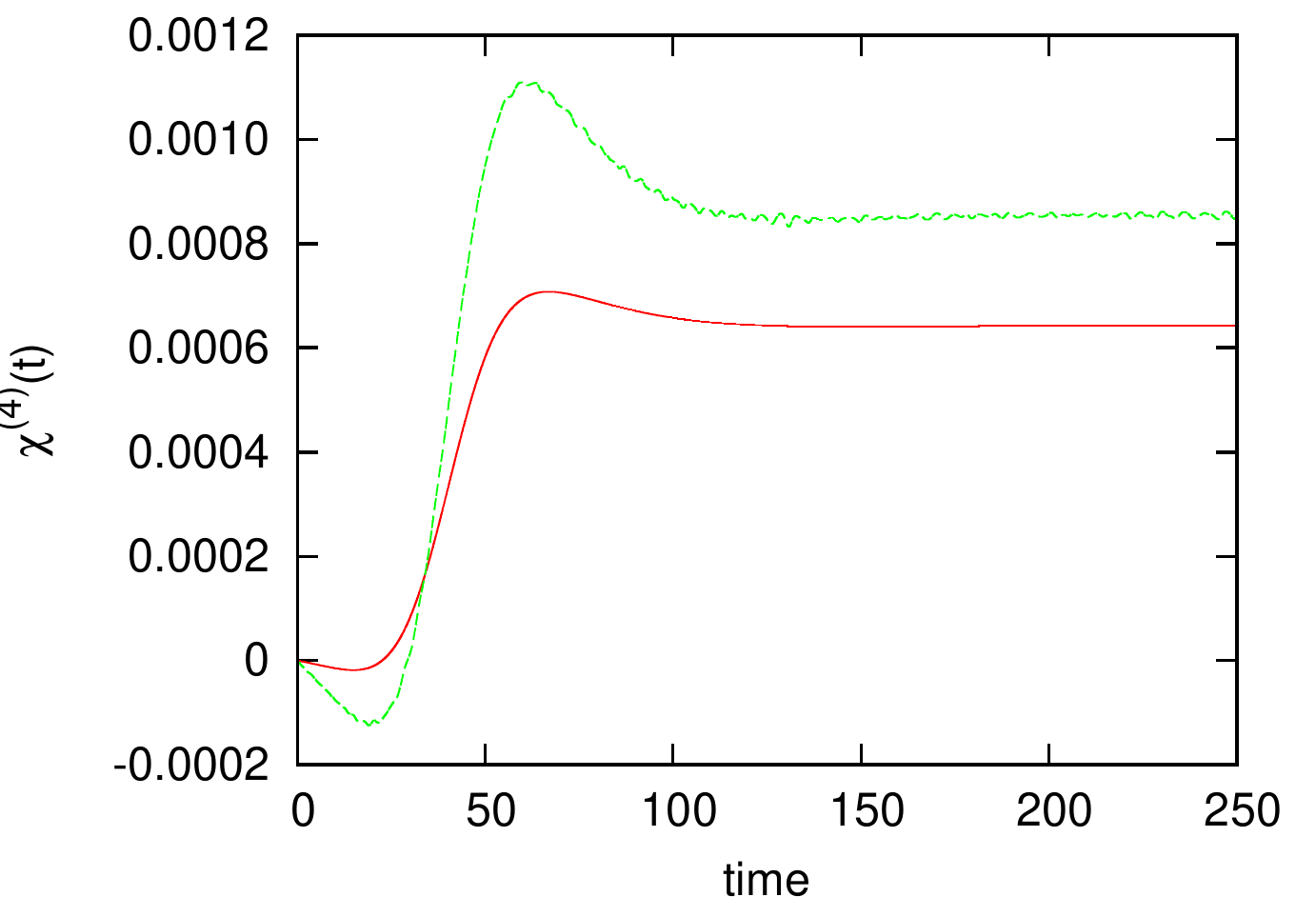}}
   \caption{The time-integrated anomaly, $\chi^{(4)}(t)$, for the soliton interactions shown in figure \ref{fig:e=pm006} with $\delta=0$ and (a) $\epsilon=0.06$, (b) $\epsilon=-0.06$;  $\delta=\frac{\pi}{4}$ and (c) $\epsilon=0.06$, (d) $\epsilon=-0.06$; $\delta=\frac{\pi}{2}$ and (e) $\epsilon=0.06$, (f) $\epsilon=-0.06$. For each plot the solid line is result of the collective coordinate approximation and the dashed line is the result of the full simulation.}
  \label{fig:ano}
\end{figure}

In figure \ref{fig:ano} we present the plots of the time-integrated anomaly for each of the trajectories shown in figure \ref{fig:e=pm006}. We note that the results are very similar although not as exact as some of the trajectories. The time integrated anomalies are most different in the case $\delta=0$ as the collective coordinate approximation shows a distinct peak when the solitons come together when compared to the results seen in the full simulation which displays only a minute deviation from zero at these points (of the order $10^{-7}$). However, when the solitons are far apart the time-integrated anomaly does return to zero as predicted in \cite{Zakr2012} when $\delta$ is an integer value of $\pi$, as this corresponds to the case when the parity symmetry described in \eqref{psi_trans} is present. When $\delta$ is not an integer multiple of $\pi$ this symmetry is not present and the integrated anomalies do not return to zero, and the collective coordinate method shows similar time-integrated anomalies to those found in the full simulation. This shows that, in addition to the trajectories, the collective coordinate approximation also does reproduce quite well the results for the anomalies.

\newpage
\section{The modified sine-Gordon model}
Next we consider the Lagrangian given by
\begin{equation}
\label{Lag_sg}
L=\int dx\, \frac{1}{2} \left( \left( \partial_t \psi\right)^2-\left( \partial_x \psi\right)^2\right)-V(\psi).
\end{equation}
For the sine-Gordon potential $V_{SG}=\frac{1}{8}\sin^2(2\psi)$ there are static one-soliton solutions of the form
\begin{equation}
\label{psi}
\psi=\text{ArcTan}\left(e^{\pm (x-x_0)}\right).
\end{equation} 
A modification on this model has been suggested in \cite{Zakr2014} by taking a change of variable $\psi \rightarrow \phi$ given by
\begin{equation}
\label{change_of_variable}
\psi(\phi)=\frac{c \phi}{\sqrt{1+\epsilon \phi (\phi -2 \gamma)}}
\end{equation}
 which has two modifying parameters $\epsilon$ and $\gamma$, and the parameter $c$ chosen to be
\begin{equation}
c=\sqrt{1+\epsilon \pi \left( \frac{\pi}{4}-\gamma\right)}
\end{equation}
such that $\phi(\psi=0)=0$ and $\phi(\psi=\frac{\pi}{2})=\frac{\pi}{2}$.

Then $\phi$, obtained by calculating $\phi=\phi(\psi)$ from \eqref{change_of_variable} and using $\psi$ given by \eqref{psi}, is a solutions of the static Euler-Lagrangian equation associated to \eqref{Lag_sg} with the potential
\begin{equation}
\label{mod_pot_sg}
V(\phi)=\left(\frac{d\phi}{d\psi}\right)^2 V_{SG}=\frac{1}{8} \frac{\left(1+\epsilon \phi \left( \phi-2 \gamma\right)\right)^3}{c^2 \left( 1-\epsilon\gamma \phi\right)^2}\sin^2\left(2 \psi(\phi)\right).
\end{equation}

\begin{figure}
  \centering
  \subfigure[]{\label{fig:use_potsg_e_005_g_0}\includegraphics[trim = 2cm 2cm 12cm 9.8cm, width=0.45\textwidth]{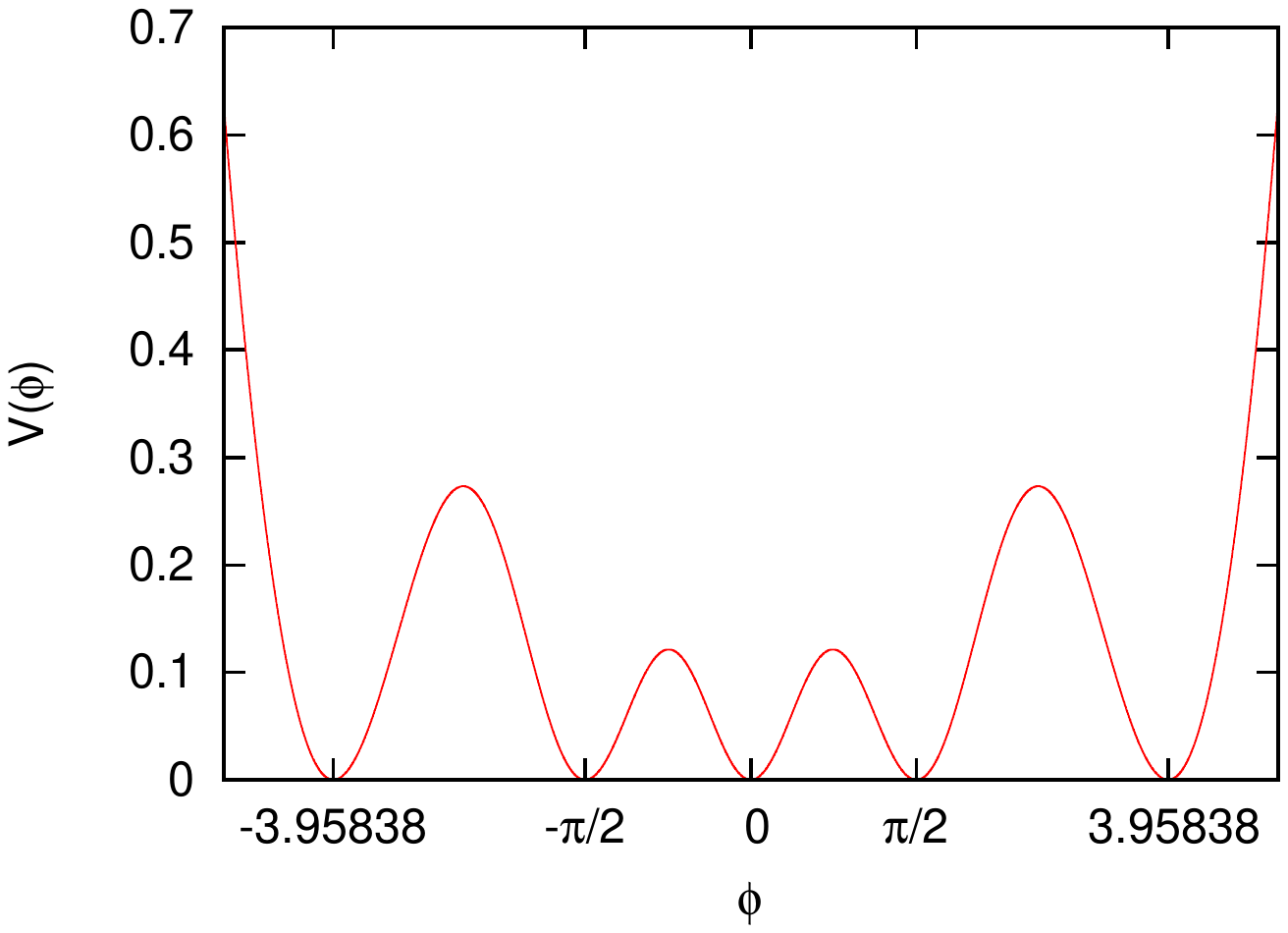}}                
  \subfigure[]{\label{fig:use_potsg_e_005_g_1}\includegraphics[trim = 2cm 2cm 12cm 9.8cm, clip, width=0.45\textwidth]{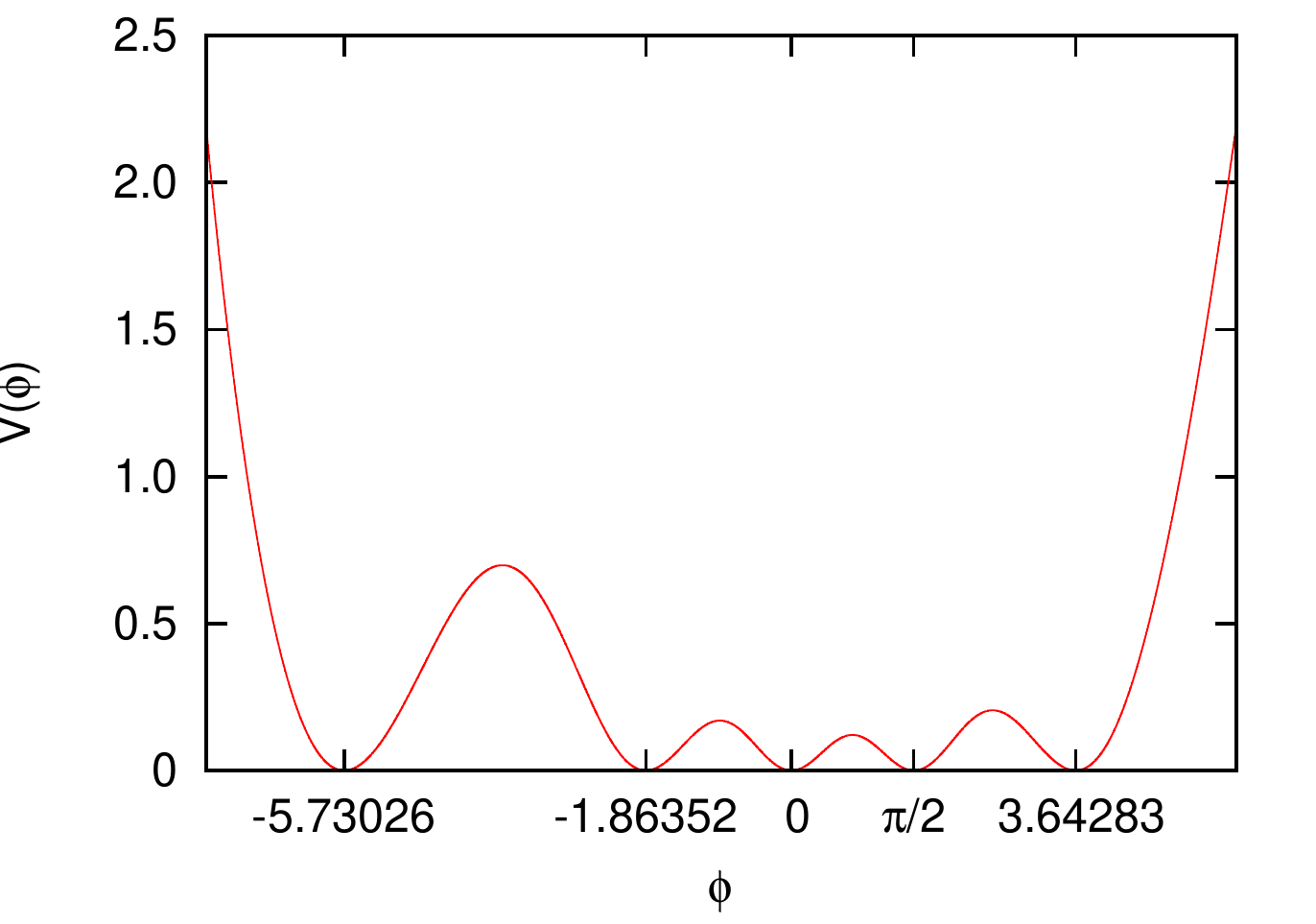}}
  \caption{The modified potential $V(\phi)$ against $\phi$ for $\epsilon=0.05$ and (a) $\gamma=0$, (b) $\gamma=1$.  }
  \label{fig:pot_sg}
\end{figure}

In the case $\epsilon=0$ the parameter $\gamma$ becomes irrelevant and the potential \eqref{mod_pot_sg} returns to the unperturbed sine-Gordon potential and $\phi=\psi$. For $\epsilon\neq 0$ and $\gamma=0$ the model has the symmetry $\phi=-\phi$, and for $\epsilon, \gamma\neq 0$ there is no symmetry. This can be seen in figure \ref{fig:pot_sg} where we plot the potential as a function of $\phi$ for $\epsilon=0.05$ and $\gamma=0$ and $\gamma=1$. By varying the parameters $\epsilon$ and $\gamma$ the effects of this symmetry on the theory can be seen. 
Note that the topological charge of $\phi(\psi)$, for $\psi$ given by \eqref{psi}, is conserved for any value of $\epsilon$ and $\gamma$.

In a similar manner to the NLS case the sine-Gordon has a set of anomalous conservation laws derived in \cite{Zakr2014} and given by:
\begin{equation}
\label{con_laws_sg}
\frac{d \tilde{Q}^{(2n+1)}}{dt}=\tilde{\beta}^{2n+1};\,\,\,\,\,\,\,\,\,\,\,\, \text{with}\,\,\, \tilde{Q}^{(2n+1)}=\int^{\infty}_{-\infty}dx\, \tilde{a}_x^{(2n+1)};
 \,\,\,\,\,\,\,\,\,\,\,\, \text{where}\,\,\, \tilde{\beta}^{(2n+1)}=\int^{\infty}_{-\infty}dx\, \tilde{X}\, \tilde{\alpha}^{(2n+1)}
\end{equation}
for $n=0,1,2,3,...$ and
\begin{equation}
\tilde{X}=\frac{i w}{2} \partial_- \phi \left[ \frac{d^2 V}{d \phi^2} +16 V -1\right]
\end{equation} 
which vanishes for the sine-Gordon potential.

If the field configuration transforms under the parity defined in \eqref{parity} as
\begin{equation}
\label{phi_trans_sg}
P(\phi)=-\phi+const.
\end{equation}
and if the potential evaluated on such a solution is even under the parity, \textit{i.e.}
\begin{equation}
\label{pot_trans_sg}
P(V)=V
\end{equation}
then we have an infinite set of conserved quantities which are conserved asymptotically, \textit{i.e.}
\begin{equation}
Q^{(2n+1)}(t=+\infty)=Q^{(2n+1)}(t=-\infty).
\end{equation}
This modified model, when $\epsilon=0$, becomes the sine-Gordon model, therefore the system is integrable and has an infinite number of conserved quantities. When $\epsilon\neq 0$ and $\gamma=0$ then the field configuration and potential transform under the parity described in \eqref{parity} as in \eqref{phi_trans_sg} and \eqref{pot_trans_sg}; therefore the system is quasi-integrable and possesses an infinite number of aymptotically conserved charges. When $\epsilon\neq 0$ and $\gamma \neq 0$ then the symmetries necessary for quasi-integrability are not present; the system is non-integrable and there are no constraints on the charges.

\subsection{The two-soliton configuration for modified sine-Gordon}
As in the NLS case we construct an appropriate two-soliton ansatz for the sine-Gordon in the collective coordinate approximation by patching together two one-kink solutions. We do this in the following way:
\begin{equation}
\label{ansatz_sg}
\text{tan}(\psi)=e^{(x-a)}-e^{-(x+a)}= 2\, \text{sinh}(x)\, e^{-a}
\end{equation}
where $a$ is our collective coordinate. When $a$ is large \eqref{ansatz_sg} represents two well separated kinks; one placed at $-a$ whose field varies between $\left( -\frac{\pi}{2},0\right)$ and one placed at $a$ which varies between $\left( 0,\frac{\pi}{2}\right)$. For energetic reasons it must be that $a>0$ for all times. This ansatz was used in \cite{Sutc1992} to test the collective coordinate approximation for the scattering of sine-Gordon kinks and was found to work remarkably well so our ansatz for our modified sine-Gordon model will be based on a generalisation of this ansatz.

To construct a modified approximation ansatz we perform the change of variable as in \eqref{change_of_variable}, for $\psi$ given by \eqref{ansatz_sg}, to get
\begin{eqnarray}
\label{anz_sg}
\phi&=&\frac{\psi^2\epsilon\gamma + \sqrt{\psi^2 c^2+\psi^4 \epsilon \lp -1+\gamma^2 \epsilon\rp}}{\psi^2\epsilon-c^2}\,\,\,\,\,\,\,\,\,\,\,\,\text{for}\,\,\,\, x<0\\
\phi&=&\frac{\psi^2\epsilon\gamma- \sqrt{\psi^2 c^2+\psi^4 \epsilon \lp -1+\gamma^2 \epsilon\rp}}{\psi^2\epsilon-c^2}\,\,\,\,\,\,\,\,\,\,\,\,\text{for}\,\,\,\, x>0\nonumber
\end{eqnarray}
and take this as our two soliton ansatz for the Euler-Lagrange equation associated to \eqref{Lag_sg} with the potential given by \eqref{mod_pot_sg}. This ansatz returns to the ansatz for the unmodified sine-Gordon in the case $\epsilon=0$. For $\epsilon\neq 0, \gamma=0$ the kinks are altered but the potential retains the symmetry $V\left( \phi\right)=V \left(-\phi\right)$, whereas for $\epsilon\neq 0, \gamma\neq 0$ this symmetry is lost due to the shift in the vacua which can be seen in figure \ref{fig:use_potsg_e_005_g_1}.

\subsection{Implementing the approximation in modified sine-Gordon}
We substitute our approximation ansatz (\ref{anz_sg}) into the Lagrangian (\ref{Lag_sg}) (with the change of variable (\ref{change_of_variable}) and modified potential (\ref{mod_pot_sg})) to find an effective Lagrangian:
\begin{equation}
\label{lag_sg}
L=\frac{g(a)}{2}\, \dot{a}^2-V(a)
\end{equation}
where the dot refers to a differentiation with respect to time. The expression for $g(a)$ is given by
\begin{equation}
g(a)=4 e^{2 a} c^4\int^{\infty}_{\infty} dx \frac{A(x,a)}{B(x,a)},
\end{equation}
where
\begin{multline}
A(x,a)=\sinh ^2 (x) \left(\epsilon ^2 \tan ^{-1} \left(2 e^{-a} \sinh (x)\right)^4  \left(\alpha  \left(8 \gamma ^2 \epsilon  \left(\gamma ^2 \epsilon
               - 1\right) + 1\right) + 4 \gamma  c^2 \left(2 - 3 \gamma ^2 \epsilon
           \right)\right)\right.\\ - 2 c^2 \epsilon  \tan ^{-1} \left(2 e^{-a} \sinh
          (x)\right)^2 \left(\alpha  \left(1 - 4 \gamma ^2 \epsilon \right) + 2
           \gamma  c^2\right)\\\left. - 4 \gamma  \epsilon ^3 \left(\gamma ^2 \epsilon 
           \left(2 \gamma ^2 \epsilon - 3\right) + 1\right) \tan ^{-1} \left(2 e^{-a}           \sinh (x)\right)^6 + \alpha  c^4\right),
\end{multline}
\begin{multline}
B(x,a)=\alpha  \left(e^{2 a}+4 \sinh ^2(x)\right)^2 \left(c^2-\epsilon  \tan ^{-1}\left(2 e^{-a} \sinh (x)\right)^2\right)^4\\ \times\left(\epsilon  \left(\gamma ^2
   \epsilon -1\right) \tan ^{-1}\left(2 e^{-a} \sinh (x)\right)^2+c^2\right).
\end{multline}

Moreover, $V(a)$ is:
\begin{equation}
V(a)=2 e^{2 a} c^4\int^{\infty}_{\infty} dx \frac{C(x,a)}{D(x,a)},
\end{equation}
where $C(x,a)$ and $D(x,a)$ are given by:
\begin{multline}
C(x,a)= \cosh (2 x) \left(\epsilon  \left(\epsilon 
   \tan ^{-1}\left(2 e^{-a} \sinh (x)\right)^4 \left(c^2
   \left(12 \gamma ^2 \epsilon  \left(4 \gamma ^2
   \epsilon -3\right)+3\right) \right.\right.\right.\\ \left.\left.\left. -2 \alpha  \gamma  \epsilon
    \left(4 \gamma ^2 \epsilon -3\right) \left(4 \gamma
   ^2 \epsilon -1\right)\right)+c^2 \tan ^{-1}\left(2
   e^{-a} \sinh (x)\right)^2 \left(4 \alpha  \gamma 
   \epsilon  \left(3-8 \gamma ^2 \epsilon \right)+3 c^2
   \left(6 \gamma ^2 \epsilon -1\right)\right) \right.\right. \\ \left.\left. +\epsilon
   ^2 \left(2 \gamma ^2 \epsilon -1\right) \left(16
   \gamma ^2 \epsilon  \left(\gamma ^2 \epsilon
   -1\right)+1\right) \tan ^{-1}\left(2 e^{-a} \sinh
   (x)\right)^6-6 \alpha  c^4 \gamma \right)+c^6\right),
\end{multline}
\begin{multline}
D(x,a)=\left(e^{2 a}+4 \sinh ^2(x)\right)^2 \left(c^2-\epsilon 
   \tan ^{-1}\left(2 e^{-a} \sinh (x)\right)^2\right)^4
   \left(-\alpha  \gamma  \epsilon \right. \\ \left. +\epsilon 
   \left(\gamma ^2 \epsilon -1\right) \tan ^{-1}\left(2
   e^{-a} \sinh (x)\right)^2+c^2\right)^2.
\end{multline}
For the clarity of the expressions we have introduced and defined $\alpha$ to be:
\begin{equation}
\alpha \equiv \sqrt{c^2 \tan ^{-1}\left(2 e^{-a} \sinh (x)\right)^2+\epsilon  \left(\gamma ^2 \epsilon -1\right) \tan ^{-1}\left(2 e^{-a} \sinh (x)\right)^4}.
\end{equation}
When $\epsilon=0$ and $\gamma=0$ the expressions for $g(a), V(a)$ revert to those given in \cite{Sutc1992}.

From the lagrangian \ref{lag_sg} we derive the equation of motion
\begin{equation}
g \ddot{a} +\frac{1}{2}\frac{dg}{da}\dot{a}^2+\frac{dV}{da}=0,
\end{equation}
which we solve using the 4th order Runge-Kutta method.
\subsection{Results for sine-Gordon}
First we analyse the scattering of our two kinks for the case $\epsilon=0$ which corresponds to the integrable sine-Gordon model. We compare the trajectories of the kinks as determined using the collective coordinate approximation and using the full numerical simulation for a range of initial velocities in order to determine the effective range of validity of our choice of approximation ansatz.

\begin{figure}
  \centering
    \subfigure[]{\label{fig:e=0,g=0}\includegraphics[trim = 2cm 2cm 12cm 9.8cm, width=0.45\textwidth]{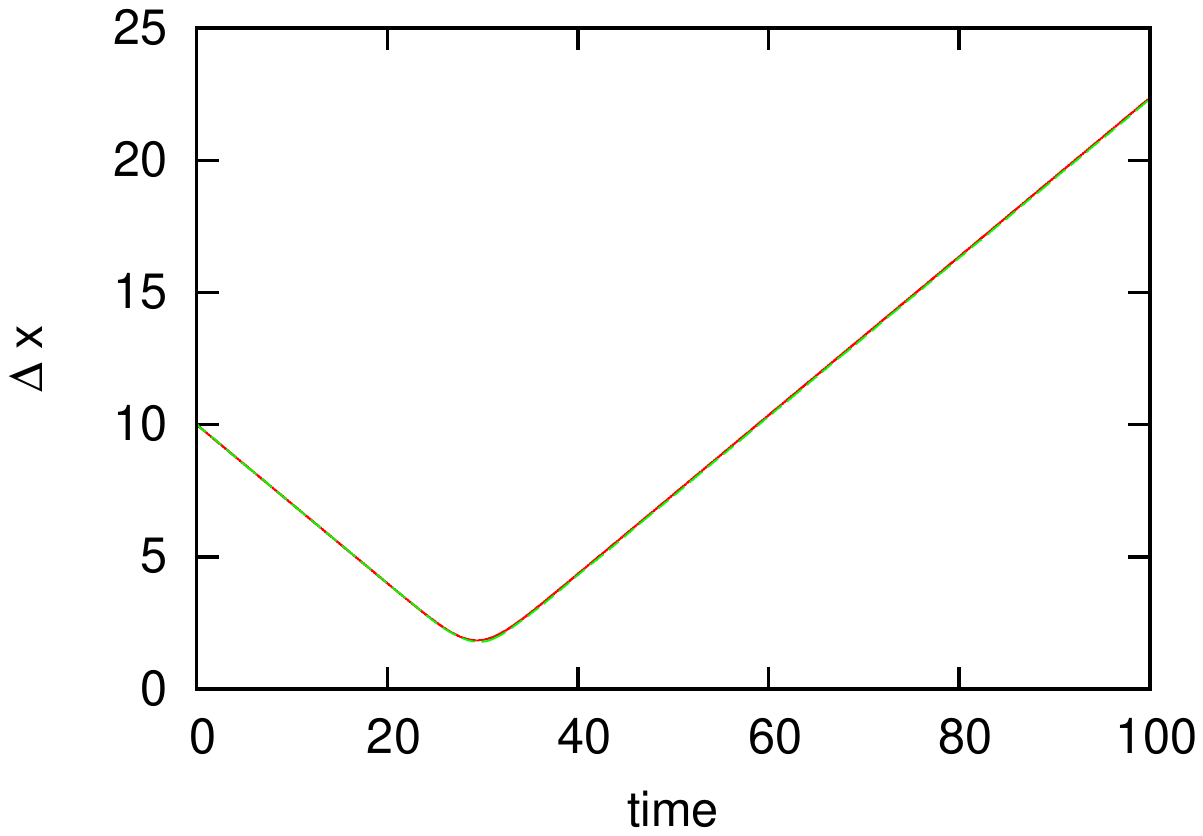}}               
    \subfigure[]{\label{fig:e=0,g=0,v=06}\includegraphics[trim = 2cm 2cm 12cm 9.8cm, width=0.45\textwidth]{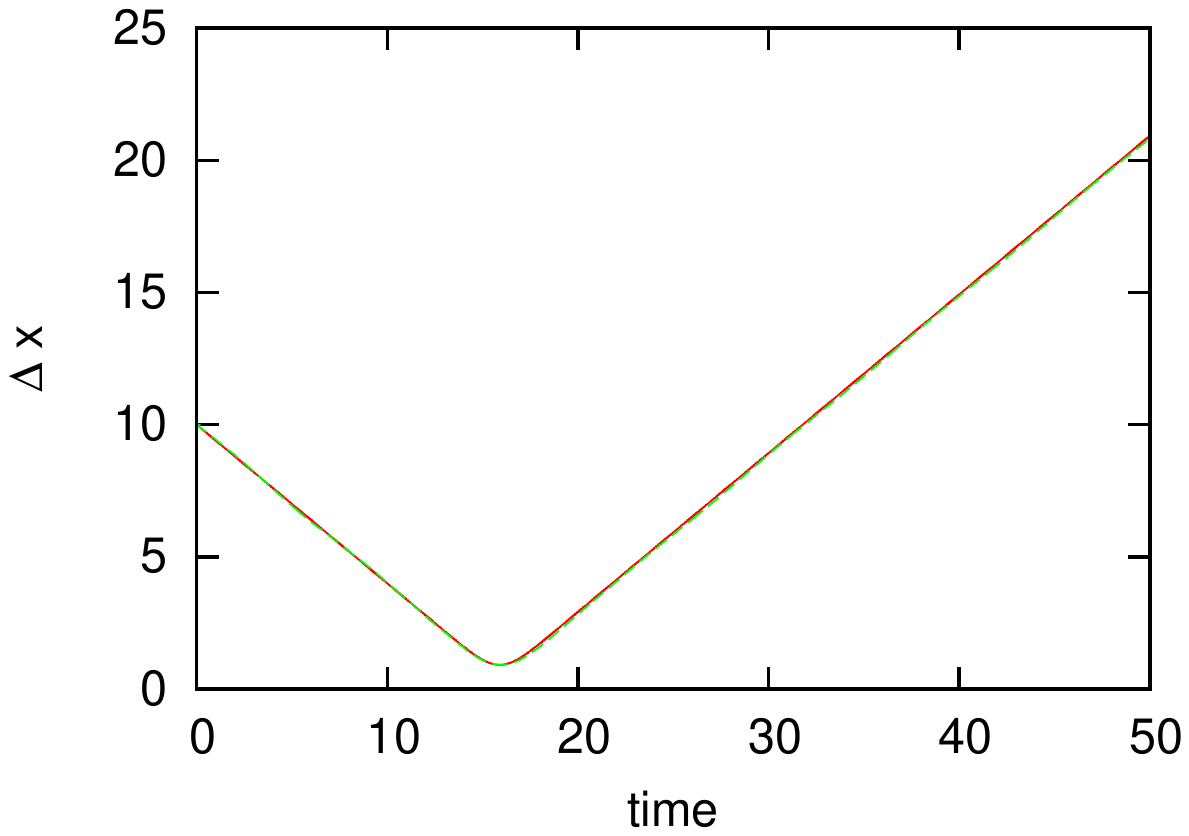}}                \caption{The distance of a soliton from the centre of mass of a system with time. The system consists of two solitons initially with $a(t)=10$, with an initial velocity towards the centre of mass of(a)   $v=0.3$ and (b)   $v=0.6$. For each plot the solid line is result of the collective coordinate approximation and the dashed line is the result of the full simulation (these are often coincident).}
  \label{fig:traj_sg_int}
\end{figure}

In figure \ref{fig:traj_sg_int} we compare the trajectories of the kinks initially placed at $a(t)=10$ and with initial approach velocities of $v=0.3$ and $v=0.6$. We see that in the integrable system the collective coordinate approximation with our choice of ansatz gives excellent agreement with the full numerical simulation up to a high velocity. This gives us confidence in our modified approximation ansatz as applied to our modified model.

\subsection{Results for modified sine-Gordon}
Next we consider the scattering of solitons when the system is no longer integrable, \emph{i.e.} $\epsilon \neq 0$, and analyse the scattering of the two kinks for various values of the parameters $\epsilon$ and $\gamma$. For each set of values we compare the trajectories of the solitons as calculated using the collective coordinate approximation and using the full numerical simulation, and in each simulation we take the initial solitons corresponding to $a=10$. In the collective coordinate approximation the positions of the kinks are equivalent to $\pm a(t)$ when $\epsilon = 0$, but when $\epsilon \neq 0$ the two are no longer equivalent and the location is taken to be the position of the maximum of the energy peak.

\begin{figure}
  \centering
  \subfigure[]{\label{fig:e=-02,g=0}\includegraphics[trim = 2cm 2cm 12cm 9.8cm, clip, width=0.45\textwidth]{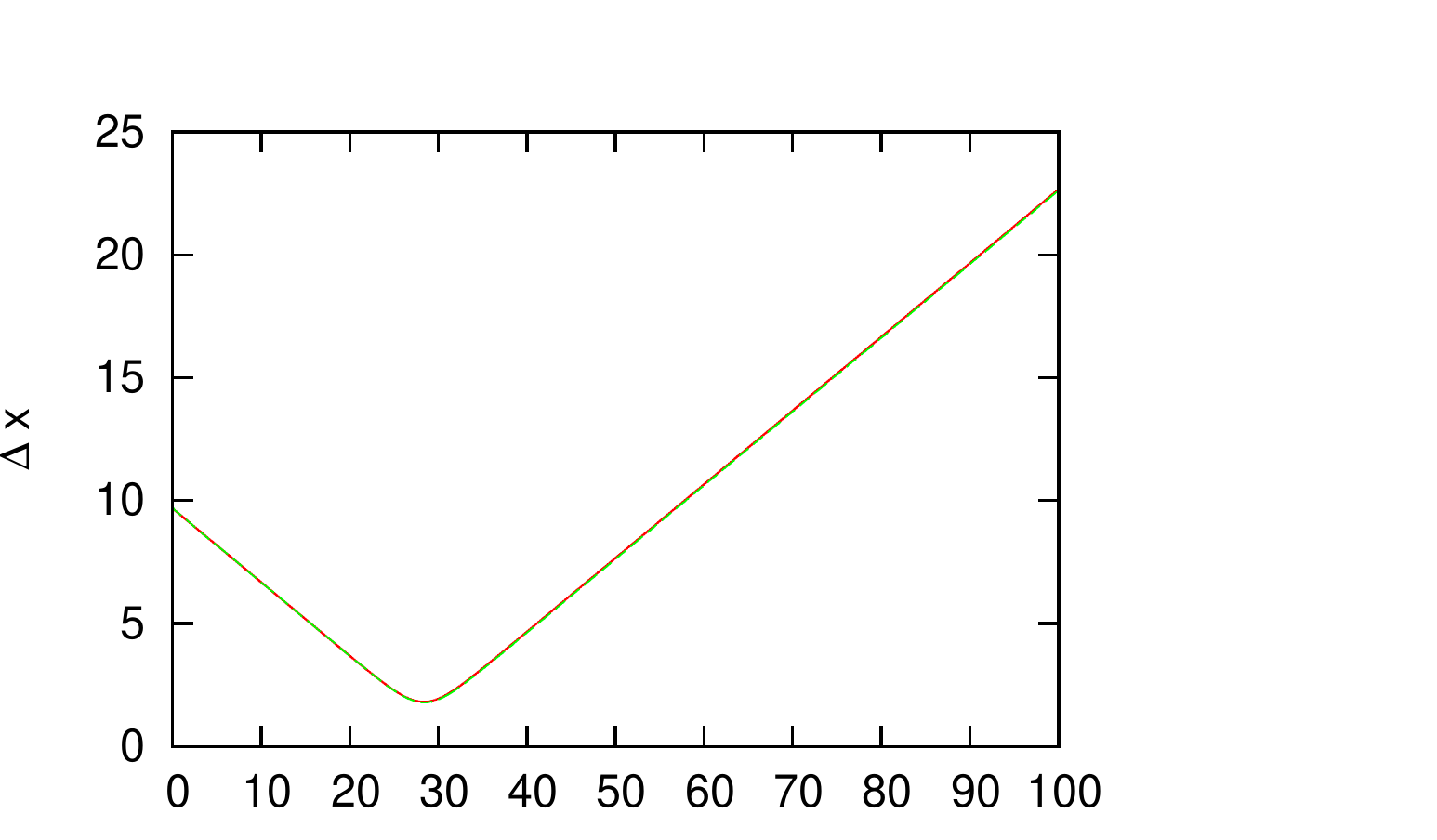}}
    \subfigure[]{\label{fig:e=04,g=0}\includegraphics[trim = 2cm 2cm 12cm 9.8cm, clip, width=0.45\textwidth]{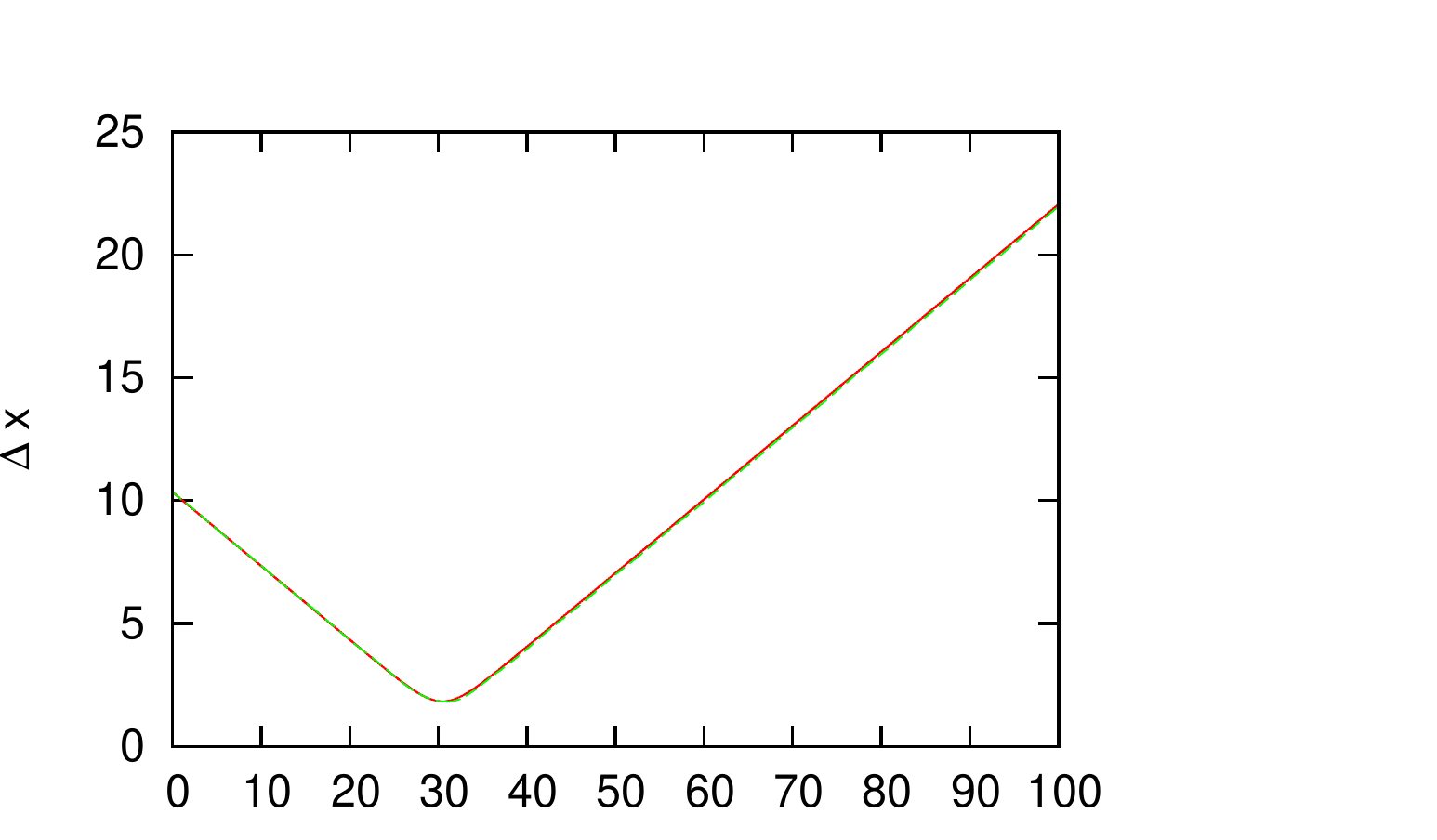}} 
       \subfigure[]{\label{fig:e=1,g=0}\includegraphics[trim = 2cm 2cm 12cm 9.8cm, clip, width=0.45\textwidth]{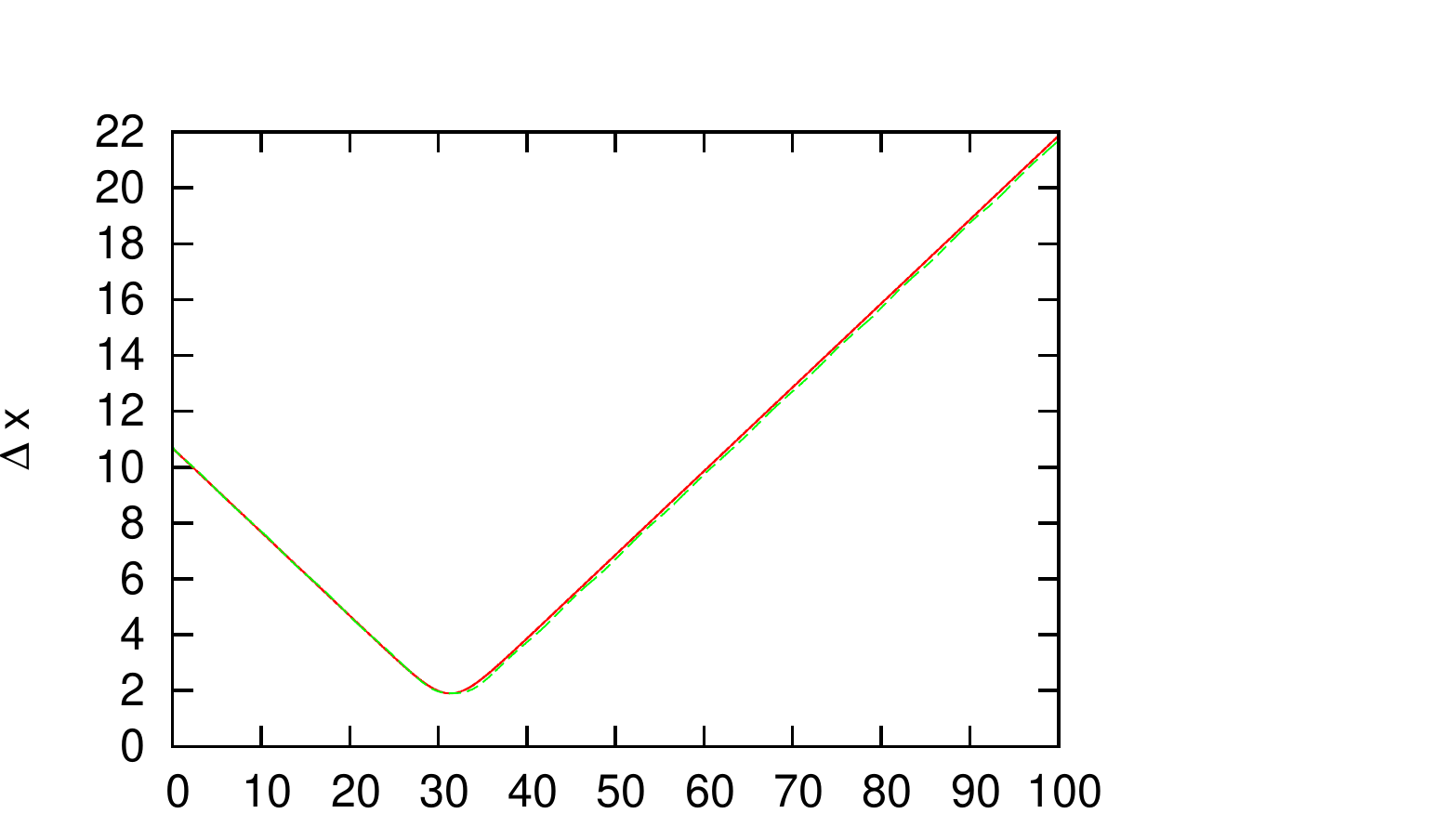}}
    \subfigure[]{\label{fig:e=04,g=01}\includegraphics[trim = 2cm 2cm 12cm 9.8cm, width=0.45\textwidth]{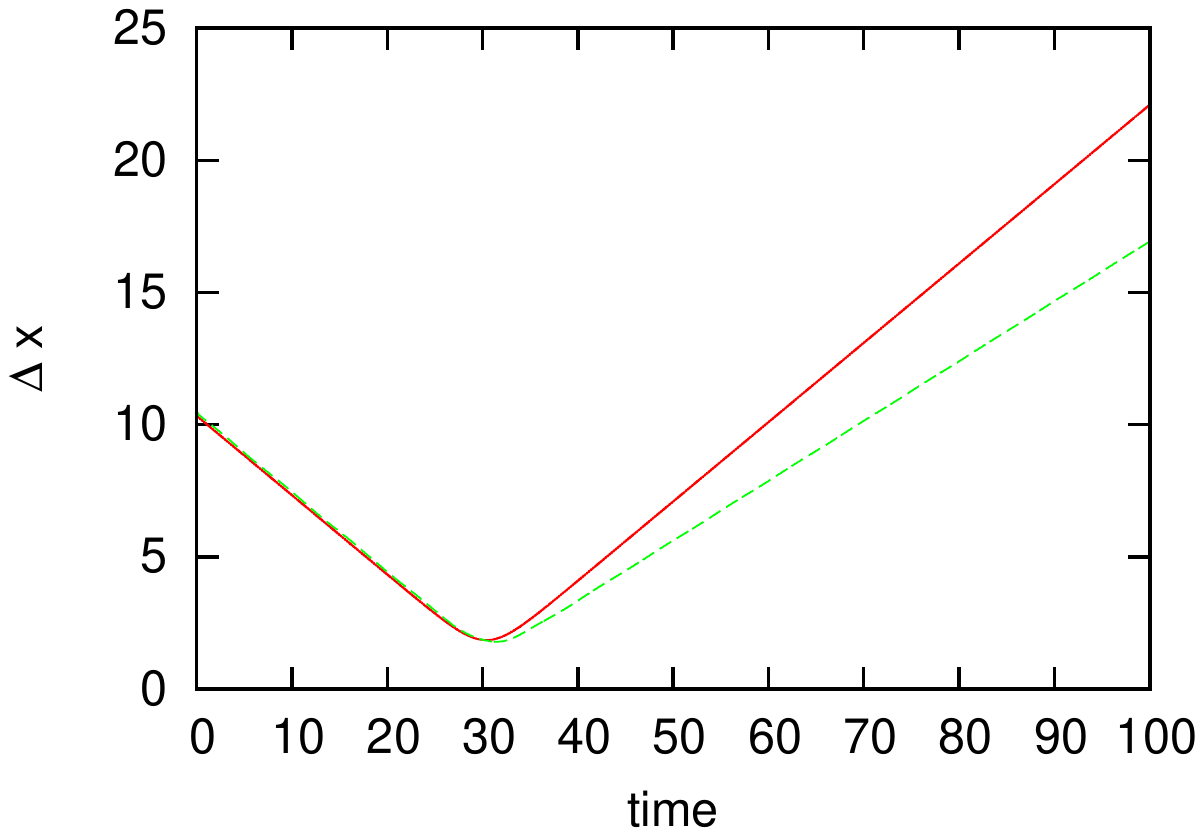}}                
  \subfigure[]{\label{fig:e=04,g=02}\includegraphics[trim = 2cm 2cm 12cm 9.8cm, clip, width=0.45\textwidth]{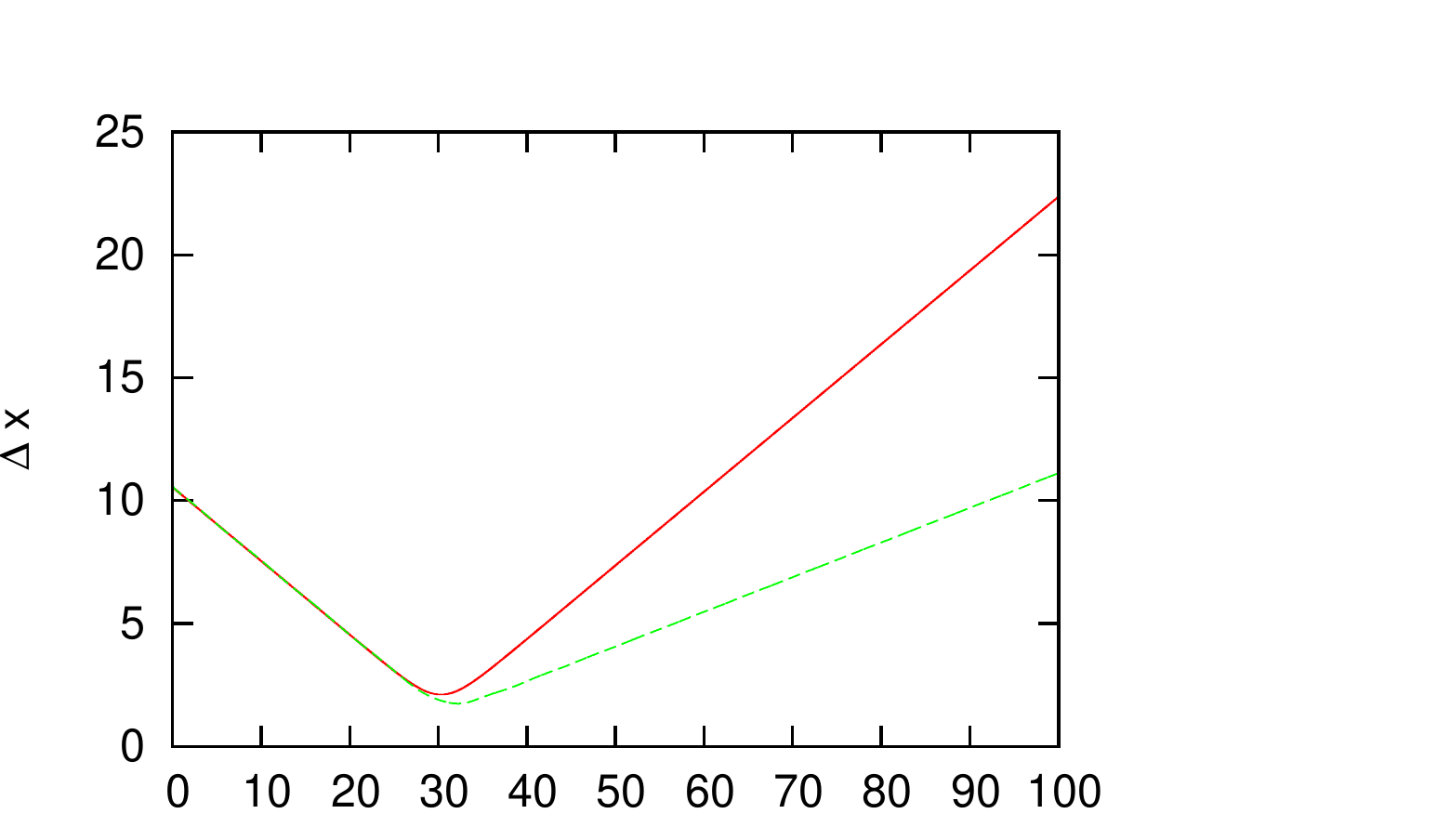}}
  \subfigure[]{\label{fig:e=04,g=-02}\includegraphics[trim = 2cm 2cm 12cm 9.8cm, clip, width=0.45\textwidth]{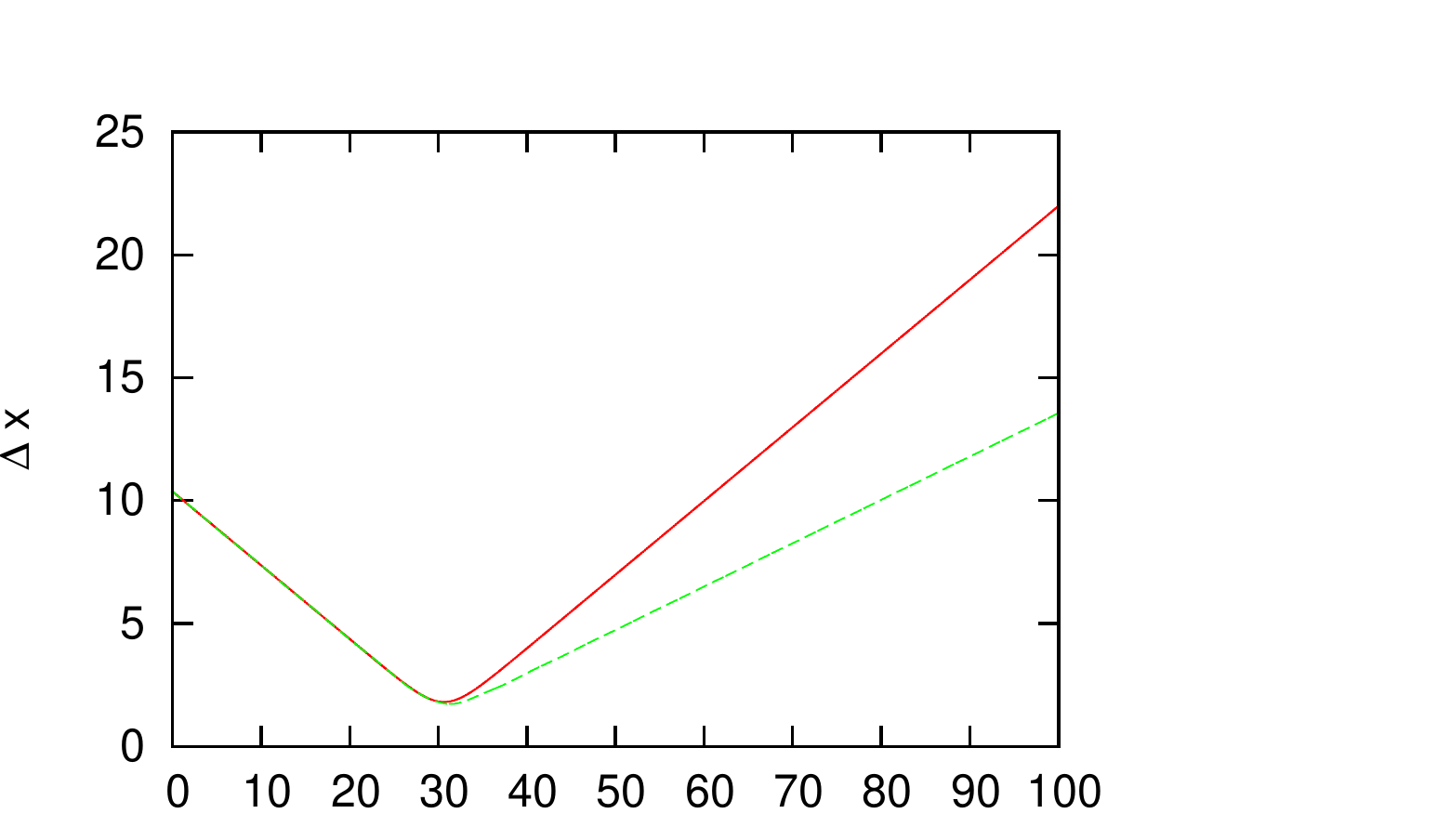}}
  \caption{The distance of a soliton from the centre of mass of a system with time. The system consists of two solitons initially with $a(t)=10$, each with an initial velocity of $0.3$ towards the centre of mass. Initial parameter are (a) $\epsilon=-0.2$, $\gamma=0$; (b)  $\epsilon=0.4$,  $\gamma=0$; (c) $\epsilon=1$, $\gamma=0$;  (d) $\epsilon=0.4$, $\gamma=0.1$; (e) $\epsilon=0.4$, $\gamma=0.2$; and (f) $\epsilon=0.4$, $\gamma=-0.2$. For each plot the solid line is result of the collective coordinate approximation and the dashed line is the result of the full simulation (these are often coincident).}
  \label{fig:traj_sg}
\end{figure}

In figure \ref{fig:traj_sg} we present a series of plots of trajectories for solitons sent together with an initial velocity of $v=0.3$ for different values of $\epsilon$ and $\gamma$. From these plots we can see that the two approaches show excellent agreement when the symmetry necessary for quasi-integrability is present, \emph{i.e.} $\epsilon \neq 0$ and $\gamma=0$. However, when the system moves away from quasi-integrability, \emph{i.e.} $\epsilon \neq 0$ and $\gamma \neq 0$, the two methods show good agreement as the solitons approach each other but the solitons scatter at slightly different distances and with different velocities. This suggests that quasi-integrability is a sufficient condition for the collective coordinate approximation to accurately model trajectories of kinks in modified sine-Gordon systems.

Next we consider the quasi-conservation of the first non-trivial charge beyond the energy itself, namely, $\tilde{Q}^{(4)}(t)$ defined in \eqref{con_laws_sg} by calculating both the anomaly $\tilde{\beta}^{(3)}$ and the time integrated anomaly which is given by:
\begin{equation}
\tilde{\chi}^{(3)}=-\frac{1}{2}\int^{t}_{t_0} dt' \,\tilde{\beta}^{(3)}=4\int^{t}_{t_0} dt'\int^{\infty}_{-\infty} dx \,\partial_- \phi \, \partial_-^2 \phi \left[ \frac{d^2 \,V}{d\, \phi^2} + 16 V-1\right]
\end{equation}
where $\partial_-=\partial_t-\partial_x$ and $t_0$ is the initial time of the simulation which is usually taken to be zero.

\begin{figure}
  \centering
    \subfigure[]{\label{fig:ano_sg,e=0.000001,g=0.00001}\includegraphics[trim = 2cm 2cm 12cm 9.8cm, width=0.32\textwidth]{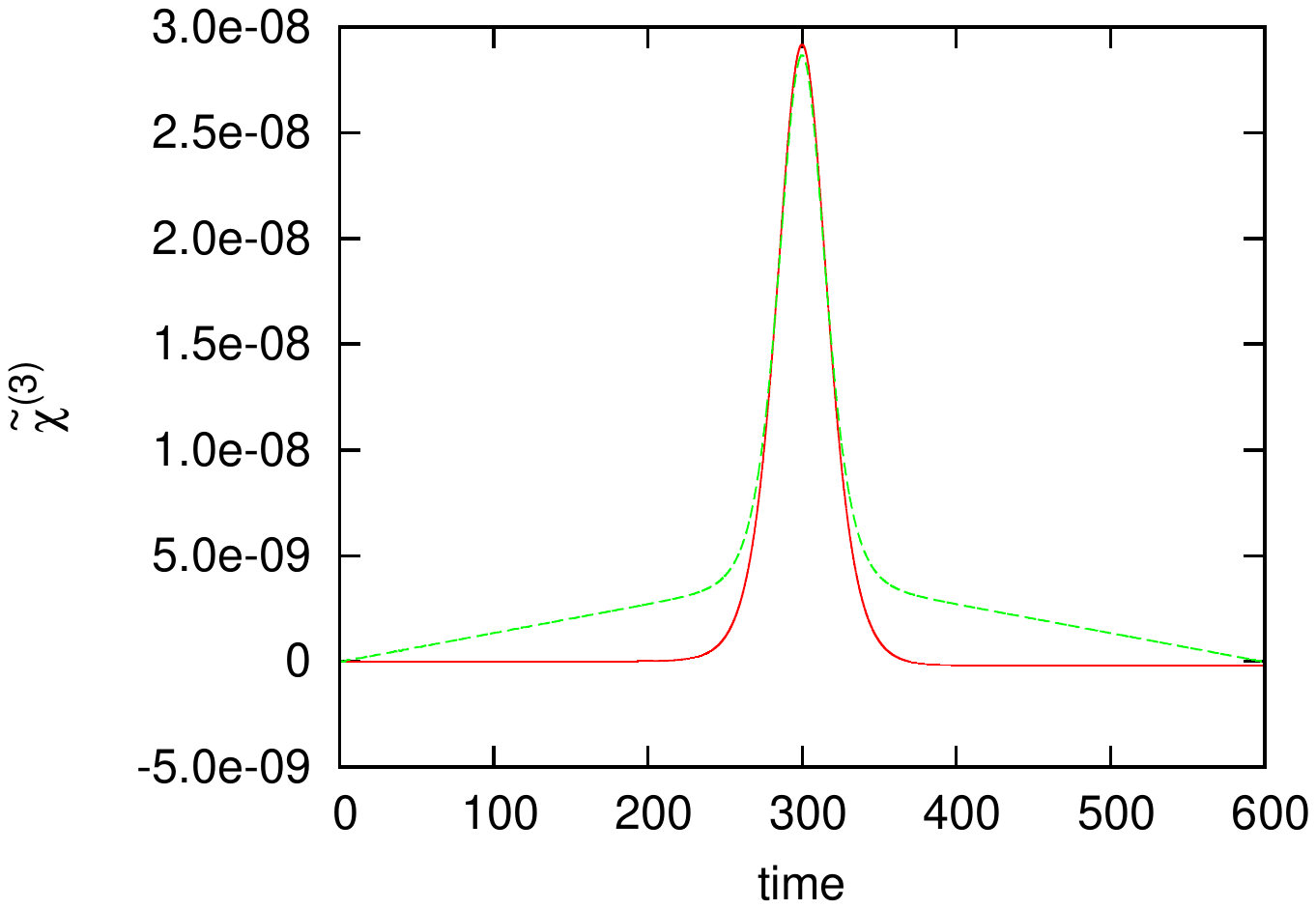}} 
    \subfigure[]{\label{fig:ano_sg,e=0.000001,g=0.002}\includegraphics[trim = 2cm 2cm 12cm 9.8cm, width=0.32\textwidth]{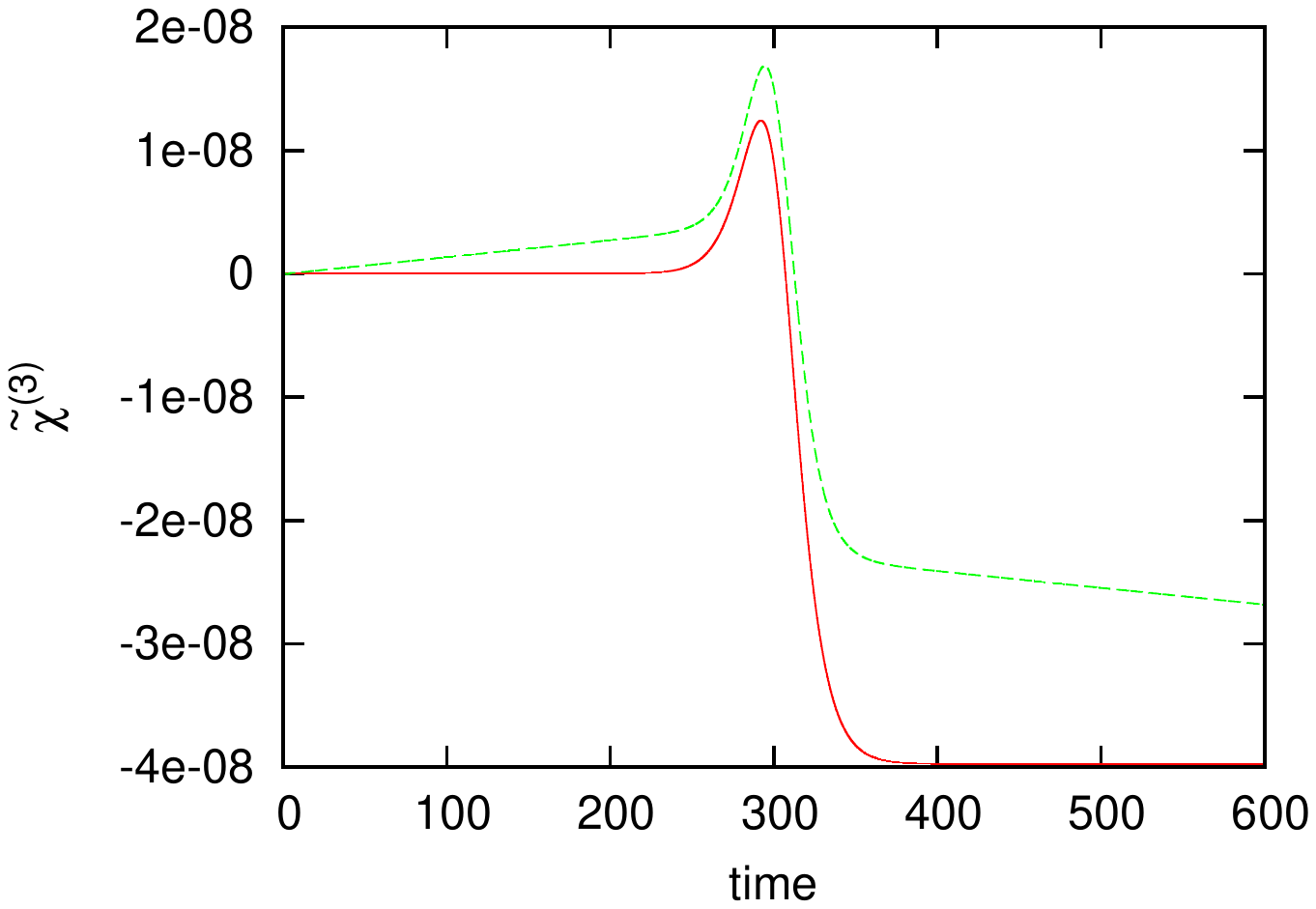}}                
  \subfigure[]{\label{fig:ano_sg,e=0.000001,g=0.004}\includegraphics[trim = 2cm 2cm 12cm 9.8cm, width=0.32\textwidth]{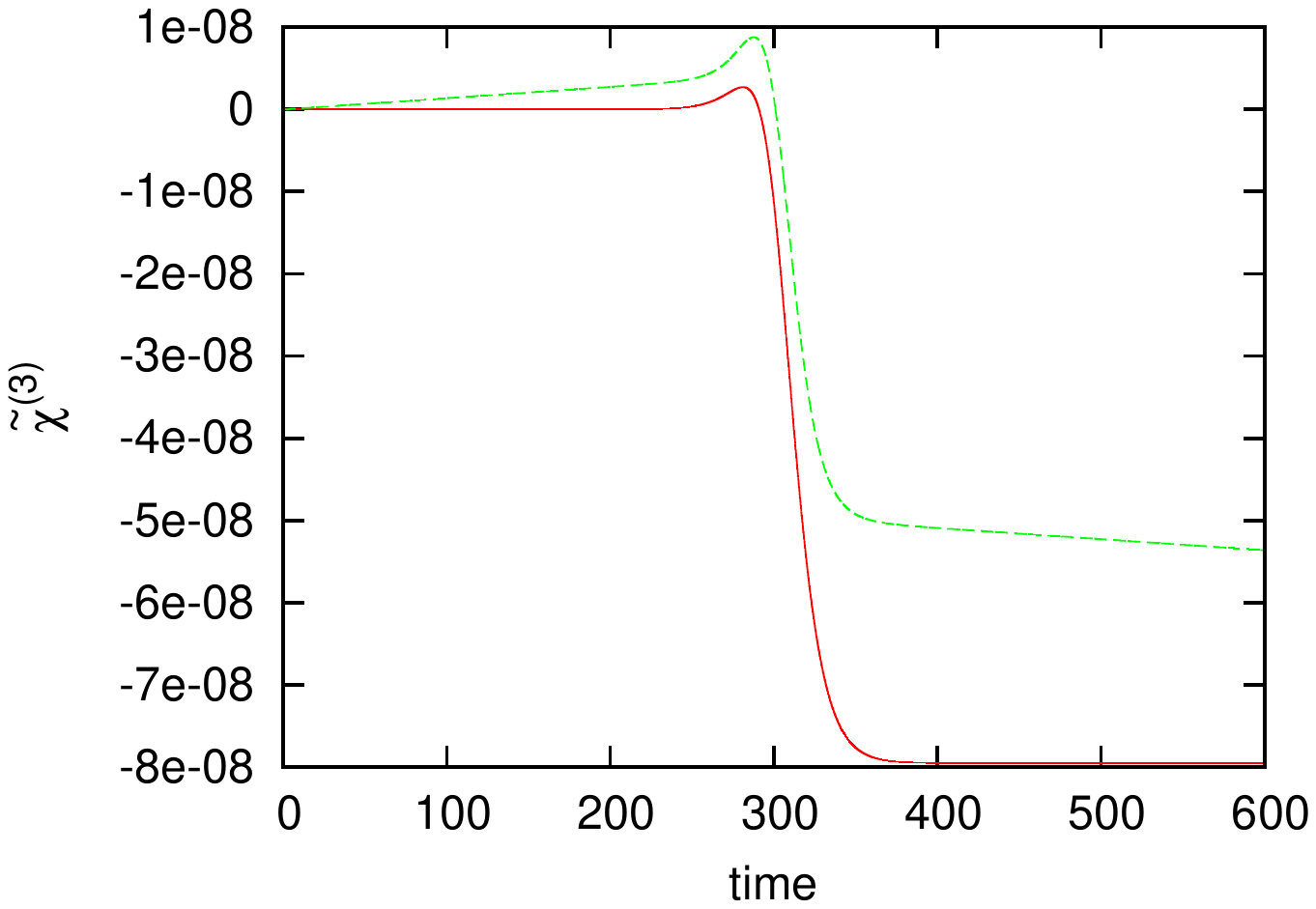}}                
  \subfigure[]{\label{fig:ano_sg,e=0.000001,g=0.1}\includegraphics[trim = 2cm 2cm 12cm 9.8cm, width=0.32\textwidth]{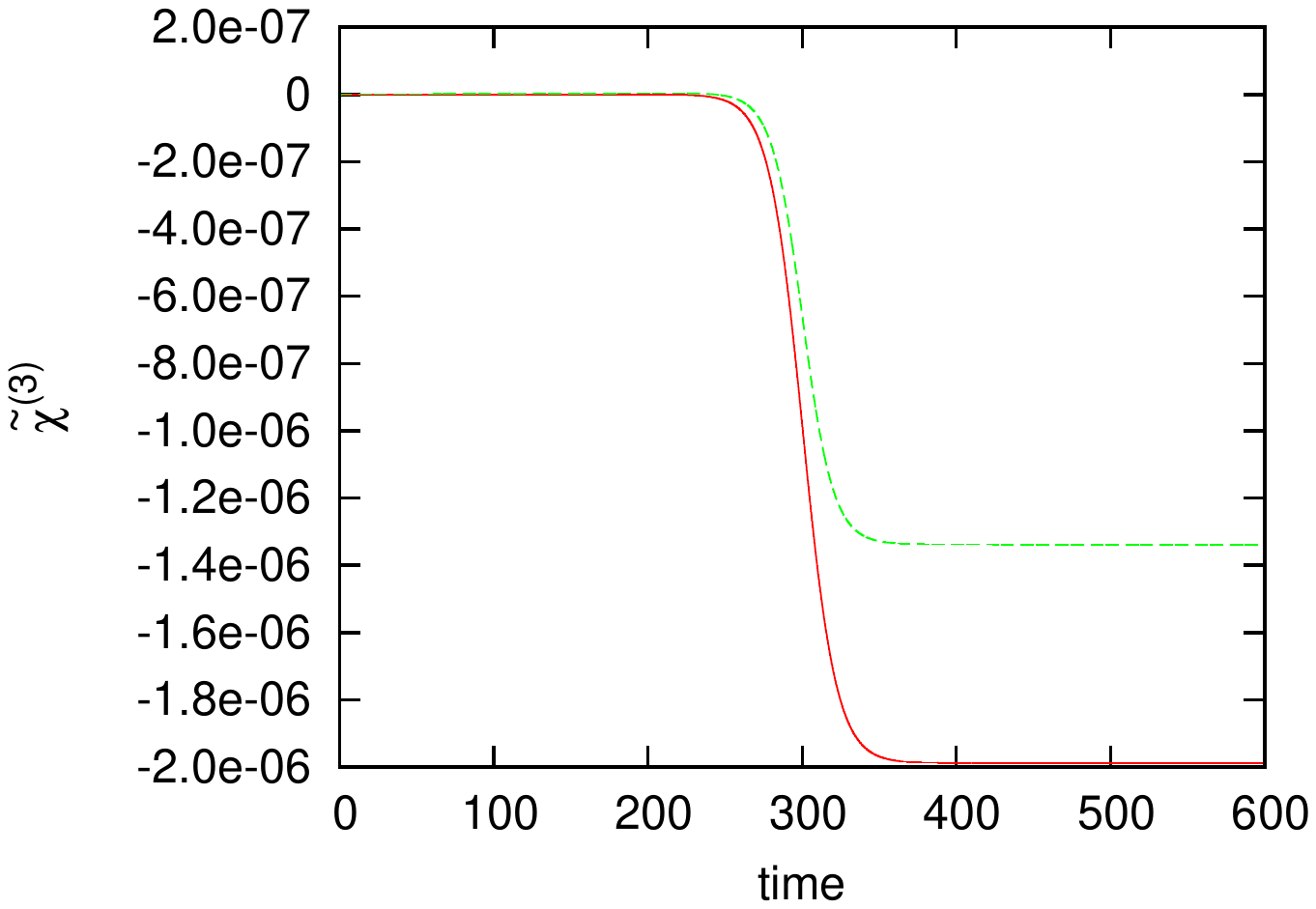}}
  \subfigure[]{\label{fig:ano_sg,e=0.000001,g=-0.002}\includegraphics[trim = 2cm 2cm 12cm 9.8cm, width=0.32\textwidth]{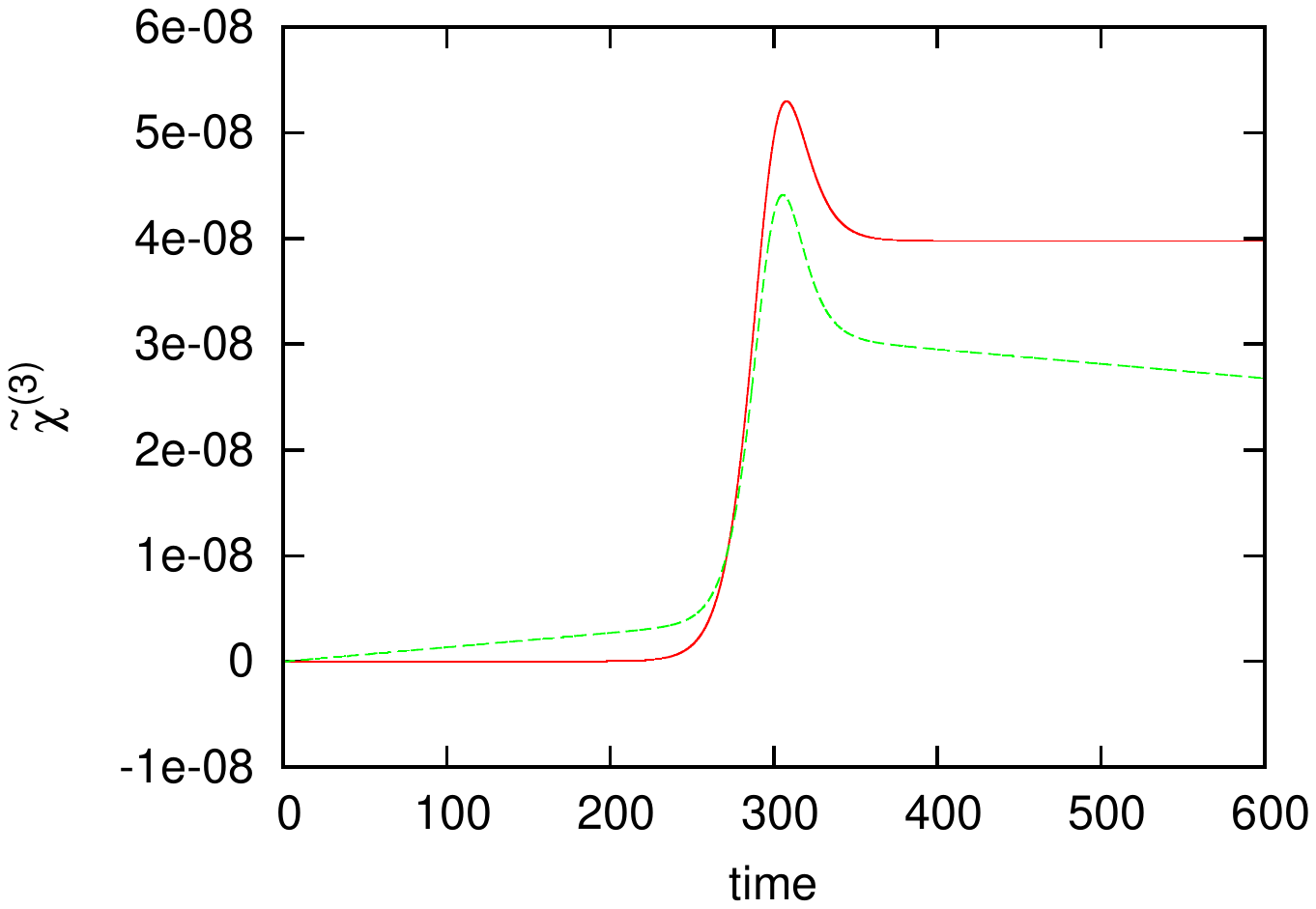}}   
  \caption{The time-integrated anomaly for solitons initially at $\pm 20$ with velocity $0.05$ towards the centre of mass and $\epsilon=0.000001$. $\gamma$ is chosen to be (a) $\gamma=0.00001$, (b) $\gamma=0.002$, (c) $\gamma=0.004$, (d) $\gamma=0.1$ and (e) $\gamma=-0.002$. The solid lines are the results for the collective coordinate approximation and the dashed lines are results for the full simulation results.}
  \label{fig:ano_sg}
\end{figure}

Figure \ref{fig:ano_sg} is the plot of the time-integrated anomaly with time for solitons placed at $\pm 20$ with initial velocity $v=0.05$, with $\epsilon=0.000001$ and various values of $\gamma$. Notice that in the full simulation the time-integrated anomaly is always slowly increasing prior to the scattering of the solitons and slowly decreasing after the scattering; this is due to slight fluctuations away from zero in the anomaly which by itself id probably a result of numerical errors rather than any physical effect. This error increases as $\epsilon$ increases and so it becomes difficult to compare the results, this is why we present plots only for a small value of $\epsilon$. We see that when $\gamma$ is small the collective coordinate approximation and the full simulation are in  excellent agreement, and far away from the scattering the time-integrated anomaly is close to zero, as expected, when $\gamma$ is small and the model is close to the symmetry described in \eqref{phi_trans_sg} and \eqref{pot_trans_sg}. When $\gamma$ is taken further from zero we move from a model with approximate symmetry to a model where this symmetry is broken. This is confirmed by our results as seen in figure \ref{fig:ano_sg} which show that the further $\gamma$ is from zero the further the time-integrated anomaly is from zero after the scattering of the solitons. Moreover, the figures \ref{fig:ano_sg,e=0.000001,g=0.002} and \ref{fig:ano_sg,e=0.000001,g=-0.002} show that the symmetry can be broken in either direction depending on the sign of $\gamma$. The collective coordinate approximation still gives a good qualitative approximation to the behaviour of the time-integrated anomaly as we move away from the symmetric case though the values, as to be expected, are not exactly the same as seen in full simulations. These observations have been checked for several values of $\epsilon$.

\section{Conclusions}

In this paper we have considered the applicability of the collective coordinate approximation to the description of the scattering of solitons in models which are deformations of the NLS or the sine-Gordon model. The deformation of these models changes their integrability properties, either completely or partially (the models become 'quasi-integrable'). Hence we have considered a modified NLS and a modified sine-Gordon system for which the trajectories were already known from full simulations studied in \cite{Zakr2012} and \cite{Zakr2014}.  Moreover, \cite{Zakr2012} and \cite{Zakr2014} have also suggested that quasi-integrability could be related to a particular symmetry of the field configurations (for configurations possessing the necessary symmetry  the anomaly terms could vanish and so 
lead to quasi-integrability), so we looked at these properties
using the collective coordinate approximation.

In the modified NLS the approximation works very well in the majority of cases and for a good range of initial conditions with a well chosen approximation ansatz. The predominant influence on the accuracy of the approximation is the time the two solitons spend in close proximity of each other during their interaction; and for simulations where the solitons do not come closer together than the width of one soliton the collective coordinate approximation accurately reproduces the scattering of the solitons and their anomaly even for an initial velocity of $v=0.2$. In these cases the trajectories of the solitons during their scattering, as calculated using the collective coordinate approximation, are often indistinguishable from the trajectories calculated using a full numerical method. Moving the system away from integrability, \emph{i.e.} increasing the deformation parameter $\epsilon$, reduces the accuracy only slightly when the solitons stay far enough apart during their scattering and the results are very good for $\epsilon$ up to at least $\epsilon=0.3$. For the vast majority of initial conditions the solitons maintain enough distance from each other during their interaction to ensure the accuracy of the approximation, and only in the most attractive cases does the approximation become less accuracte.

 When the solitons come closer together than the width of one soliton during their interaction the general behaviour of the solitons (both their trajectory and anomaly) is still reproduced but the accuracy of the approximation is reduced. We think this is probably because in the full simulation the solitons deform one another away from the form given by \eqref{ansatz} when they are in close proximity, and when $\epsilon\neq 0$ they also radiate out energy, and the collective coordinate approximation does not allow this to happen. This effect is exacerbated as the system moves away from integrability because this radiation/deformation effect increases with increasing $\epsilon$.

 The effect of quasi-integrability in the modified NLS is impossible to assess as the field configurations possess the necessary symmetries for quasi-integrability when the solitons are most attractive or repulsive (\emph{i.e.} when $\delta$ is an integer value of $\pi$) and the effect of the solitons proximity during their interaction eclipses any effect of quasi-integrability.

In the modified sine-Gordon case the approximation very accurately describes the trajectories and anomalies of scattering kinks when the system is either integrable or quasi-integrable (\emph{i.e.} $\gamma=0$) up to initial velocities of $v=0.6$ and for values of $\epsilon=1$. However, when the field configuration moves away from the symmetry necessary for quasi-integrability (\emph{i.e.} when $\gamma$ moves away from $0$) the collective coordinate approximation becomes less accurate for both the trajectories and the anomalies.

These observations suggest that the collective coordinate approximation is a suitable method to study various properties 
of the scatterings of solitons and so can be used also to  investigate quasi-integrability in other perturbations of integrable models. In modified sine-Gordon models the presence of the symmetries necessary for quasi-integrability seem to be a sufficient condition to ensure accuracy, but in any model care should be taken if the solitons have the opportunity to strongly deform each other.

\section{Acknowledgements}
HB is supported by an STFC studentship. 
\appendix
\section{Appendix}
\label{appendix}

\begin{multline}
I_{\xi_1}=\int^{\infty}_{\! - \!\infty} dx \frac{1}{4}  \left(\frac{a_1 \sqrt{\epsilon \! + \!2} }{\text{cosh}(\omega_ 1)}\right)^{\frac{1}{\epsilon \! + \!1}} \left(
\! -  4 i a_1 \left(e^{2 i   \theta_ 1}\! - \!e^{2 i \theta_ 2}\right) \tanh   (\omega_1) \left(\frac{a_2 \sqrt{\epsilon \! + \!2}}{\text{cosh}(\omega_2)}\right)^{\frac{1}{\epsilon \! + \!1}}\right. \\ \left. \! +  \mu_1 \left(2 e^{i   (\theta_ 1\! + \!\theta_ 2)} \left(\frac{a_1 \sqrt{\epsilon \! + \!2}}{\text{cosh}(\omega_1)}\right)^{\frac{1}{\epsilon \! + \!1}}\! + \!\left( e^{2 i   \theta_ 1}\! + \!e^{2 i \theta_ 2}\right)
 \left(\frac{a_2 \sqrt{\epsilon \! + \!2} }{\text{cosh}(\omega_ 2)}\right)^{\frac{1}{\epsilon \! + \!1}} \right)\right)
\end{multline}
\begin{multline}
I_{\xi_2}=\int^{\infty}_{\! - \!\infty} dx \frac{1}{4} \left(\frac{a_2 \sqrt{\epsilon \! + \!2} }{\text{cosh}(\omega_ 2)}\right)^{\frac{1}{\epsilon \! + \!1}} \left(4   i a_2 \left(e^{2 i \theta_ 1}\! - \!e^{2 i \theta_ 2}\right)   \tanh (\omega_2) \left(\frac{a_1 \sqrt{\epsilon \! + \!2}}{\text{cosh}(\omega_1)}\right)^{\frac{1}{\epsilon \! + \!1}} \right. \\ \left.\! +  \mu_2 \left( \left( e^{2 i   \theta_ 1}\! + \!e^{2 i \theta_ 2}\right) \left(\frac{a_1 \sqrt{\epsilon \! + \!2} }{\text{cosh}(\omega_ 1)}\right)^{\frac{1}{\epsilon \! + \!1}}
\! + \!2 e^{i (\theta_ 1\! + \!\theta_ 2)} \left(\frac{a_2 \sqrt{\epsilon \! + \!2}}{\text{cosh}(\omega_2)}\right)^{\frac{1}{\epsilon \! + \!1}}\right)\right)
\end{multline}
\begin{multline}
I_{\mu_1}=\int^{\infty}_{\! - \!\infty} dx \frac{(\xi_ 1\! + \!2 x)}{4}  \left(\frac{a_1 \sqrt{\epsilon \! + \!2}}{\text{cosh}(\omega_1)}\right)^{\frac{1}{\epsilon \! + \!1}} \left(2   e^{i (\theta_ 1\! + \!\theta_ 2)} \left(\frac{a_1 \sqrt{\epsilon \! + \!2}}{\text{cosh}(\omega_1)}\right)^{\frac{1}{\epsilon \! + \!1}} \right. \\ \left.\! + \!\left(e^{2 i   \theta_ 1}\! + \!e^{2 i \theta_ 2}\right) \left(\frac{a_2 \sqrt{\epsilon \! + \!2} }{\text{cosh}(\omega_ 2)}\right)^{\frac{1}{\epsilon \! + \!1}} \right)
\end{multline}
\begin{multline}
I_{\mu_2}=\int^{\infty}_{\! - \!\infty} dx \frac{(\xi_ 2\! + \!2 x)}{4} \left(\frac{a_2 \sqrt{\epsilon \! + \!2}}{\text{cosh}(\omega_2)}\right)^{\frac{1}{\epsilon \! + \!1}}   \left( \left(e^{2 i \theta_ 1}\! + \!e^{2 i   \theta_ 2}\right) \left(\frac{a_1 \sqrt{\epsilon \! + \!2}}{\text{cosh}(\omega_1)}\right)^{\frac{1}{\epsilon \! + \!1}} \right. \\ \left. \! +  2 e^{i (\theta_ 1\! + \!\theta_ 2)} \left(\frac{a_2 \sqrt{\epsilon \! + \!2}}{\text{cosh}(\omega_2)}\right)^{\frac{1}{\epsilon \! + \!1}}\right)
\end{multline}
\begin{multline}
I_{a_1}=\int^{\infty}_{\! - \!\infty} dx \frac{1}{a_1(\epsilon \! + \!1)} \left(\frac{a_1 \sqrt{\epsilon \! + \!2}}{\text{cosh}(\omega_1)}\right)^{\frac{1}{\epsilon \! + \!1}} \left(i   \left(e^{2 i \theta_ 1}\! - \!e^{2 i \theta_ 2}\right)   \left(\frac{a_2 \sqrt{\epsilon \! + \!2} }{\text{cosh}(\omega_ 2)}\right)^{\frac{1}{\epsilon \! + \!1}}\right.  \\  \left.\! - 2   a_1^2 \,t\, (\epsilon \! + \!1) \left(2 e^{i (\theta_ 1\! + \!\theta_ 2)}   \left(\frac{a_1 \sqrt{\epsilon \! + \!2} }{\text{cosh}(\omega_ 1)}\right)^{\frac{1}{\epsilon \! + \!1}} \! + \!\left(e^{2 i \theta_ 1}\! + \!e^{2 i \theta_ 2}\right) \left(\frac{a_2 \sqrt{\epsilon \! + \!2}}{\text{cosh}(\omega_2)}\right)^{\frac{1}{\epsilon \! + \!1}}\right) \right. \\ \left. \! - i a_1 (\epsilon \! + \!1) \left(e^{2   i \theta_ 1}\! - \!e^{2 i \theta_ 2}\right) (\xi_1\! + \!x) \tanh (\omega_1) \left(\frac{a_2 \sqrt{\epsilon \! + \!2}}{\text{cosh}(\omega_2)}\right)^{\frac{1}{\epsilon \! + \!1}} \right)
\end{multline}
\begin{multline}
I_{a_2}=\int^{\infty}_{\! - \!\infty} dx \frac{1}{a_2   (\epsilon \! + \!1)} \left(\frac{a_2 \sqrt{\epsilon \! + \!2}}{\text{cosh}(\omega_2)}\right)^{\frac{1}{\epsilon \! + \!1}} \left( \! - i \left(e^{2 i \theta_ 1}\! - \!e^{2 i \theta_ 2}\right) \left(\frac{a_1 \sqrt{\epsilon \! + \!2}}{\text{cosh}(\omega_1)}\right)^{\frac{1}{\epsilon \! + \!1}} \right. \\ \left.
\! - 2   a_2^2\,t\, (\epsilon \! + \!1) \left(\left(e^{2 i \theta_ 1}\! + \!e^{2 i \theta_ 2}\right) \left(\frac{a_1 \sqrt{\epsilon \! + \!2}}{\text{cosh}(\omega_1)}\right)^{\frac{1}{\epsilon \! + \!1}}\! + \!2 e^{i (\theta_ 1\! + \!\theta_ 2)} \left(\frac{a_2 \sqrt{\epsilon \! + \!2}}{\text{cosh}(\omega_2)}\right)^{\frac{1}{\epsilon \! + \!1}}\right) \right. \\ \left.\! +  i   a_2 (\epsilon \! + \!1) \left(e^{2 i \theta_ 1}\! - \!e^{2 i \theta_ 2}\right) (\xi_ 2\! + \!x) \tanh (\omega_2) \left(\frac{a_1 \sqrt{\epsilon \! + \!2}}{\text{cosh}(\omega_1)}\right)^{\frac{1}{\epsilon \! + \!1}} \right)
\end{multline}
\begin{equation}
I_{\lambda_1}=\int^{\infty}_{\! - \!\infty} dx  \left(\frac{a_1 \sqrt{\epsilon \! + \!2}}{\text{cosh}(\omega_1)}\right)^{\frac{1}{\epsilon \! + \!1}} \left(\! - 2   e^{i (\theta_ 1\! + \!\theta_ 2)} \left(\frac{a_1 \sqrt{\epsilon \! + \!2}}{\text{cosh}(\omega_1)}\right)^{\frac{1}{\epsilon \! + \!1}}\! - \!\left(e^{2 i   \theta_ 1} \! + \!e^{2 i \theta_ 2}\right)\left(\frac{a_2 \sqrt{\epsilon \! + \!2} }{\text{cosh}(\omega_ 2)}\right)^{\frac{1}{\epsilon \! + \!1}}\right)
\end{equation}
\begin{equation}
I_{\lambda_2}=\int^{\infty}_{\! - \!\infty} dx \left(\frac{a_2 \sqrt{\epsilon \! + \!2}}{\text{cosh}(\omega_2)}\right)^{\frac{1}{\epsilon \! + \!1}}   \left(\! - \! \left(e^{2 i \theta_ 1}\! + \!e^{2 i   \theta_ 2}\right) \left(\frac{a_1 \sqrt{\epsilon \! + \!2}}{\text{cosh}(\omega_1)}\right)^{\frac{1}{\epsilon \! + \!1}}\! - \!2 e^{i (\theta_ 1\! + \!\theta_ 2)} \left(\frac{a_2 \sqrt{\epsilon \! + \!2}}{\text{cosh}(\omega_2)}\right)^{\frac{1}{\epsilon \! + \!1}}\right)
\end{equation}
\begin{multline}
 V=\int^{\infty}_{\! - \!\infty} dx \left[ \left(a_1^2 \left(\frac{a_1 \sqrt{\epsilon \! + \!2}}{ \text{cosh}(\omega_1)}\right)^{\frac{1}{\epsilon \! + \!1}} \left(2e^{i (\theta_1\! + \!\theta_2)} \left(\frac{a_1 \sqrt{\epsilon \! + \!2}}{ \text{cosh}(\omega_1)}\right)^{\frac{1}{\epsilon \! + \!1}}\! + \!\left(e^{2 i\theta_1}\! + \!e^{2 i \theta_2}\right) \left(\frac{a_2 \sqrt{\epsilon \! + \!2}}{ \text{cosh}(\omega_2)}\right)^{\frac{1}{\epsilon \! + \!1}}\right) \right. \right. \\ \left. \left. a_2^2 \left(\frac{a_2\sqrt{\epsilon \! + \!2}}{\text{cosh}(\omega_2)}\right)^{\frac{1}{\epsilon \! + \!1}}\left( \left(e^{2 i \theta_1}\! + \!e^{2 i\theta_2}\right) \left(\frac{a_1 \sqrt{\epsilon \! + \!2}}{ \text{cosh}(\omega_1)}\right)^{\frac{1}{\epsilon \! + \!1}}\! + \!2 e^{i (\theta_1\! + \!\theta_2)} \left(\frac{a_2 \sqrt{\epsilon \! + \!2}}{ \text{cosh}(\omega_2)}\right)^{\frac{1}{\epsilon\! + \!1}}\right)\right)  \right. \\ \left.  \! - \! \left( \left( \frac{i\mu_1}{2} \! - \!a_1 \tanh (\omega_1) \right) e^{i \theta_2} \left(\frac{a_1 \sqrt{\epsilon \! + \!2}}{ \text{cosh}(\omega_1)}\right)^{\frac{1}{\epsilon\! + \!1}}\! + \!\left( \frac{i\mu_2}{2}\! - \! a_2  \tanh (\omega_2) \right)e^{i \theta_1} \left(\frac{a_2\sqrt{\epsilon \! + \!2}}{\text{cosh}(\omega_2)}\right)^{\frac{1}{\epsilon \! + \!1}}  \right)   \right. \\ \left. \times\left(\left(i e^{i \theta_1} \mu_1\left(\frac{a_1 \sqrt{\epsilon \! + \!2}}{ \text{cosh}(\omega_1)}\right)^{\frac{1}{\epsilon \! + \!1}}\! + \!i e^{i \theta_2}\mu_2 \left(\frac{a_2 \sqrt{\epsilon \! + \!2}}{ \text{cosh}(\omega_2)}\right)^{\frac{1}{\epsilon \! + \!1}}\right)\right.  \right. \\ \left. \left.\! - a_1 e^{i \theta_1}\tanh (\omega_1) \left(\frac{a_1 \sqrt{\epsilon \! + \!2}}{ \text{cosh}(\omega_1)}\right)^{\frac{1}{\epsilon \! + \!1}} \! + \!a_2 e^{i\theta_2} \tanh (\omega_2) \left(\frac{a_2 \sqrt{\epsilon \! + \!2}}{ \text{cosh}(\omega_2)}\right)^{\frac{1}{\epsilon\! + \!1}}\right)  \right. \\ \left.\! - \frac{1}{\epsilon \! + \!2}\left(  e^{i \left(\theta_1\! + \!\theta_2\right)}\left(\left(\frac{a_1 \sqrt{\epsilon \! + \!2}}{ \text{cosh}(\omega_1)}\right)^{\frac{2}{\epsilon \! + \!1}}\! + \!\left(\frac{a_2\sqrt{\epsilon \! + \!2}}{\text{cosh}(\omega_2)}\right)^{\frac{2}{\epsilon \! + \!1}}\right)  \! + \!\left(e^{2 i  \theta_1} \! + \!e^{2i \theta_2}\right)
 \left(\frac{2 a_1 a_2 (\epsilon \! + \!2)}{ \text{cosh}(\omega_1)\text{cosh}(\omega_2)}\right)^{\frac{1}{\epsilon \! + \!1}}\right)^{\epsilon\! + \!2} \right]
\end{multline}

\end{document}